\documentclass[12pt,a4paper]{article}
\usepackage{jheppub}
\usepackage[FIGTOPCAP]{subfigure}
\usepackage{amsmath}
\usepackage{amsfonts}
\usepackage{amssymb}
\usepackage{tensor,epigraph}
\usepackage{hyperref}
\usepackage{cancel}
\usepackage[normalem]{ulem} 
\usepackage{musixtex}
\usepackage{booktabs}
\usepackage{empheq}
\usepackage{amsthm}
\usepackage{comment}
\usepackage{psfrag}
\usepackage[table]{xcolor}
\usepackage{floatrow}
\usepackage{caption}
\usepackage{graphicx}
\usepackage{color}
\usepackage{bbm,bm}

\definecolor{klgreen}{rgb}{0.0, 0.5, 0.0}

\definecolor{color1}{rgb}{1,0,0}
\definecolor{color2}{rgb}{0,0,1}
\definecolor{color3}{rgb}{0,1,0}
\definecolor{color4}{rgb}{0.6,0.4,0.2}
\definecolor{color5}{rgb}{1,0,1}
\definecolor{color6}{rgb}{1,0.5,0}
\definecolor{color7}{rgb}{0.666667,0.666667,1}
\definecolor{color8}{rgb}{0.5,0,0.5}
\definecolor{color9}{rgb}{0.666667,0.666667,0}
\definecolor{green2}{rgb}{0.0, 0.5, 0.0}

\newcommand{\exclude}[1]{}

\newcommand{\beq}{\begin{equation}}
\newcommand{\eeq}{\end{equation}}
\newcommand{\bea}{\begin{eqnarray}}
\newcommand{\eea}{\end{eqnarray}}
\long\def\/*#1*/{}

\newcommand{\dd}{\mathrm{d}}

\usepackage{tikz}

\newcommand{\junk}[1]{}

\title{\Large A Holographic Superfluid Symphony}
\author[1,2]{Daniel Are\'an,}
\author[3]{Matteo Baggioli,}
\author[1,2]{Sebastian Grieninger,}
\author[1]{and Karl Landsteiner}
\affiliation[1]{Instituto de F\'isica Te\'orica UAM/CSIC, Calle Nicol\'as Cabrera 13-15, 28049 Madrid, Spain}
\affiliation[2]{Departamento de F\'isica Te\'orica, Universidad Aut{\'o}noma de Madrid, Campus de Cantoblanco, 28049 Madrid, Spain}
\affiliation[3]{Wilczek Quantum Center, School of Physics and Astronomy, Shanghai Jiao Tong University, Shanghai 200240, China \& Shanghai Research Center for Quantum Sciences, Shanghai 201315.}
\preprint{IFT-UAM/CSIC-21-80}

\emailAdd{daniel.arean@uam.es}
\emailAdd{b.matteo@sjtu.edu.cn}
\emailAdd{sebastian.grieninger@gmail.com}
\emailAdd{karl.landsteiner@csic.es}

\abstract{
We study the hydrodynamic excitations of backreacted holographic superfluids by computing the full set of quasinormal modes (QNMs) at finite momentum and matching them to the existing hydrodynamic theory of  superfluids.
Additionally, we analyze the behavior of the low-energy excitations in real frequency and complex momentum, going beyond the standard QNM picture.
Finally, we carry out a novel type of study of the model by computing the support of the hydrodynamic modes across the phase diagram. We achieve this by determining the support of the 
corresponding QNMs on the different operators in the dual theory, both in complex frequency and complex momentum space. From the support, we are able to reconstruct the hydrodynamic dispersion relations using the hydrodynamic constitutive relations. Our analysis rules out a role-reversal phenomenon between first and second sound in this model, contrary to
results obtained in a weakly coupled field theory framework.
}

\begin{document}

\maketitle
\section{Introduction}

The holographic superfluid model \cite{Hartnoll:2008vx,Hartnoll:2008kx,Herzog:2009xv,Hartnoll:2009sz}, often mislabelled as holographic
superconductor,\footnote{See \cite{Domenech:2010nf} for a nice discussion about the differences in the context of field theory and holography.} is one of the most simple and widely studied setups in the context of applied holography \cite{ammon2015gauge,zaanen2015holographic,Hartnoll:2016apf,Baggioli:2019rrs}.
Its potential application in the understanding of
high-T$_c$ superconductors~\cite{Zaanen2004}
has made this system the object of many thorough analyses
in the years since its inception (see~\cite{Horowitz:2010gk,Cai:2015cya} for a review).
Thus, the lack of a complete study of the quasinormal mode excitations corresponding to hydrodynamic
modes of the holographic superfluid~\cite{Hartnoll_2008} is surprising. 
Until now, the study of the hydrodynamic modes has been limited to the so-called probe limit, where all the gravitational degrees of freedom (and in particular energy and momentum) are kept frozen \cite{Amado:2009ts,Amado:2013xya,Amado:2013aea,Esposito:2016ria}.\\ 
In this work we fill this gap in the literature by
carrying out the first complete study of the quasinormal modes
of the holographic s-wave superfluid with backreaction \cite{Hartnoll_2008}.\footnote{The p-wave superfluid model \cite{Gubser:2008wv,Ammon:2008fc,Ammon:2009fe,Cai:2013pda,Cai:2013aca} with backreaction may be found in \cite{Ammon:2009xh} and the analysis of the associated transport/hydrodynamics in \cite{Erdmenger:2011tj,Erdmenger:2012zu}. At the same time, a holographic model for d-wave superconductivity was introduced in \cite{Chen:2010mk,Benini:2010pr,Kim:2013oba} and with backreaction explored in \cite{Ge:2012vp}.
}
In particular, we numerically compute the spectrum of the
lowest lying quasinormal modes for the original
setup~\cite{Hartnoll_2008}, and also for the modified
system~\cite{Andrade:2014xca,Kim:2015dna,Baggioli:2015zoa,Baggioli:2015dwa} where momentum dissipation is introduced
using the so-called holographic axion model \cite{Andrade:2013gsa,Baggioli:2021xuv}.
We successfully compare our results to the predictions of superfluid hydrodynamics \cite{Herzog:2011ec} extending the results of \cite{Herzog:2009md,Yarom:2009uq}.

This study of the QNMs will allow us to carry out a novel
analysis of the support of the hydrodynamic modes of the
holographic superfluid.
Superfluids display a characteristic coexistence of two different sound modes \cite{doi:10.1063/1.3248499} which can be easily captured within the two-fluid Tisza-Landau model \cite{tisza1938transport,landau1941theory}. In addition to the standard density fluctuations, giving rise to the common first sound excitation, superfluids support also a second sound mode which is driven by temperature
fluctuations. First sound can be visualized as the in-phase motion of the normal fluid component and the superfluid one; on the contrary, second sound is displayed as the out-of-phase collective dynamics of the two. 
In~\cite{Schmitt:2014eka}, a field theoretical study
of the nature of first and second sound as density or entropy waves or a mixture of the two was performed.
Even though in superfluid $^4$He, the nature of first and second sound is preserved for all temperatures, this seems not to be the case for ultra-cold atoms and weakly-interacting bosons \cite{Schmitt:2014eka}.
Here, we carry out a similar analysis within the holographic superfluid which stands as a putative toy model for a strongly coupled superfluid system:
we determine the support of the hydrodynamic modes 
of the system on excitations of the different operators
present in the dual theory.

In order to analyze the nature of the hydrodynamic
excitations of the holographic superfluid we 
define a quantitative measure of the support that a
quasinormal mode excitation has on the different dual
operators. We then apply this procedure to the sound and diffusive modes present in our gravity dual.
Our results rule out any role reversal between first and second sound and place the holographic model in the same phenomenological class of standard superfluid $^4$He. This is in contrast to the weakly coupled bosonic field theory results of \cite{Schmitt:2014eka} described above.
Naturally, an obvious difference between both setups
is given by the inherently strongly coupled nature
of the holographic description. Moreover, \cite{Schmitt:2014eka} is neglecting dissipating effects in their description.

In a final part of this work we study the response of the system from a different perspective.
One is typically
interested in the poles of (the Fourier-transformed) Green's functions in the complex frequency and real momentum space.
These are the QNMs.
Nevertheless one can also consider the poles in the space
of real frequency and complex momentum. These modes
carry information about the response of the system to a 
perturbation of the form
$e^{-i \omega t+ i k x}$ where $\omega$ now is a real number and $k$ a complex one. Although this approach is not usual in holography, it is a standard practice in the study of propagation of waves, in particular in relation to absorption and reflection properties of media (\textit{e.g.} EM waves in a medium \cite{jackson_classical_1999}). Once momentum is decomposed as $k=\mathrm{Re}(k)\,+\,i\,\mathrm{Im}(k)$, its real part defines the propagation wavelength in real space while its imaginary part the so-called penetration length $\lambda\equiv 1/\mathrm{Im}(k)$. In other words, this second setup corresponds to considering the decay of a wave in space rather than in time.

The curves $\omega(k)$ described by the QNMs in the
complex plane feature a complicated structure
(\textit{e.g.} the presence of different sheets,
and singularities). Hence, 
\textit{a priori} it is not clear that 
the behavior of the modes at real frequency and complex momentum can be obtained straightforwardly from that of the standard QNMs. Indeed, notice that these two types of modes encode
responses to perturbations with different boundary
conditions. Consider for instance
an elastic rod immersed in a viscous liquid. Considering complex frequency and real momentum corresponds to keeping the extreme of the rod fixed and creating an oscillating wave on it. The wave in that case will not propagate in space but just oscillate (and decay) in time. On the contrary, one could imagine exciting a wave at one of the extremes of the rod and studying its propagation along the rod.
That is equivalent to considering real frequency and complex momentum.

In the context of holography, the modes at real frequency and complex momentum have been overlooked and only few works considered this second setup \cite{Amado:2007pv,Amado:2008ji,Forcella:2014dwa,Blake:2014lva,Sonner:2017jcf}.
On a different note, recently, several studies  of the modes in the complex frequency/complex momentum plane have been performed in relation to the so-called pole-skipping phenomenon \cite{Grozdanov:2017ajz} and the convergence radius of the linearized hydrodynamics expansion \cite{Grozdanov:2019kge}.
In the last part of our work, we study the excitations of a holographic superfluid at real frequency and complex momentum and discuss their physical interpretation in detail.

\subsubsection*{Structure of the paper.} In section \ref{sec1}, we first summarize the main results of the
hydrodynamic description of a 
relativistic superfluid. We then introduce the
holographic model and set the framework of the analysis
we carry out in the remainder sections.
In section \ref{sec:transp}, we compute the quasinormal modes of the backreacted holographic superfluid model numerically and use the results to match
holographic and hydrodynamic 
predictions for the system.
In section \ref{sec:tomog}, we perform a new analysis of the support of the various hydrodynamic modes in terms of the dual field theory operators. 
Finally, in section \ref{sec4}, we compute the excitations of the system and their main properties by considering real frequency and complex momentum. 
We conclude in section \ref{sec5}
with a summary of our results and some comments for the future.

\section{Superfluids in hydrodynamics and holography}\label{sec1}
In this section, we first review the main features of the hydrodynamic description of a conformal superfluid. Next, we
introduce the holographic dual of a superfluid
and set up the tools for the analysis performed in the 
remaining of this work.

\subsection{Superfluid hydrodynamics, a brief summary}
\label{sechydro}
We shall briefly summarize the main features of superfluid hydrodynamics \cite{doi:10.1063/1.1703944,Schmitt:2014eka,1974anh.....3.....P,Nicolis:2011cs,Son:2002zn,Bhattacharya:2011tra},
with a focus on the properties of the different
hydrodynamic modes.
For connections between superfluid hydrodynamics, holography and effective field theory see \cite{Herzog:2011ec,Herzog:2009md,Yarom:2009uq,Son:2002zn,Andersen:2002nd,Nicolis:2011cs,Sonner:2010yx,Escobedo:2010uv,Berezhiani:2020umi,Landry:2020ire,Landry:2021kko,Pajer:2018egx,Delacretaz:2019brr}.\\

Let us start by considering a relativistic charged fluid with conserved momentum \cite{Kovtun:2012rj,Son:2006em}. The longitudinal spectrum contains two sets of hydrodynamic modes. First, charge and energy conservation imply the presence of a thermoelectric diffusive mode $\omega=-i D_q k^2$ whose diffusion constant  is explicitly given later on in eq.~\eqref{eq:normaldiff}. Additionally, we have a pair of propagating first sound modes $\omega=\pm c_1\,k-\,i\,\Gamma_1\,k^2$ which result from the interplay of longitudinal momentum and energy. In a conformal system, their propagation speed is given by
\begin{equation}
    c_1^2=\frac{\partial p}{\partial \epsilon}=\frac{1}{2}\label{speed1}
\end{equation}
where $\epsilon$ and $p$ are the energy density and the thermodynamic pressure, respectively. Additionally, the attenuation constant reads
\begin{equation}
    \Gamma_1=\frac{\eta}{2\,\chi_{\pi\pi}}\,,
    \label{att1}
\end{equation}
with $\eta$ being the shear viscosity and $\chi_{\pi\pi}=\epsilon+p$ the momentum susceptibility.\\
Below the critical temperature, a superfluid may be viewed in the two-fluid picture as composed of two components: 
a normal component as the (relativistic) fluid just described; and a condensed
or superfluid component that flows without dissipation.
Each component has its own velocity and density.
In the normal phase, the interplay of charge,
    energy and longitudinal momentum gives rise to a propagating mode: first sound, and a diffusive mode.
    In the superfluid phase $T<T_c$, first sound remains in the spectrum, the diffusive mode gets a purely
    imaginary gap, and the Goldstone fluctuations, coupled to those of charge and energy give rise to a new
    propagating mode: second sound $\omega=\pm c_2\,k-\,i\,\Gamma_2\,k^2$, known as second sound \cite{doi:10.1063/1.3248499}. Its speed and attenuation constant are given by 
\begin{align}
    &c_2^2\,=\,\frac{\rho_s\,\xi^2}{\left(\mu \,\rho_n +s T\right)\,\frac{\partial \xi}{\partial T}}\,,\label{speed2}\\
    &\Gamma_2\,=\,\frac{\mu\,\rho_s}{2\,\chi_{\pi\pi}\,\left(\mu\,\rho_n+sT\right)}\,\eta\,+\,\frac{\mu\,\chi_{\pi\pi}}{2\,T^3\,\rho^2\,(\partial \xi/ \partial T)_{|_\mu}}\,\kappa\,+\,\frac{\rho_s\,\chi_{\pi\pi}}{2\,\mu\left(\mu \rho_n+ sT \right)}\,\zeta\,.\label{att2}
\end{align}
Here we have defined: the superfluid density $\rho_s$, the normal density $\rho_n=\rho-\rho_s$, the reduced entropy $\xi=s/\rho$, the diffusivity $\kappa$ and the superfluid diffusivity $\zeta$.\\
Notice the only difference with the results presented in \cite{Herzog:2011ec}, and valid for $d=4$, is the numerical constant in front of the term proportional to the shear viscosity $\eta$.\\

In the normal phase, $\rho_s=0$, the first sound remains in the spectrum and the second sound becomes the thermoelectric diffusive mode \cite{Hartnoll:2014lpa,Kovtun:2014nsa}
\begin{equation}
    \omega\,=\,-i D_q k^2\,=-i\,k^2\,\frac{\kappa\,\mu\,\chi_{\pi\pi}}{T^3\,\rho^2\,(\partial \xi/ \partial T)_{|_\mu}}\,.\label{eq:normaldiff}
\end{equation}
Once momentum conservation is broken, the first sound in the normal phase gets substituted by the energy diffusion mode $\omega=-i D_e\,k^2$ whose diffusion constant obeys the standard Einstein relation. 
At the same time, in the superfluid phase, the second sound mode is replaced by the so-called fourth sound first predicted by~\cite{PhysRev.73.608,PhysRev.113.962} where we keep the normal component stationary. The dispersion relation of fourth sound reads
$\omega=\pm c_4\,k-\,i\,\,\Gamma_4\,k^2$, where the velocity and attenuation constant are given by
\begin{align}
   & c_4^2\,=\,\frac{\rho_s}{\mu\,\left(\frac{\partial_\rho}{\partial_\mu}\right)}\,,\qquad \Gamma_4\,=\,\frac{\kappa}{2\,T\,\chi_{qq}}\,+\,\frac{\rho_s}{2\,\mu}\,\zeta\,. \label{four}
\end{align}
For completeness, we list here all the Kubo formulas used to extract the various transport and thermodynamic parameters entering in the dispersion relations above
\begin{align}
    &\eta= \lim_{\omega \rightarrow 0}\frac{1}{\omega}\mathrm{Im}\,\mathcal{G}_{T_{xy}T_{xy}}(\omega,0)\,,\\
    &\frac{\kappa}{T}=\lim_{\omega \rightarrow 0}\lim_{k \rightarrow 0}\frac{1}{k}\mathrm{Im}\,\mathcal{G}_{J_{x}J_{t}}(\omega,k)\,,\\
      &\zeta=\lim_{\omega \rightarrow 0}\omega\,\mathrm{Im}\,\mathcal{G}_{\phi\phi}(\omega,0)\,,
\end{align}
together with the expression for the electric conductivity
\begin{equation}
    \sigma(\omega)=\frac{i}{\omega}\mathcal{G}_{J_xJ_x}(\omega,0)=\left(\frac{i}{\omega}+\delta(\omega)\right)\left[\frac{\rho_n^2}{\mu\,\rho_n+s\,T}+\frac{\rho_s}{\mu}\right]+\sigma_0\,.\label{eq:rhos1}
\end{equation}
For slow momentum dissipation this two-point function is modified to
\begin{equation}
    \sigma(\omega)\,=\,\frac{i}{\omega}\,\mathcal{G}_{J_xJ_x}(\omega,0)\,=\,\frac{\rho_n^2}{\mu\,\rho_n+s\,T}\,\frac{1}{\Gamma\,-\,i\,\omega}\,+\frac{i}{\omega}\,\frac{\rho_s}{\mu}+\sigma_0\,.\label{eq:rhos2}
\end{equation}
\\
In this discussion, $\phi$ refers to the superfluid Goldstone mode which is dual to the phase of the complex bulk scalar (see next section for the details). Also, $\mathcal{G}_{AB}$ is the retarded Greens function of the operators $A$ and $B$. These Kubo formulas will be used in 
section~\ref{sec:transp} when we analyze the transport properties of holographic superfluids.

\subsection{The holographic model}
\label{ssec:holom}
We consider the standard bulk action for a s-wave holographic superfluid \cite{Hartnoll:2008vx,Hartnoll_2008}
\begin{equation}
\label{eq:action}
S\,=\,\frac{1}{2\kappa_4^2}\int \dd^{d+1}x\, \sqrt{-g}\left[R-2\Lambda-\frac{1}{4}F_{mn}F^{mn}-|D\psi|^2-M^2|\psi|^2\right]\,,
\end{equation}
where $\psi$ is a complex scalar, $F\equiv \dd A$, and the covariant derivative is defined as $D_m\psi\equiv (\partial_m -i\, q\, A_m)\, \psi $. We consider the 4-dimensional bulk case $d=3$, and fix $\Lambda=-3$ and $2\kappa_4^2=1$. The AdS radius $L$ has also been set to one for convenience.\\
The resulting equations of motion are
\begin{subequations}
\begin{align}
&R_{mn}-\frac{1}{2}\left(R-2\Lambda-\frac{1}{4}F^2-|D\psi|^2-M^2|\psi|^2\right)g_{mn}\nonumber\\
&=\frac{1}{2}F_{mk}F_{n}^{\;\;k}+
\frac{1}{2}\left[D_m\psi (D_n\psi)^\star +D_n\psi (D_m\psi)^\star\right]\,,\\
&\frac{1}{\sqrt{-g}}\,\partial_m\left(\sqrt{-g}\, F^{mn}\right)-iq(\psi^\star D^n\psi-\psi D^n\psi^\star)=0\,,\\
&\frac{1}{\sqrt{-g}}\,D_m \left(\sqrt{-g}\,D^m\psi\right)-M^2\psi=0\,.
\end{align}
\label{eq:eoms}
\end{subequations}
We consider the following ansatz for the background
\begin{align}
&\dd s^2=\frac{1}{z^2}\left[-u(z)\,\dd t^2-2\, \dd t \dd z+g(z)\, (\dd x^2+\dd y^2)\right],\\ &A=A_t(z)\,\dd t\,,\qquad \psi=\psi_1(z)-i \,\psi_2(z)\,,
\end{align}
in Eddington-Finkelstein coordinates $\{t,z,x,y\}$ with the AdS boundary located at $z=0$ and the black brane horizon at $z_h=1$.\\
Assuming this ansatz, the temperature and chemical potential of the dual field theory are given as
\begin{equation}
    T\,=\,-\frac{u'(1)}{4\pi}\,,\qquad \mu\,=\, A_t(0)- A_t(1)\,.
\end{equation}
For numerical convenience, we impose $A_t(1)=0$.  Furthermore, we choose the value of the background condensate to be real by setting $\psi_2^{(2)}$ to zero (see \eqref{eq:onepoints} for the definition of the expectation values)\footnote{We add a factor of 2 to the expectation values of the scalars $\psi$ to account for the non-canonical normalization in the action~\eqref{eq:action}.}
\begin{equation}
    \langle \mathcal O^\psi_\text{cond}\rangle =2\,\psi^{(2)}_1.
\end{equation}

The analysis in sections \ref{sec:transp}
and \ref{sec:tomog} will rely on the
study of the fluctuations about the background discussed above.
In particular, we switch on the full set of longitudinal perturbations
$ g_{mn}=\hat g_{mn}+\epsilon\, h_{mn}/z^2,\, A_m=\hat A_m+\epsilon\, a_m,\, \psi_i=\hat\psi_i+\epsilon\, \delta\psi_i$
\begin{equation}
    \{h_{tt},\,h_{tx},\,h_{xx},\,h_{yy},\,a_t,\,a_x,\,\delta \psi_1,\, \delta \psi_2\}\,,
    \label{eq:flucts}
\end{equation}
where we choose the momentum to be aligned with the $x$-direction without loss of generality. From now on, we will add to all symbols referring to background quantities a hat $\hat A$. Moreover, we assume the radial gauge\footnote{Note that since we work in Eddington-Finkelstein coordinates, the bulk solution for the scalar field is complex even though we impose the $\psi^{(2)}_2$ to be zero.}
\begin{equation}
    h_{zi}=0=a_z.
\end{equation}
We decompose all our fluctuations as
\begin{equation}
    \mathfrak{f}(t,z,x)\,=\,e^{-i \omega t + i k x}\,\mathfrak{f}(z)\,.
    \label{eq:harmonic}
\end{equation}
The quasinormal modes and all the transport coefficients can be obtained following standard methods. To compute the thermodynamic derivatives accurately, for example the one appearing in eq.~\eqref{speed2}, we re-write the derivative in terms of dimensionless quantities; we then compute the background on a Chebychev grid in that variable (for example $T/\mu$) and use the spectral differentiation matrices to compute the thermodynamic derivative (see appendix B.4 in~\cite{Ammon:2020rvg} for more details). \\

In order to introduce momentum dissipation and study the properties of fourth sound, we supplement the action
\eqref{eq:action}
by a linear axion term \cite{Andrade:2013gsa}
\begin{equation}
    S_{2}\,=\,-\frac{1}{2\kappa_4^2}\,
\int \dd^{d+1}x\,\sqrt{-g}\, \partial_m \phi^I \partial^m \phi^I
\end{equation}
where the index $I$ runs only along the spatial directions $I=x,y$, and the background profile for the scalars is
\begin{equation}
    \hat\phi^I\,=\,\alpha\,x^I\,.
\end{equation}
In the dual field theory, this corresponds to switching on an operator which breaks translational invariance explicitly.
For more details about this model we refer to the recent review \cite{Baggioli:2021xuv}. In presence of momentum dissipation, we add the fluctuation of the longitudinal axion field $\delta\phi^x$ to the fluctuations in eq.~\eqref{eq:flucts}.\\

Finally, throughout all the manuscript we set $q=1$ and $M^2=-2$.
\section{Transport \& hydrodynamic modes}
\label{sec:transp}
In this section, we explicitly verify that the hydrodynamic framework described in the previous section (see \cite{Herzog:2011ec} for the original reference, albeit in one dimension more) is indeed the correct hydrodynamic description of the holographic superfluid model \cite{Hartnoll:2008vx}. Surprisingly, even though partial checks may be found in the literature \cite{Herzog:2011ec,Yarom:2009uq,Herzog:2009md}, a complete matching of the two pictures considering the full backreaction \cite{Hartnoll:2008kx} in the gravitational model has not appeared yet. 
\begin{figure}[ht!]
     \centering
     \includegraphics[width=0.49\linewidth]{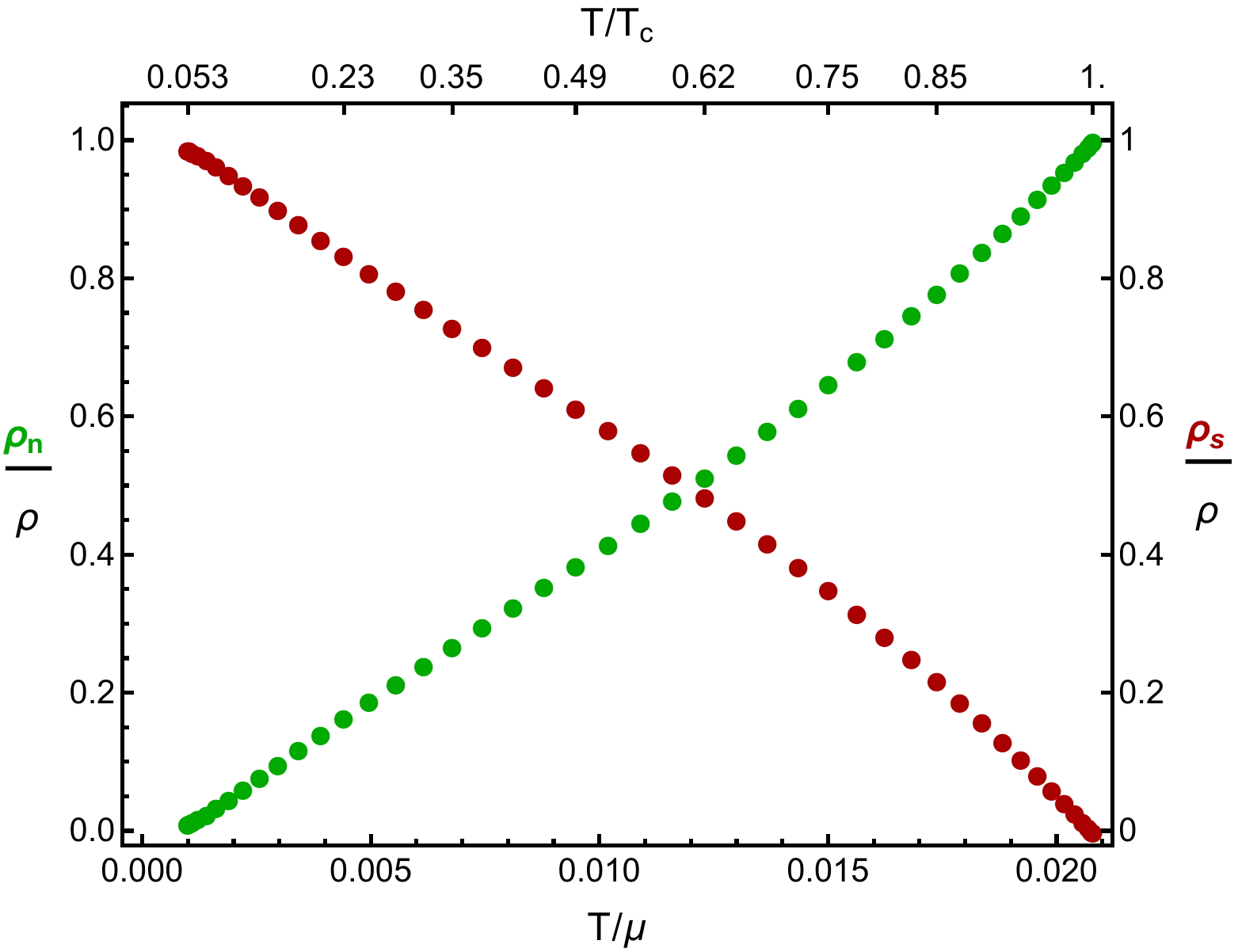}
       \includegraphics[width=0.49\linewidth]{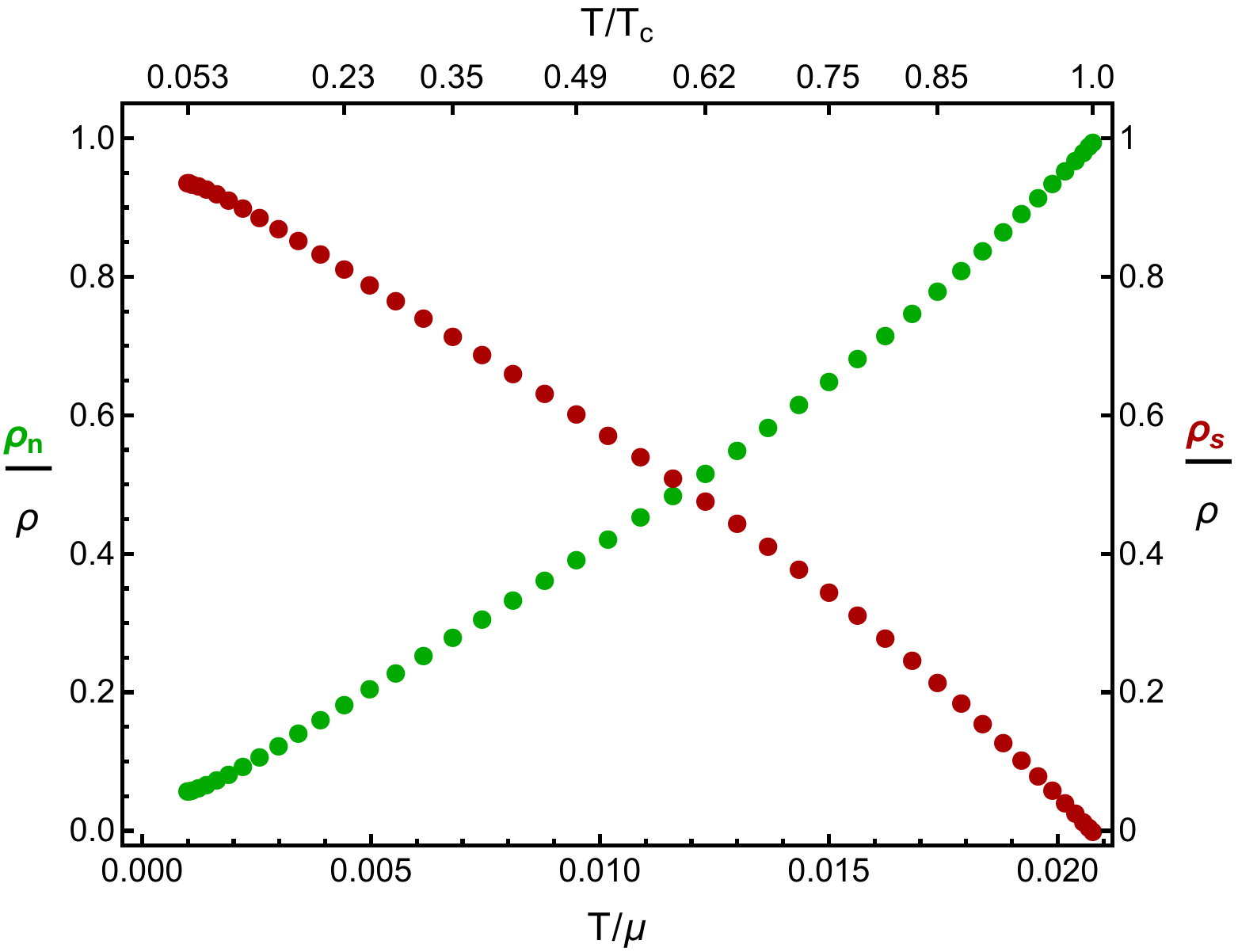}
     \caption{The normal and superfluid densities $\rho_n$ and $\rho_s$ as function of the temperature. \textbf{Left:} Translational invariant theory. \textbf{Right:} Model with broken translations, $\alpha/\mu=1/10$.}
     \label{fig:densities}
 \end{figure}\\
 We start by plotting the normal and superfluid densities as function of $T/T_c$ in the superfluid phase. The normal and superfluid density, respectively, may be extracted numerically from eq.~\eqref{eq:rhos1} (translationally invariant case) and eq.~\eqref{eq:rhos2} (broken translations). While in the latter case, the superfluid density is simply given by the $i/\omega$ pole in the conductivity, we have to use that $\rho_n=\rho-\rho_s$ and solve eq.~\eqref{eq:rhos1} for $\rho_s$ in the translationally invariant case. The results are shown in fig.~\ref{fig:densities} for both the translational invariant model and the model where translations are broken explicitly using axion fields \cite{Baggioli:2021xuv,Baggioli:2016rdj}. As it should, the superfluid density vanishes at
 the critical temperature for both models.
In the translational invariant case, we also
 observe that the normal density vanishes at zero temperature.
This last feature holds in superfluid Helium
  and BCS superconductors, while a residual normal density at zero temperature has been reported in some high-Tc superconductors~\cite{PMID:27535534}. Interestingly, 
  ~\cite{Gouteraux:2019kuy} has shown that in quantum
  critical superfluids the nature of the IR fixed
  point of the theory is crucial for the fate of
  the normal density: it vanishes for fixed points with an emergent  Lorentz symmetry
  as in our translation invariant case~\cite{Horowitz:2009ij}
  , while some
  quantum critical points of the hyperscaling-Lifshitz type
  can support a nonvanishing normal
  density~\cite{Gouteraux:2020asq}.
The model breaking translations
  (right panel of fig.~\ref{fig:densities}) shows a slower decay of the normal density for $T\to0$ and a fit of the numerical data shows a scaling of $\rho_n/\rho\sim 0.05265+3.14843 (T/\mu)^{1.2583}$ for $T/\mu\in[5\cdot10^{-7},5\cdot10^{-6}]$ indicating a non-zero normal density in the zero temperature limit. It would be interesting to characterize 
  the IR geometry of our model and 
  analyze it in the light of the results of~\cite{Gouteraux:2019kuy}. We leave a more accurate analysis of this feature for the future.
\\

\begin{figure}[ht!]
     \centering
     \includegraphics[width=0.47\linewidth]{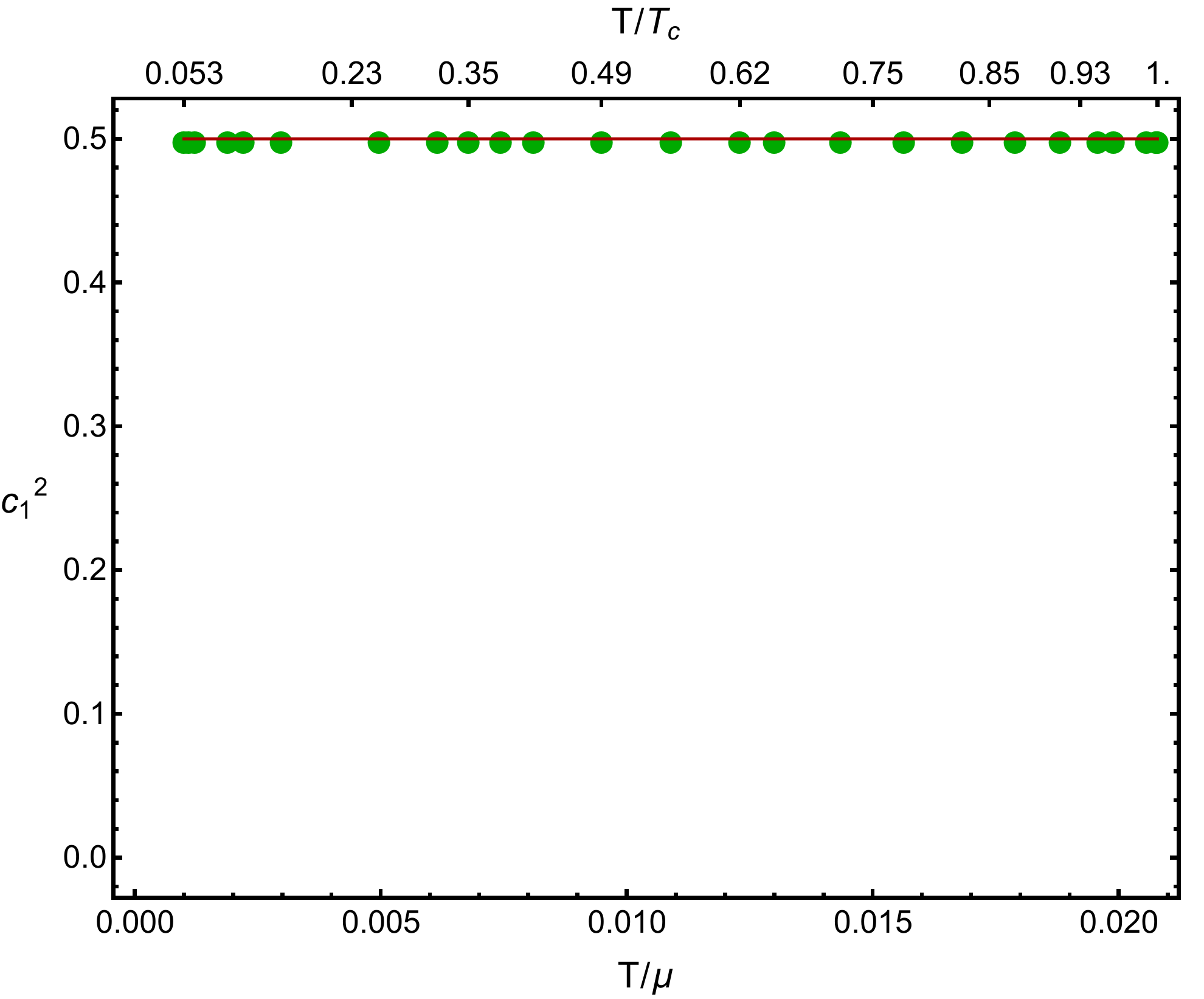}
       \includegraphics[width=0.49\linewidth]{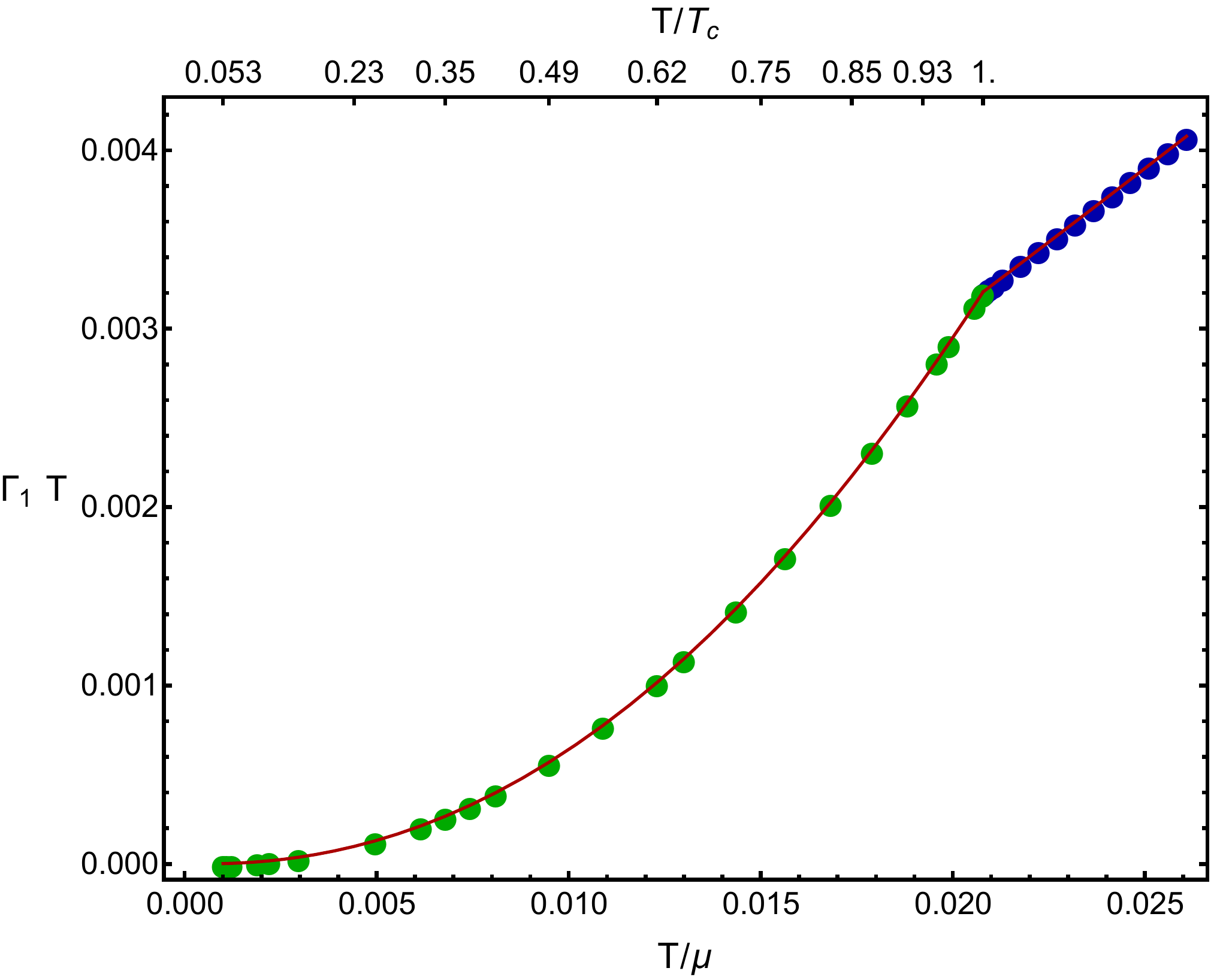}
     \caption{The propagation speed and the attenuation constant of first sound as function of $T/T_c$ above and below the critical point. The red lines are the hydrodynamic formulas \eqref{speed1}, \eqref{att1} and the colored circles the numerical data extracted from the QNMs. Blue color indicates the normal phase. The plot for the speed is not extended into the normal phase since the speed is trivially constant in the full range of $T$ due to the conformal symmetry.}
     \label{fig:firstsound}
 \end{figure}

\begin{figure}[ht!]
     \centering
     \includegraphics[width=0.474\linewidth]{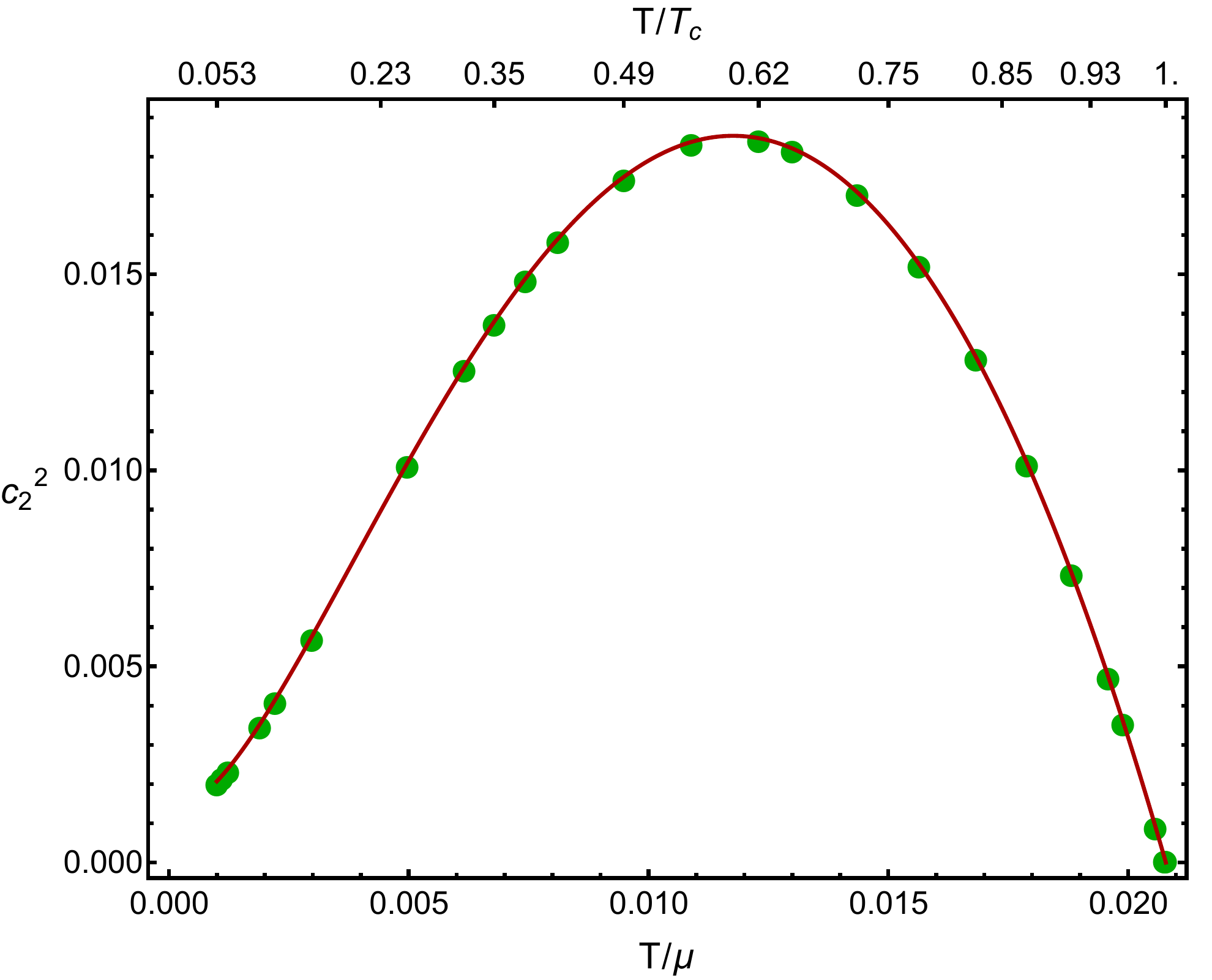}
       \includegraphics[width=0.49\linewidth]{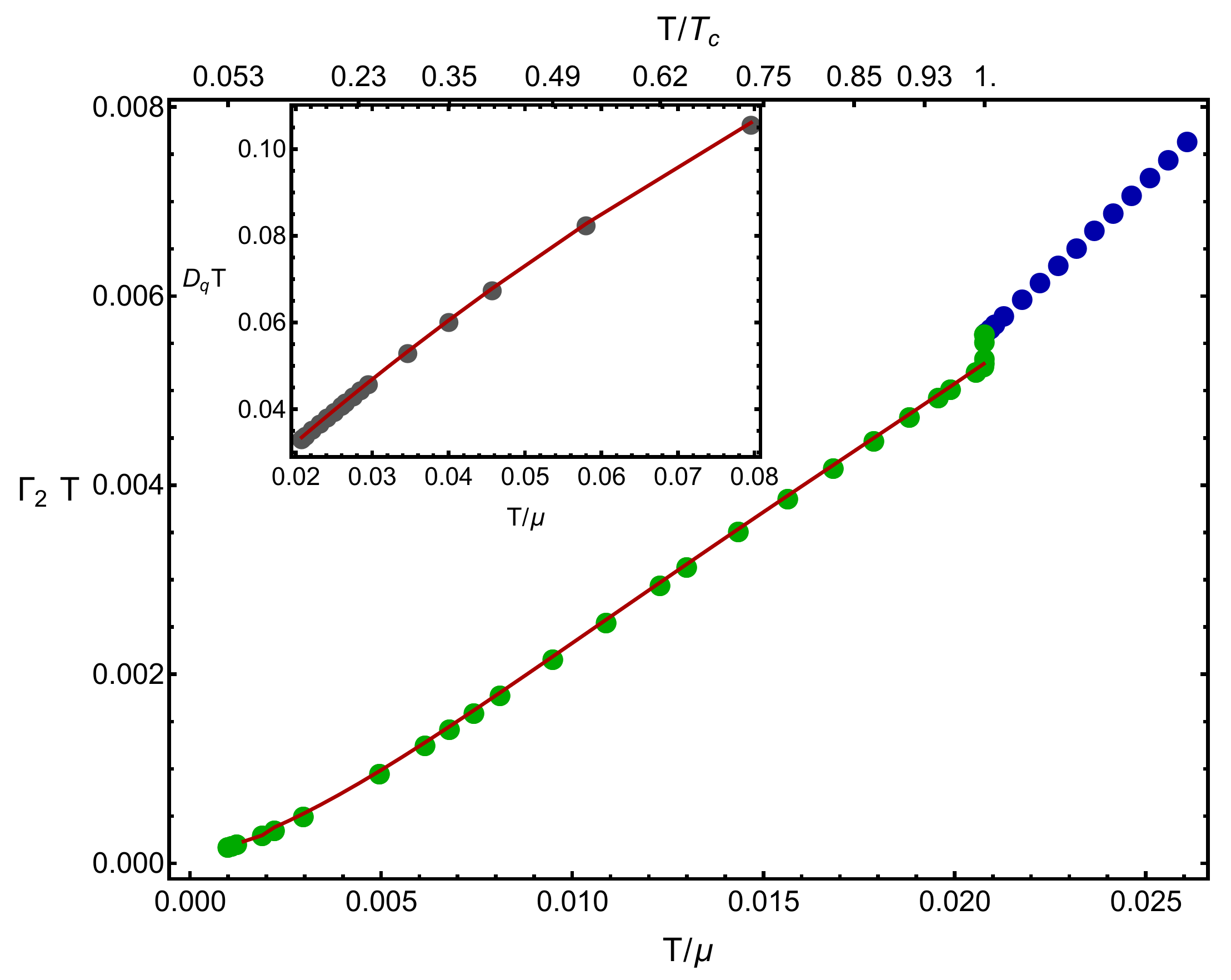}
     \caption{The propagation speed and the attenuation constant of second sound as function of $T/T_c$ above and below the critical point. The red lines are the hydrodynamic formulas \eqref{speed2}, \eqref{att2} and the colored circles the numerical data extracted from the QNMs. Blue color indicates the normal phase. The inset in the right panel shows the diffusion constant of the purely diffusive mode in the normal phase together with its hydrodynamic expression.}
     \label{fig:secondsound}
 \end{figure}

 We then move to the study of the sound modes. 
 On the gravity side these correspond to quasinormal
 modes of the black hole geometry. As we review in  section~\ref{sec:qnm}, with further technical details in appendix~\ref{app:numerics}, we obtain them by numerically solving the linearized equations of motion for the fluctuations~\eqref{eq:flucts}.
 This will allow us to compute the speed and attenuation
 of the different sound modes and compare them against
 the hydrodynamic predictions of section~\ref{sechydro}.
 The propagation speed and attenuation constant of the first sound mode are shown in fig.~\ref{fig:firstsound}. 
  The propagation speed is fixed by conformal symmetry to the expected value $c_1^2=1/2$ and it is continuous across the phase transition at $T=T_c$. Its dimensionless attenuation constant $\Gamma_1 T$ is shown in the right panel of fig.~\ref{fig:firstsound} and it is in perfect agreement with the hydrodynamic formula in eq.~\eqref{att1} in which all the parameters are independently computed using thermodynamic relations and transport Kubo formulas. One should notice that the dimensionless attenuation constant vanishes at zero temperature, indicating that at $T=0$ first sound is an infinite living excitation. This is tantamount to saying that the shear viscosity of the system vanishes at $T=0$. 
 Finally, we observe that the attenuation constant of first sound is continuous at $T=T_c$ even though its first derivative is not.
  \\
 We perform an analogous analysis for second sound in fig.~\ref{fig:secondsound}. We remind
 the reader that the second sound mode exists only in the superfluid phase below the critical temperature $T<T_c$. Interestingly, we observe a very non-monotonic behavior for the second sound speed which vanishes both at the critical temperature $T=T_c$ and at zero temperature $T=0$, with a maximum around $T=0.6\,T_c$. Apparently, this behavior is not universal and it rather depends on the charge and the conformal dimension of the dual scalar operator \cite{Herzog:2009md}. Again the hydrodynamic formula \eqref{speed2} fits the numerical data very well.\\
 
 The attenuation constant of second sound is more interesting. First, it vanishes at zero temperature. Second, across the phase transition it is connected to the imaginary part of the gapped scalar modes present in the normal phase \cite{hydroSC} and responsible for the superfluid instability at $T=T_c$. Nevertheless, the data display a clear and sharp jump around $T=T_c$ where our hydrodynamic description is no longer applicable as is evident in the right panel of fig.~\ref{fig:secondsound}. Moreover, the attenuation constant of second sound is not continuously connected to the diffusion constant which we present in the inset in the same figure \ref{fig:secondsound}. Our numerical observation was confirmed by the authors of~\cite{Donos:2021pkk} in a simpler setup (without chemical potential) using analytic arguments.

\begin{figure}[ht!]
     \centering
     \includegraphics[width=0.474\linewidth]{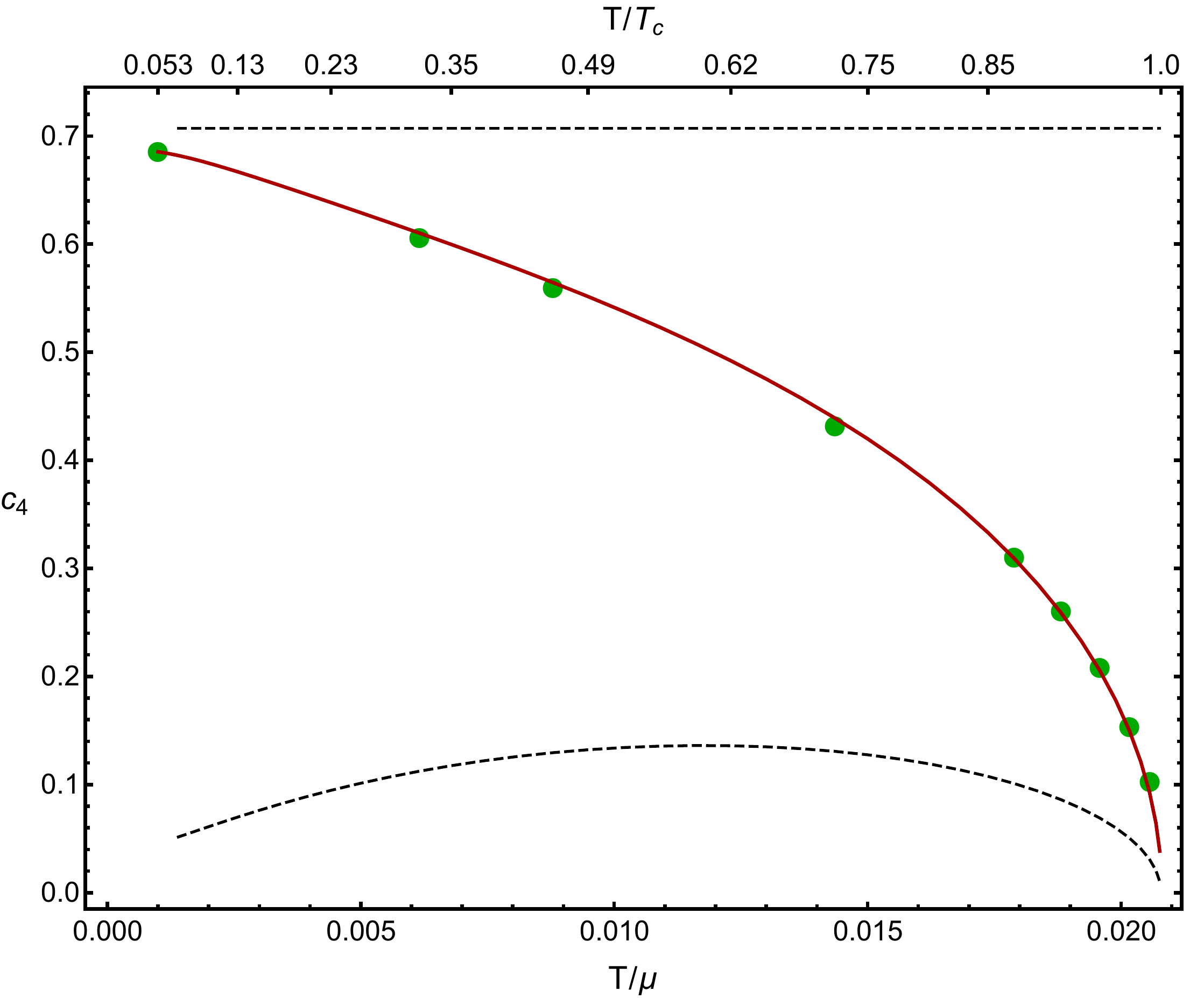}
       \includegraphics[width=0.49\linewidth]{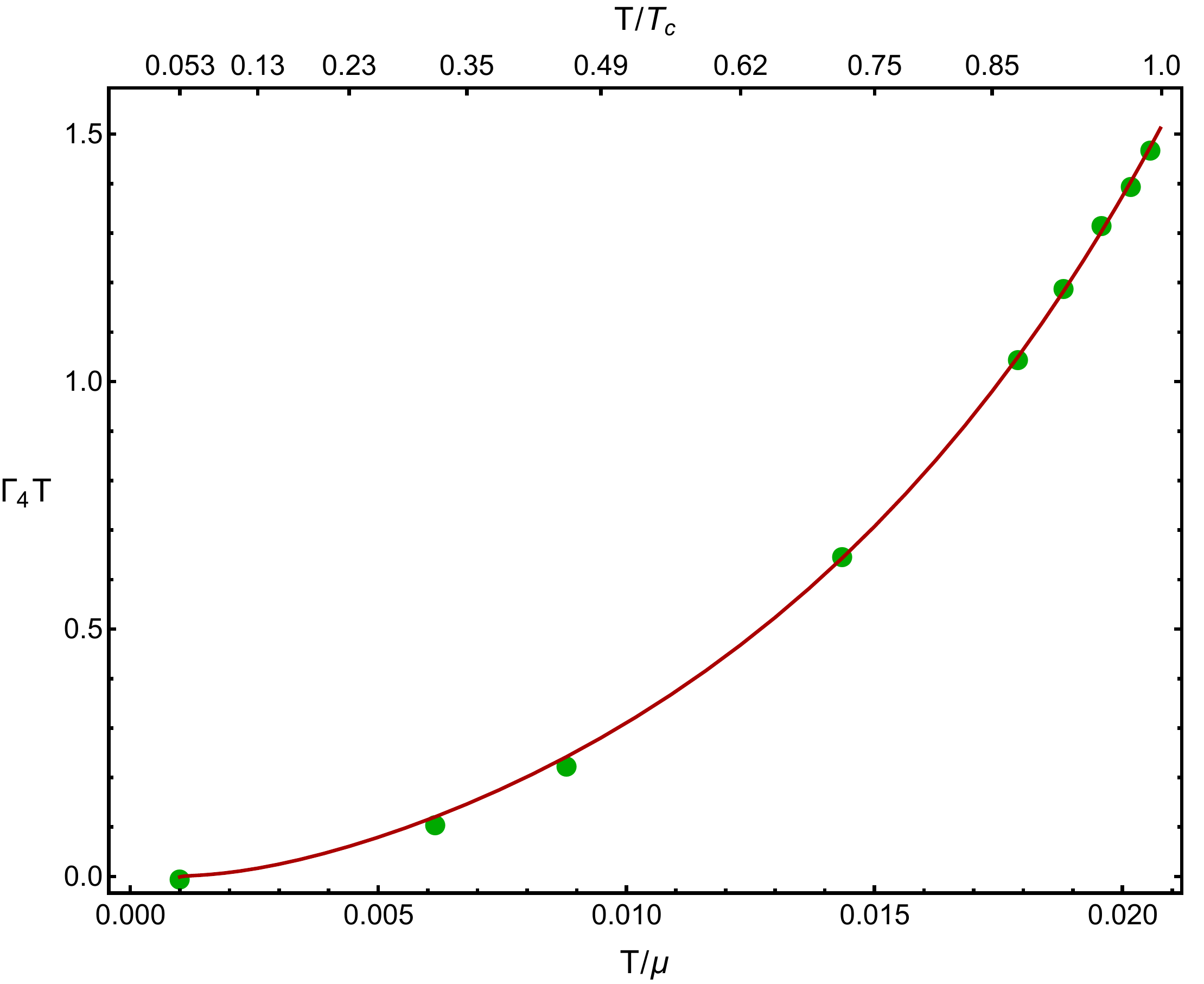}
     \caption{The propagation speed and the attenuation constant of fourth sound in function of $T/T_c$ below the critical temperature for $\alpha/\mu=0.1$. The red lines are the hydrodynamic formulas \eqref{four} and the colored circles the numerical data extracted from the QNMs. The dashed lines in the left panel are the speed of first and second sound for $\alpha/\mu=0$.}
     \label{fig:fourthsound}
 \end{figure}

 Finally, when momentum is dissipated and the normal component is prevented from flowing, a new excitation called fourth sound appears in the superfluid \cite{PhysRev.113.962,PhysRev.73.608}. We obtained the speed of propagation and the attenuation constant numerically in fig.~\ref{fig:fourthsound} and compared them with the hydrodynamic expressions in eq.~\eqref{four}. As expected the speed of fourth sound interpolates between the speed of second sound at the critical temperature $T_c$ to the speed of first sound for $T \rightarrow 0$ \cite{PhysRev.113.962,PhysRev.137.A1383}. Additionally, notice how the attenuation constant of fourth sound is much larger than those of first and second sounds. Finally, we observe that the speed and attenuation constant of fourth sound have the same qualitative behavior as those of the sound mode obtained in the probe limit in \cite{Amado:2013xya}. We do expect the two to coincide in the limit of very fast momentum dissipation, $\alpha/T \gg 1$.\\
 
 To conclude this first part, let us iterate that the hydrodynamic description presented in the previous section perfectly matches the numerical results obtained from the holographic superfluid model both in the normal and superfluid phases with and without momentum dissipation.
\section{Tomography of superfluid sound modes}
\label{sec:tomog}
After verifying the hydrodynamic theory in our holographic model explicitly, we now move on to investigate the so-called support of the hydrodynamic modes.
\subsection{Theory}
\label{sec:qnm}
A quasinormal mode is a solution to the linearized equations of motion in a black hole background with in-going boundary conditions on the horizon and normalizable ones at the  boundary of AdS \cite{Berti:2009kk,Nollert_1999}. In general, such a solution will consist of a collection of fields which we denote by $\Phi^I$. For a single QNM all these fields oscillate with the same frequency and decay exponentially with the same decay rate. At late times and close to equilibrium, we may write a generic QNM solution as a superposition of modes 
\begin{equation}\label{eq:qnmsol}
\phi^I(z,t,\vec x) = \int\frac{\dd^dk}{(2\pi)^d}\sum_n A^I_n(z,\vec k) e^{-i\Omega_n(\vec k) t} e^{-\Gamma_n(\vec k) t} e^{i \vec k \vec x}\,.
\end{equation}
The bulk wave functions of a QNM $A^I_n(\vec k,z)$ do not contain the leading (non--normalizable) terms of the asymptotic boundary condition. We can therefore multiply the field $\phi^I$ with the appropriate power of $z$ such that $A_n^I(0,\vec k)$ is the coefficient of the non-normalizable mode. According to the holographic dictionary the boundary value of $\Phi_n^I$ encodes the expectation value of an operator $\langle O^I(t,\vec x)\rangle $ (see appendix \ref{app1}).  Since the
solution (\ref{eq:qnmsol}) evaluated at the boundary represents a propagating and attenuated sound wave in the dual field theory, it is instructive 
to think about the quasi-normal mode as corresponding to a hydrodynamic sound wave. One should
notice that sound waves consist of more than one perturbation, for example the perturbations of the energy density $\epsilon$ and the pressure $p$ (as is clear from the thermodynamic formula for the speed of sound $c_s^2 = \partial p/\partial \epsilon$). Hence, we expect the QNM solution 
dual to a sound mode to consist of more than one field $\Phi^I$ and taking the form of eq.~\eqref{eq:qnmsol}.
\\

To simplify the analysis further, we assume that rotational symmetry is not broken. Then the quasinormal frequencies are function of $|\vec k|^2$ only. From now on we write $k$ for $|\vec k|$. Since the collection $\Phi^I$ includes vector fields and metric components the dependence of the amplitudes may be more complicated in general. \\

Demanding that the fields $\Phi^I$  be real results
in the following condition for the quasi-normal frequencies and amplitudes
\begin{equation}
(\phi^I)^* = \phi^I = \int\frac{\dd^dk}{(2\pi)^d}\sum_m A_m^{I*}(z,-\vec k) e^{i\Omega_m(k) t} e^{-\Gamma_m(k) t} e^{i \vec k \vec x}\label{eq:qnmcomplex}\,.
\end{equation}
Combining this constraint with (\ref{eq:qnmsol}) leads to the condition that the quasinormal fequencies come in pairs $(n,m)$ with
\begin{align*}
\Omega_m = - \Omega_n\,,\qquad
\Gamma_m = \Gamma_n\,,\qquad
A_m^{I*}(z,-\vec k) = A_n(z,\vec k)\,,
\end{align*}
or are purely damped with vanishing $\Omega_n$ and $A_n^{I*}(z,-\vec k) = A_n^I(z,\vec k)$.

In view of this structure, the minimal way of constructing a real QNM solution
for $\Phi^I$ is to excite the mode $n$ at wave number $\vec k$ and the mirror mode $m$ with $\Omega_m=-\Omega_n$ at wave number $-\vec k$.
\begin{equation}
\Phi^I(z,t,\vec x) = e^{-\Gamma_n t}\left[ A^I(z,\vec k)\, e^{-i \Omega t + i\vec k\vec x} + A^{I}(z,\vec k)^* e^{i \Omega t - i\vec k\vec x}\right]\,.
\end{equation}
If we evaluate this at the boundary, we get the space-time evolution of the set of operators $O^I$ as
\begin{equation}\label{eq:opqnm}
\langle O^I(t,\vec x)\rangle = e^{-\Gamma_n t} \langle O^I(0,\vec k)\rangle\cos(\Omega_n t - \vec k \vec x + \alpha^I_n)\,,
\end{equation}
where we take $\langle O^I(0,\vec k)\rangle$ to be positive since
the phase $\alpha^I_n$ accounts for the
sign of $\langle O^I(t,\vec x)\rangle$. 
We need to take into account, however, that (\ref{eq:opqnm}) represents a solution to the linearized equations of motion. This means that we can multiply it with an arbitrary complex number and obtain another valid solution. Therefore only relative amplitudes and phases contain physical information.
Moreover, since a generic solution will involve
operators $O^I$ with different dimensions, we
should only compare dimensionless expectation values
that can be easily obtained by normalizing with
the corresponding power of the temperature, namely
\begin{equation}
\mathcal{O}^I(t,\vec x) = T^{-\Delta} \langle O^I(t,\vec x) \rangle\,. 
\end{equation}
We also denote the dimensionless absolute values of the amplitudes as $R_n^I(\vec k) = T^{-\Delta} |\langle O^I_n(0,\vec k)\rangle |$
(note that all $R^I_n \in \mathbb{R}^+_0$).
We can then write our normalized operators as
\begin{equation}
\mathcal{ O}^I(t,\vec x) = e^{-\Gamma_n t} R_n^I(\vec k) \cos(\Omega_n t - \vec k \vec x + \alpha^I_n)\,.
\end{equation}

A measure of the relative support on the different operators $O^I$ of a given QNM is thus given by the ratio of their amplitudes $R_n^I$. Namely, $r_n^{JI}=R^J_n/R^I_n$ measures the relative support
of a mode on the operator $O^I$ with respect to the
operator $O^J$. In addition, the relative phases $\phi^{IJ} = \alpha^I-\alpha^J$ encode the phase delay between the excitations of operators $O^I$ and $O^J$.~\footnote{We could also consider taking the logarithm of the ratios $r^{IJ}_n$ and defining $\tilde r^{IJ}_n = \log(R^J_n)-\log(R^I_n)$. This has the advantage that the $\tilde r^{JI}_n$ are antisymmetric  in $JI$ as are the relative phases $\phi^{IJ}$.}
\\

A useful way of thinking about these ratios and
relative phases of the expectation values, would be to consider
a (thought) experiment where a device couples directly to the operator $O^I$ and only to that one. This
device would act as a source of $O^I$ alone among
all other operators. But the operator mixing inherent
to the system would result in a state where
operators other than $O^I$ are also excited.
Then, the ratios $r_n^{JI}$ and phases $\phi^{JI}$
will precisely characterize the response of the
operator $O^J$ when a source is switched on for the
operator $O^I$.

A (thorough) study of relative amplitudes and phases
of QNMs is lacking in the literature, although
closely related analyses of the residues of 
holographic Green's functions were carried out in
\cite{Amado:2007yr,Amado:2008ji,Davison:2011uk}.
We will do this here for a particularly interesting example, namely the first and second sound modes of a holographic superconductor. 
Notice that 
a similar study of relative phases and amplitudes for sound modes of a weakly coupled superfluid has been
carried out in \cite{Alford:2013koa}, and we will eventually compare our results to it.\\

Finally, we shall comment on the difference between the ratios of  amplitudes that we will determine here, and the residues of a certain mode $\omega(k)$.
The residue of a mode quantifies the contribution
of that mode to a given response:
\textit{i.e.} the weight of its corresponding pole
in the spectral representation of a given Green's
function. Instead, the amplitudes we discuss here
represent the weight of a given operator in the
collective excitation corresponding to a specific
mode. In a sense, these observables provide
complementary information about the system.

\subsection{Results}
\label{ssec:amps}
For concreteness, we study the longitudinal spectrum of  excitations, to which first and second sounds belong.
This amounts to solving for the set of  fluctuations in eq.~\eqref{eq:flucts}, as we describe at the end
of section~\ref{ssec:holom}. Moreover, we will
consider solutions corresponding to QNMs:
these are dual to configurations in which 
all the sources of the dual operators 
vanish, while (some of) the expectation values do not. 
In appendix~\ref{app1}, we work out the map between
fluctuations and dual operators. As a result, in the dual
theory we will be dealing with the following set of
operators corresponding to the expectation values of the fluctuations
\begin{align}
   & \tikz\draw[red,fill=red] (0,0) circle (.5ex);\,\,\langle \delta T^{tt} \rangle\,\,:\,\,\,\text{energy density},\nonumber\\
   &\tikz\draw[blue,fill=blue] (0,0) circle (.5ex);\,\, \langle \delta T^{tx} \rangle\,:\,\,\,\text{longitudinal momentum},\nonumber\\
   &\tikz\draw[green,fill=green] (0,0) circle (.5ex);\,\, \langle \delta T^{xx} \rangle:\,\,\,\text{pressure in the $x$ -- direction},\nonumber\\
   &\tikz\draw[black,fill=black] (0,0) circle (.5ex);\,\, \langle\delta  T^{yy} \rangle:\,\,\,\text{pressure in the $y$ -- direction},\nonumber\\
   & \tikz\draw[color4,fill=color4] (0,0) circle (.5ex);\,\, \langle\delta J^t \rangle\,\,\,:\,\,\,\text{charge density},\nonumber\\
   &\tikz\draw[color5,fill=color5] (0,0) circle (.5ex);\,\, \langle \delta J^x \rangle\,\,:\,\,\,\text{longitudinal current},\nonumber\\
   &\tikz\draw[color6,fill=color6] (0,0) circle (.5ex);\,\,\langle \delta\mathrm{H} \rangle\,\,\,\,:\,\,\,\text{Higgs mode (condensate fluctuations)},\nonumber\\
   & \tikz\draw[color7,fill=color7] (0,0) circle (.5ex);\,\,\langle\,\partial_t \varphi \rangle\,\,:\,\,\,\text{time derivative of the phase fluctuation (Goldstone)}\,,\nonumber\\
  & \tikz\draw[yellow,fill=yellow] (0,0) circle (.5ex);\,\,\langle\,\eta_1 \rangle\,\,\,\,\,:\,\,\,\text{real part of the charged scalar operator}\,,\nonumber\\
   & \tikz\draw[color8,fill=color8] (0,0) circle (.5ex);\,\,\langle\,\eta_2 \rangle\,\,\,\,\,:\,\,\,\text{imaginary part of the charged scalar operator}\,,\nonumber\\
   &\tikz\draw[color9,fill=color9] (0,0) circle (.5ex);\,\,\langle\,\delta \phi^x \rangle\,:\,\,\,\text{longitudinal axion operator}\,. \label{legends}
\end{align}
In eq.~\eqref{eq:onepoints} we express these
expectation values in terms of the asymptotics of the 
fluctuations~\eqref{eq:flucts}. 
Notice that in the superfluid phase 
we are rewriting the operator dual to the fluctuation of the complex scalar in terms of its modulus and phase, which we denote by $\delta H$ and $\varphi$ respectively.
Moreover, for the model breaking translations, we switch on an extra
fluctuation, $\delta\phi^x$.
The meaning of all of these quantities is well known apart from $\delta\phi^x$, corresponding to the longitudinal axion operator. The colored circles in \eqref{legends} display the color scheme used in all the figures in this section.

In the remainder of this section we will present results
for the amplitudes and phases of the expectation values~\eqref{legends} for the different QNMs of the system. These are all computed in Fourier space (see eq.~\eqref{eq:harmonic}), at finite frequency and momentum.
Finally, we restrict our analysis of the amplitudes and phases to the following hydrodynamic modes
\begin{enumerate}
    \item first and second sounds in the superfluid phase (with preserved translations);
    \item fourth sound and energy diffusion in the superfluid phase with broken translations;
    \item first sound and a purely diffusive mode in the normal phase (with preserved translations);
    \item thermo-electric diffusion modes (energy and charge diffusion in the decoupling limit) in the normal phase with broken translations.
\end{enumerate}

Additionally, in the normal phase we shall also consider
the gapped scalar mode that drives the superfluid phase
transition at $T=T_c$.

Finally, let us explain our normalization of the amplitudes. We may formulate the linearized equations for the fluctuations in terms of a generalized eigenvalue problem (see appendix~\ref{app:numerics} for more details)
\begin{equation}
    (\bm{A}(k)\,\omega-\bm{B}(k)\,)\,\bm{x}(k)=0,
\end{equation}
where $\bm{A}$ and $\bm{B}$ are differential operators and $\bm{x}$ is a vector consisting of the fluctuations. Solving this generalized eigenvalue problem numerically, we find a set of eigenfunctions $\bm{x}_n$ (consisting of the fluctuations) for each eigenvalue $\omega_n$ (the quasinormal frequency). We will refer to the set of eigenfunctions as eigenvector from now on. First, we make all expectation values of the eigenvectors consisting of the fluctuations~\eqref{legends} dimensionless by dividing by the appropriate power of the temperature (depending on the conformal dimension of the corresponding operator).
Then, we normalize the sum of all absolute values of the dimensionless expectations values 
of the eigenvectors to one.
This means that an amplitude of $0.5$ implies for example that $50\%$ of the support of such a mode is controlled by this type of fluctuation. 
With this normalization, the amplitudes are invariant
under the scaling symmetry inherent to the linear analysis which acts on all fluctuations as $\langle \delta f^i\rangle \rightarrow \lambda \langle \delta f^i\rangle$.\\
 \begin{figure}[ht!]
     \centering
     \includegraphics[width=0.326\linewidth]{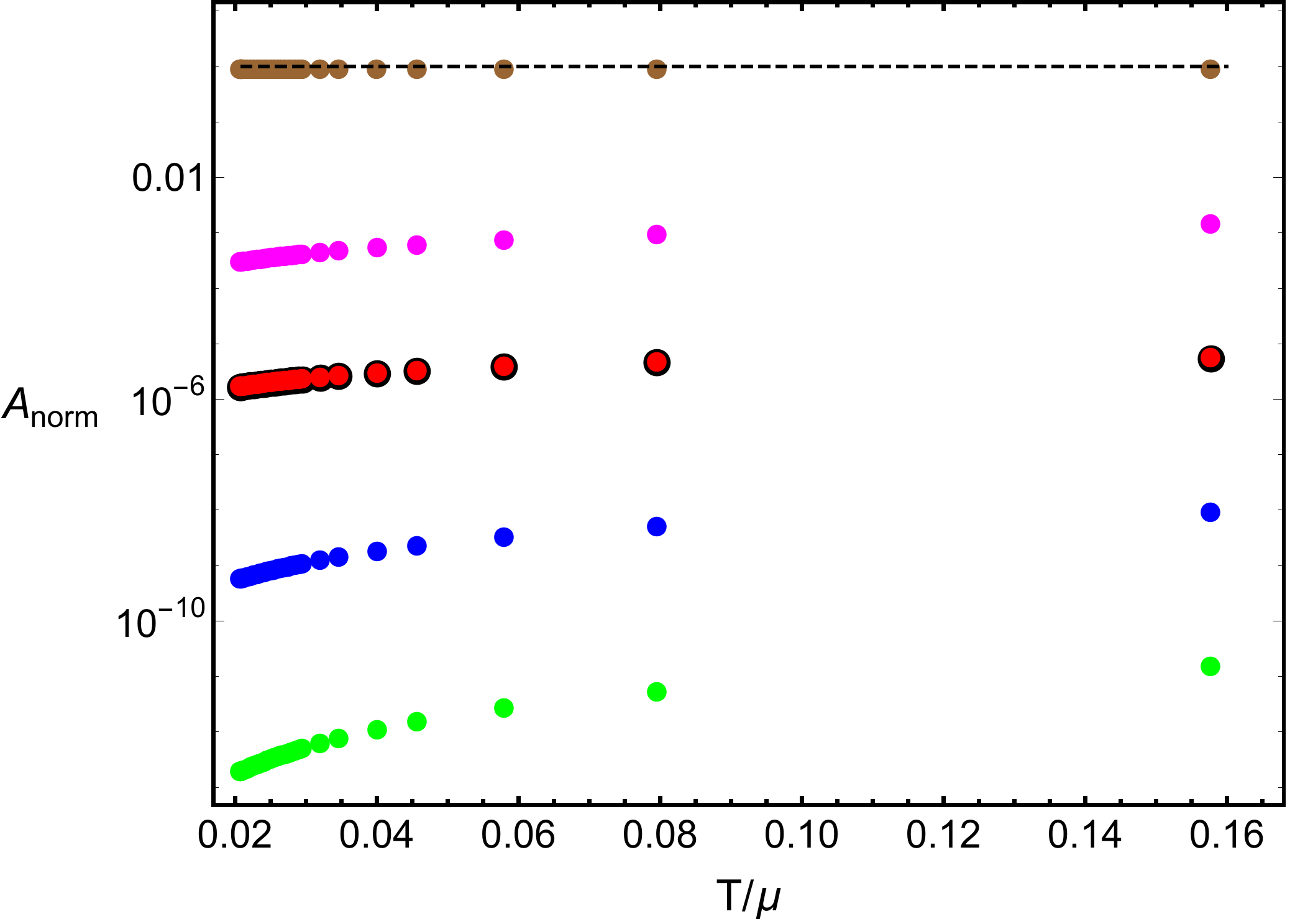}
       \includegraphics[width=0.326\linewidth]{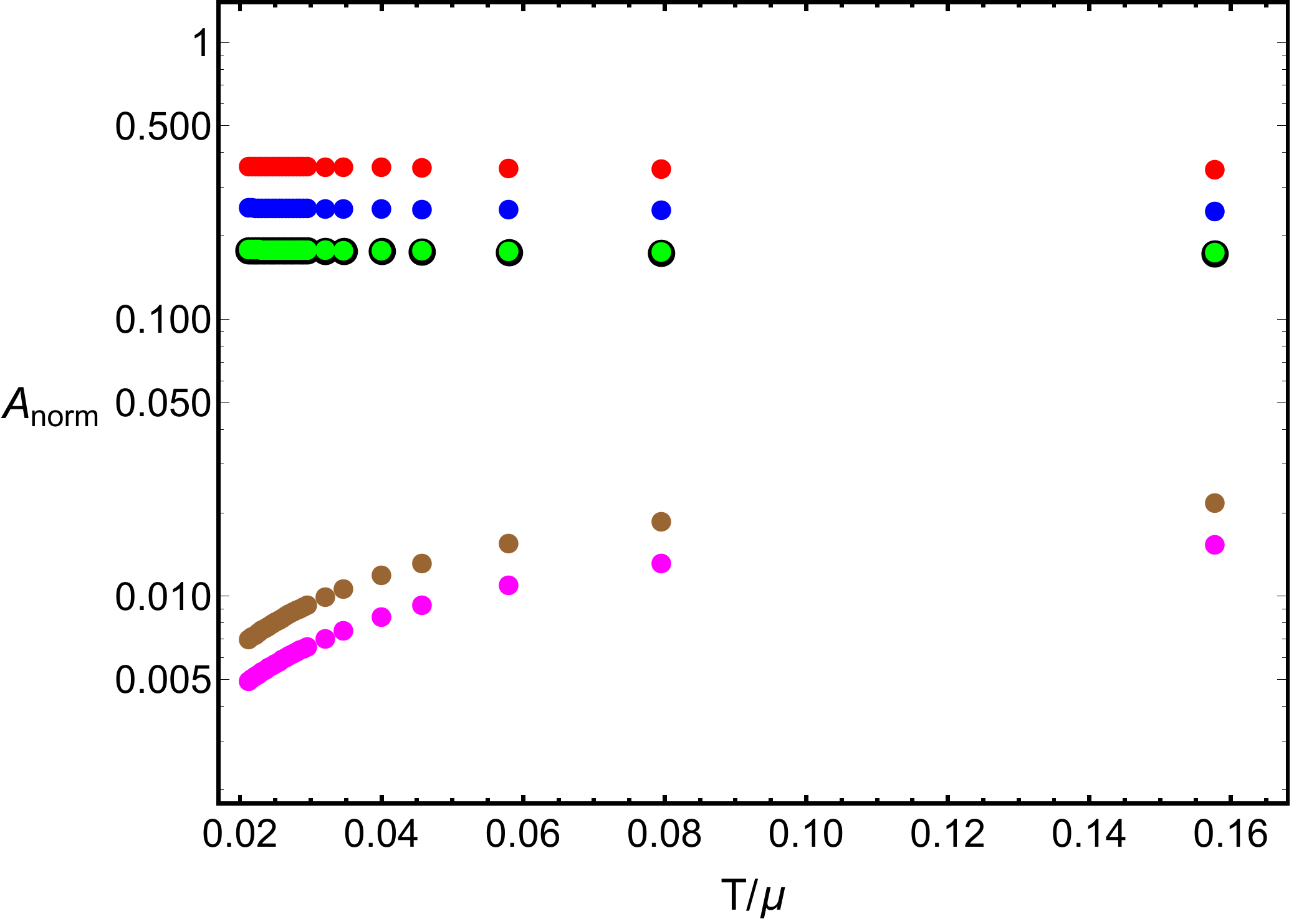}
       \includegraphics[width=0.326\linewidth]{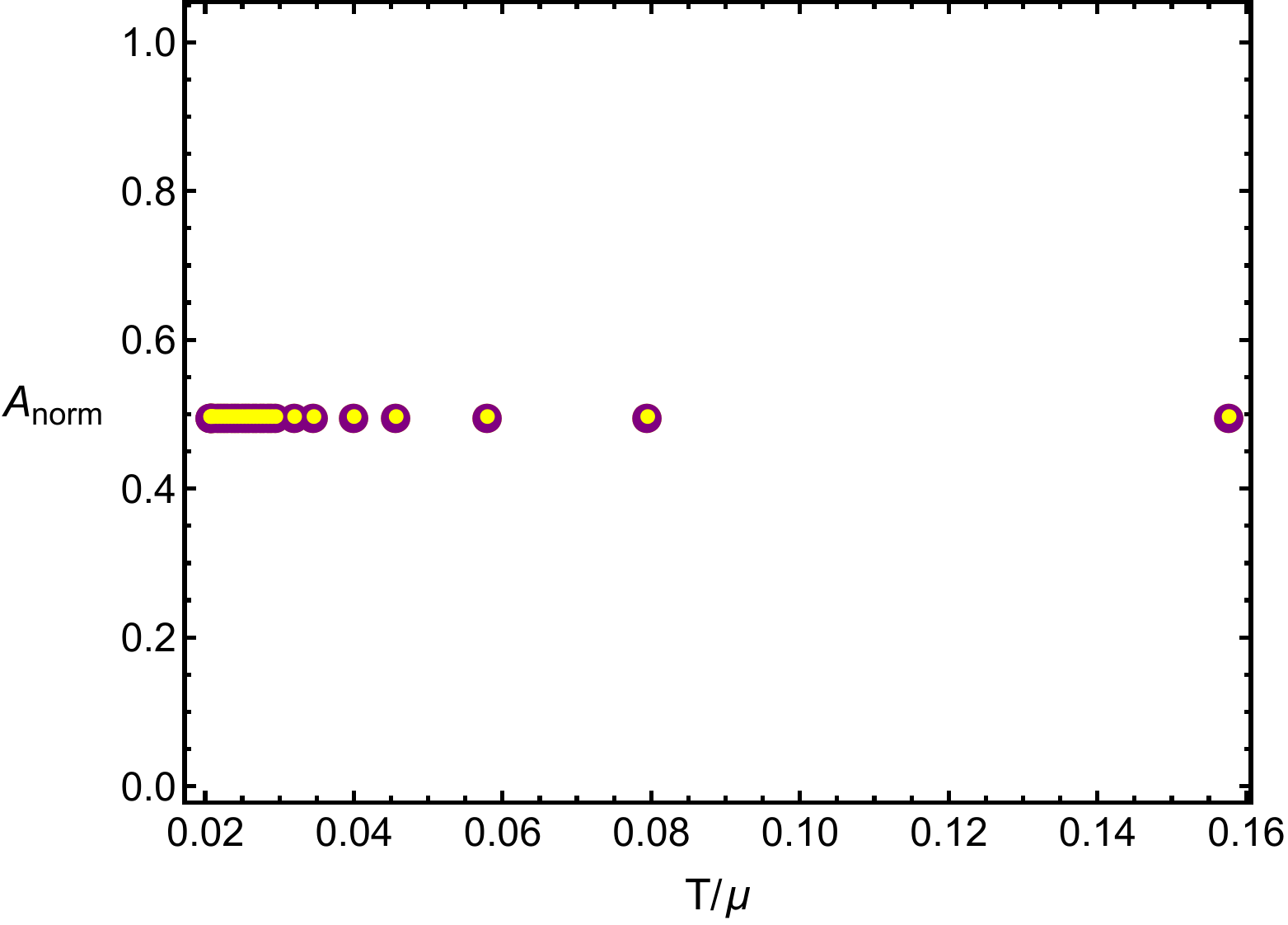}
     \caption{The amplitudes of the various field theory operators \eqref{legends} in the normal phase ($T>T_c$) at fixed $k/T=0.01$ for the diffusive mode (\textbf{left}), first sound (\textbf{center}), and the gapped scalar modes (\textbf{right}). The color scheme is that illustrated in \eqref{legends}.}
     \label{fig:ampnormal}
 \end{figure}\\
 In order to make the reader familiar with the language, we start our analysis in the normal phase which corresponds to considering a standard relativistic charged fluid in the field theory dual. There, the two hydrodynamic modes are first sound and the purely diffusive mode. Nevertheless, for completeness we will consider also the gapped scalar modes which are the responsible for the superfluid instability at $T\leq T_c$ \cite{hydroSC}. Notice that in the normal phase the background value of the scalar is zero and therefore the fluctuations of the phase are not a well defined object. Thus, in the normal phase, we will consider the expectation value of the fluctuations of the imaginary part of the scalar operator dual to the complex bulk field $\psi$.\\

 The results for the amplitudes of the normal fluid are shown in fig.~\ref{fig:ampnormal}. As expected, we find that the gapped scalar mode is solely carried by the fluctuations of the scalar charge operator. The purely diffusive mode (left panel of fig.~\ref{fig:ampnormal}) is carried by the fluctuations of the charge density $\langle \delta J^t\rangle$, as one can derive analytically from hydrodynamics (see appendix~\eqref{eq:eigenv3}). The hydrodynamic treatment shows that in first order hydrodynamics this mode consists solely of $\langle\delta J^t\rangle$ (and of the by a Ward identity related $\langle\delta J^t\rangle$). Moreover, the second order correction includes $\delta T^{tt}$ with a relative amplitude of $\langle\delta T^{tt}\rangle/(\langle\delta J^t\rangle\,T)=k^2/T\,\rho/(\mu\rho+s\,T)$. 
 We shall point out that our result that the diffusive mode is solely carried by $\langle\delta J^t\rangle$ is not
 in contradiction with the findings
 of \cite{Hartnoll:2014lpa,Davison:2015taa} where
 it was observed that the diffusive mode is carried
 by the `incoherent charge'
 $\delta Q^\text{diff}\equiv\delta J^t-\rho/(\mu\rho+sT)\,\delta T^{tt}$ \footnote{We thank Blaise Gout\'eraux and Richard Davison for discussions about this point.}.
 As we detail in appendix \ref{appendixtodo},
 our computation describes the situation
 where only the diffusive mode is excited in the system. The authors of~\cite{Hartnoll:2014lpa,Davison:2015taa}
 consider a generic perturbation in which all modes are excited instead. We show in that appendix
 that one can relate both situations via the
 hydrodynamic framework and match them to the holographic computation. 
 
 Finally, first sound is dominated by the fluctuations of energy density and 
 is almost entirely supported on the
 fluctuations of energy density, pressure and longitudinal momentum.
 The ratio between pressure and energy fluctuations is approximately $0.5$ as we expect from the Ward identities that constrain the expectation values given in eq.~\eqref{eq:wardrel1}. Note that the pressure fluctuations contribute approximately equally via $\langle \delta T^{xx}\rangle$ and $\langle\delta T^{yy}\rangle$ to first sound as is obvious from the tracelessness of the energy-momentum tensor and the Ward identities~\eqref{eq:wardrel1}. We furthermore observe that the amplitudes of the fluctuation of longitudinal momentum are approximately 0.7 time the energy fluctuations (or 1.4 times the pressure fluctuations)
which may be seen from eq.~\eqref{eq:wardrel1} or equivalently from eq.~\eqref{eq:speedfromamplitude} using the conformal speed. From the inset of fig.~\ref{fig:secondsound} we know that in the range of temperatures shown there, the diffusion constant
 takes values in the range $D_q\, T\approx\{0.04,0.1\}$.
Using the Ward identity~\eqref{eq:wardrel2}, we find that the fluctuations of the longitudinal current in the purely diffusive mode should be approximately $\{4\cdot 10^{-4}, 10^{-3}\}$ of the charge fluctuation as evident in the left panel in fig.~\ref{fig:ampnormal} and in eq.~\eqref{diff:cons}. The simple relation eq.~\eqref{diff:cons} is Fick's law for diffusion. This discussion shows that we can use the amplitudes and the Ward identities to compute the diffusion constant in the hydrodynamic dispersion relation and vice versa. In the case of first sound a similar analysis holds albeit it is a bit more complicated. We refer the interested reader to appendix \ref{sec:dispaampl}. In the normal phase, the scalar sector decouples and the fluctuations of the scalar operator do not enter into either of the sound modes in the normal phase.
  \begin{figure}[ht!]
     \centering
     \includegraphics[width=0.43\linewidth]{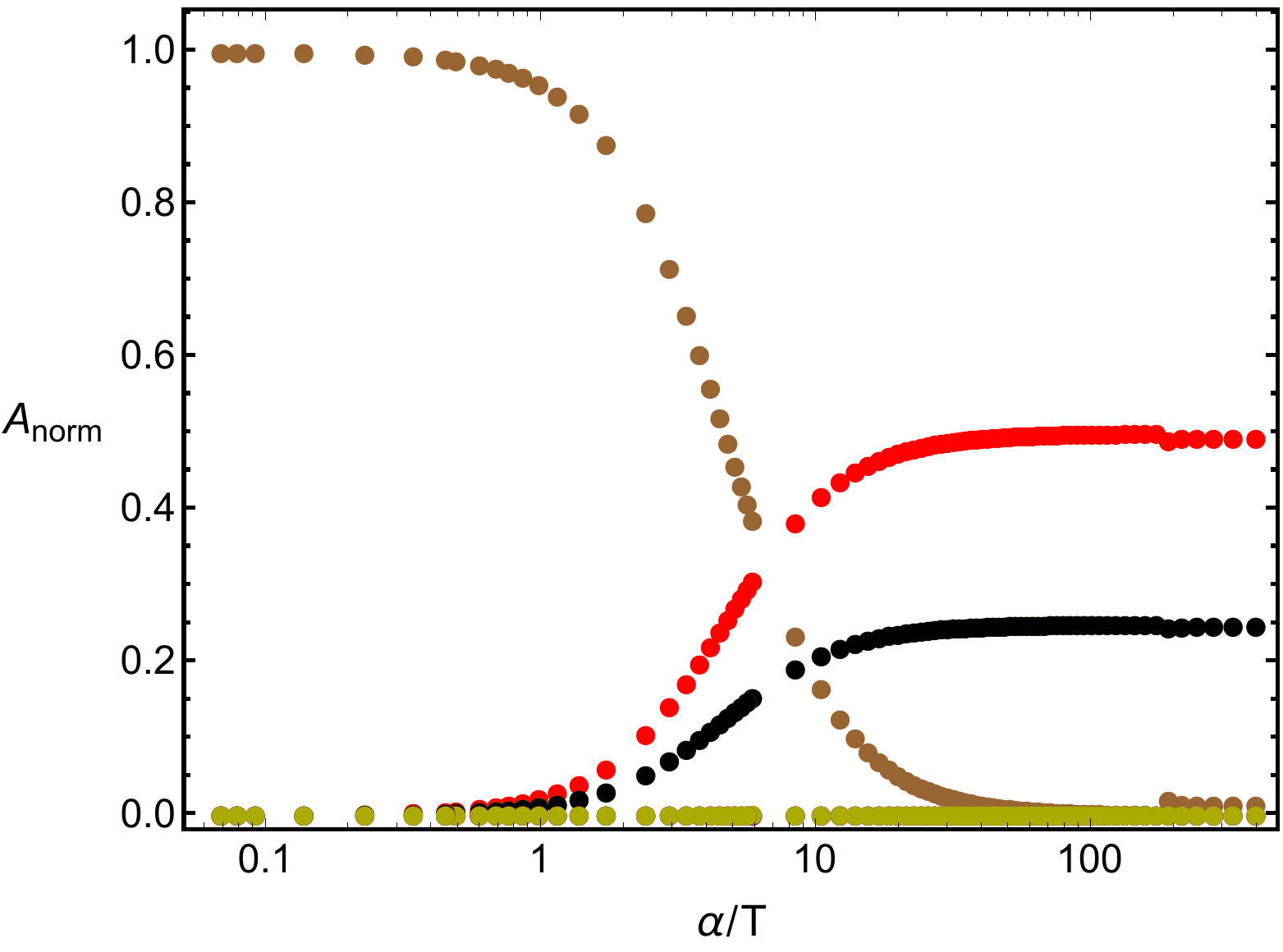}\qquad
       \includegraphics[width=0.43\linewidth]{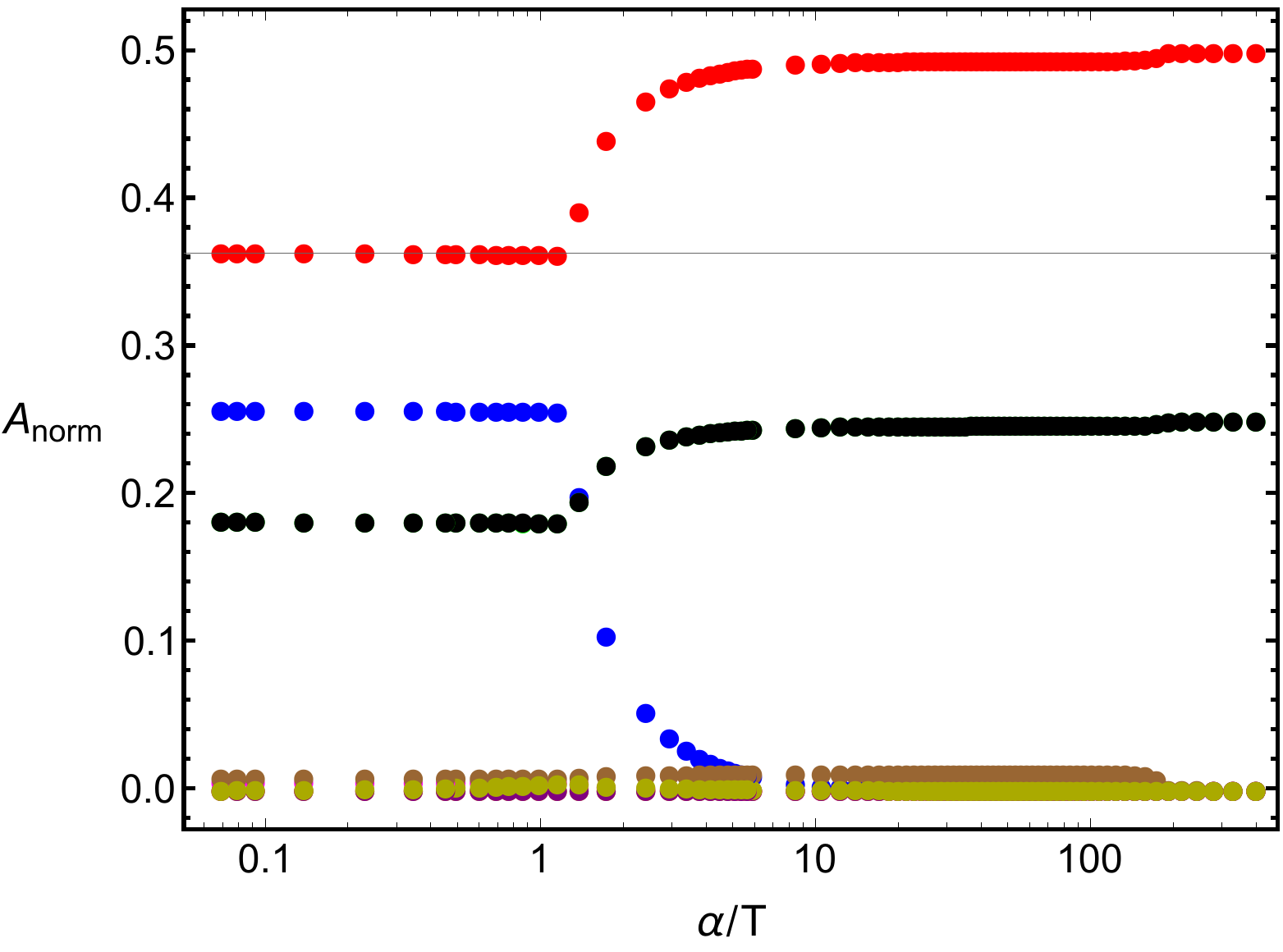}
     \caption{The amplitudes of the various operators \eqref{legends} for the thermo-electric diffusive modes in the normal phase with momentum dissipation. We fixed $k/T=0.01$ and $T/\mu=0.025$. The color scheme is that illustrated in \eqref{legends}.}
     \label{fig:axnor}
 \end{figure}\\
 As already mentioned, when the axion fields are switched on and momentum is not a conserved quantity anymore, first sound is destroyed at large wavelength and the only two hydrodynamic modes in the normal phase are the thermo-electric diffusive modes \cite{Kovtunbound}. We show the amplitudes for these two modes in fig.~\ref{fig:axnor}. At $\alpha=0$, the left panel corresponds to the diffusive mode shown in fig. \ref{fig:ampnormal}. Indeed, for $\alpha/T\ll1$, such a mode is totally dominated by charge fluctuations. Interestingly, around $\alpha/T \approx 8$, we observe a smooth crossover to a regime where the original diffusive mode is now dominated by energy fluctuations. Moreover,  the ratio between pressure and energy fluctuations is approximately $0.5$ for all $\alpha/T$, as we already observed for the first sound mode which is a consequence of the tracelessness of the energy-momentum tensor (since $\langle \delta T^{tt}\rangle=\langle \delta T^{xx}\rangle+\langle \delta T^{yy}\rangle$).
 Notice that in that large $\alpha/T\gg1$ regime
 the contribution from charge fluctuations drops to zero indicating a drastic change in the nature of the diffusive mode. 
 
As shown on the right panel of fig.~\ref{fig:axnor}, the situation is different for the other diffusive mode
 present in the normal phase of the model with
 momentum dissipation. 
For small values of momentum dissipation, $\alpha/T\lesssim2$, the momentum chosen for the extraction of the amplitudes, $k/T=0.01$, is too large and the mode is not diffusive anymore but already sound-like, \textit{i.e.} propagating. There, we find the same features as for first sound in fig.~\ref{fig:ampnormal}. Around $\alpha/T \approx 2$, momentum is not a well-defined quantity anymore, and its contribution to the mode drops rapidly to zero. At very large value of momentum dissipation, the mode in the right panel of fig.~\ref{fig:axnor} is totally decoupled from everything else and it is dominated by energy and pressure. Note that for $\alpha/T\gtrsim200$, we enter the regime where the scalar modes are unstable since the critical temperature (which is a function of the momentum dissipation strength) is now larger the temperature we chose. Similarly to the diffusive mode, we observe that the ratio between energy and pressure fluctuations is approximately 0.5 for all $\alpha/T$.
 \begin{figure}[ht!]
     \centering
     \includegraphics[width=0.43\linewidth]{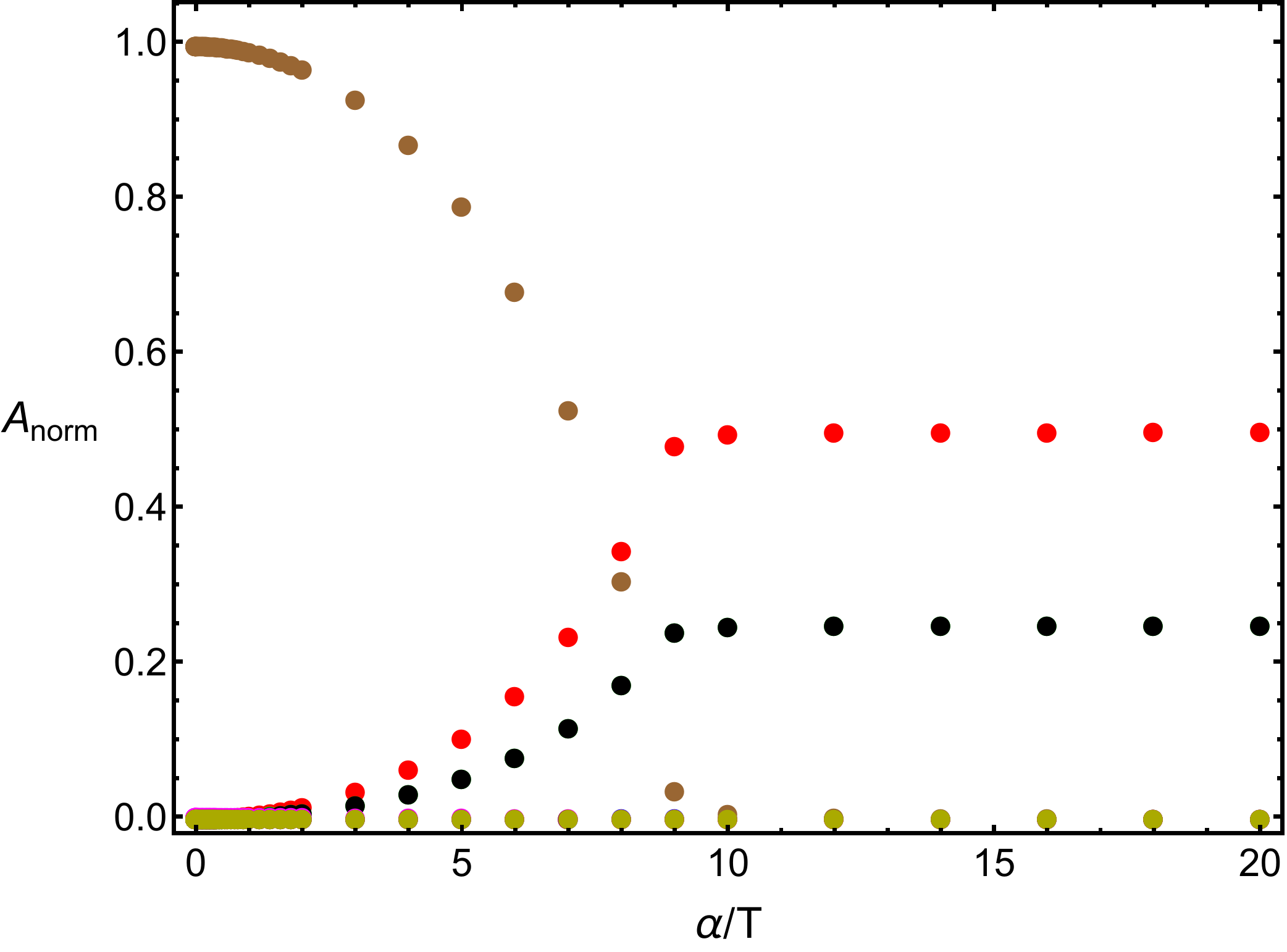}\qquad
      \includegraphics[width=0.43\linewidth]{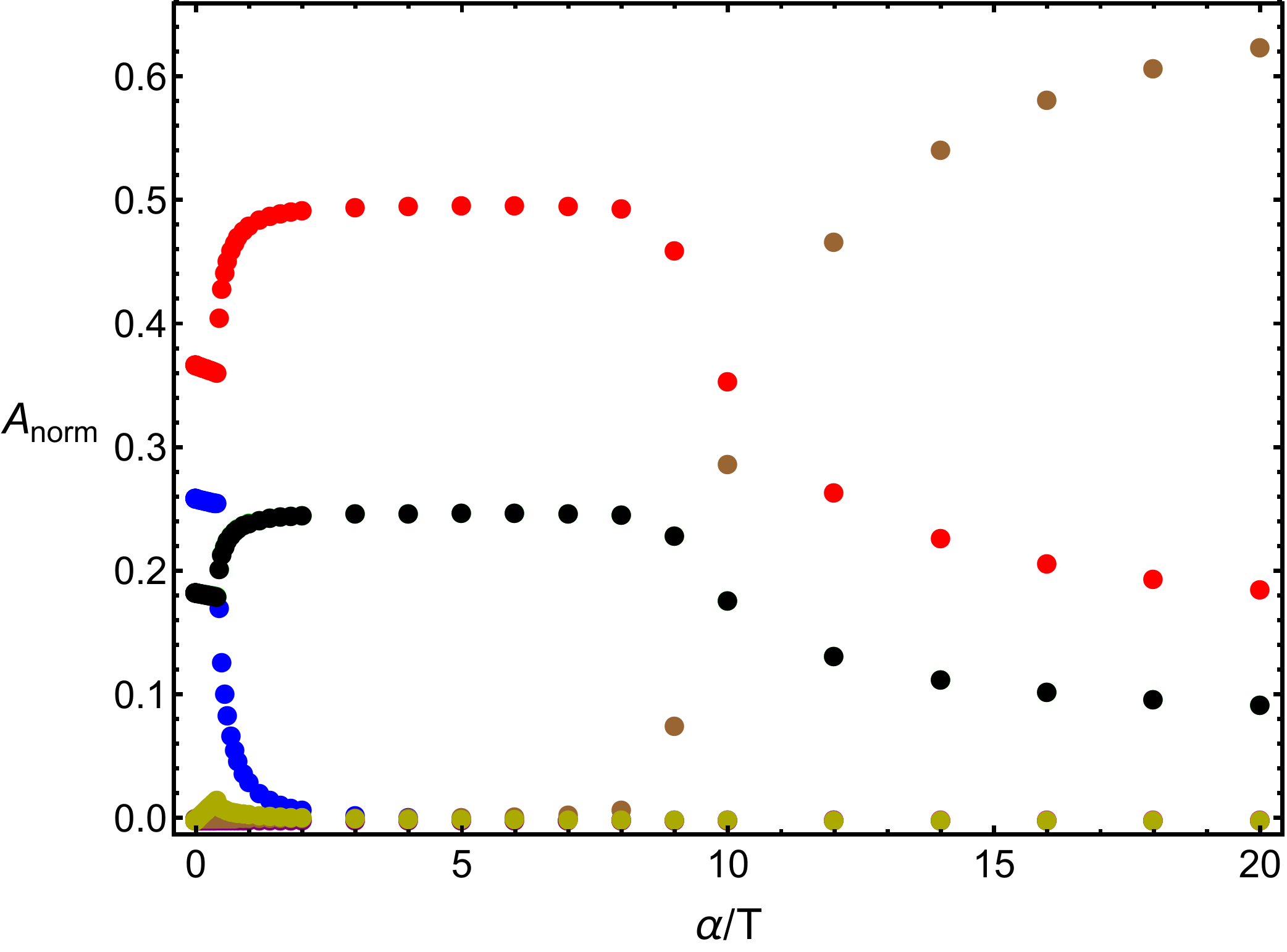}
     
     \vspace{0.2cm}
     
        \includegraphics[width=0.43\linewidth]{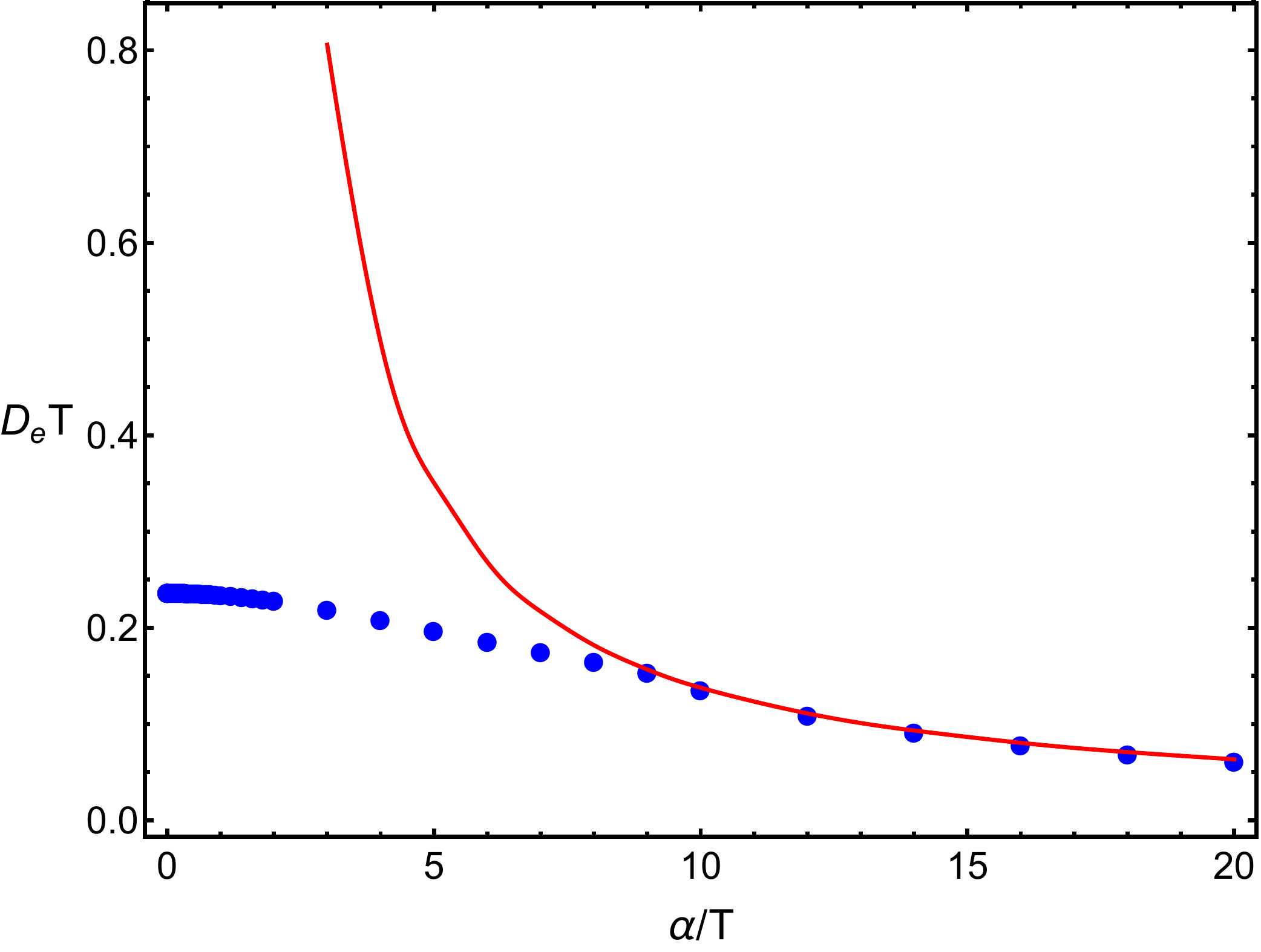}\qquad
      \includegraphics[width=0.43\linewidth]{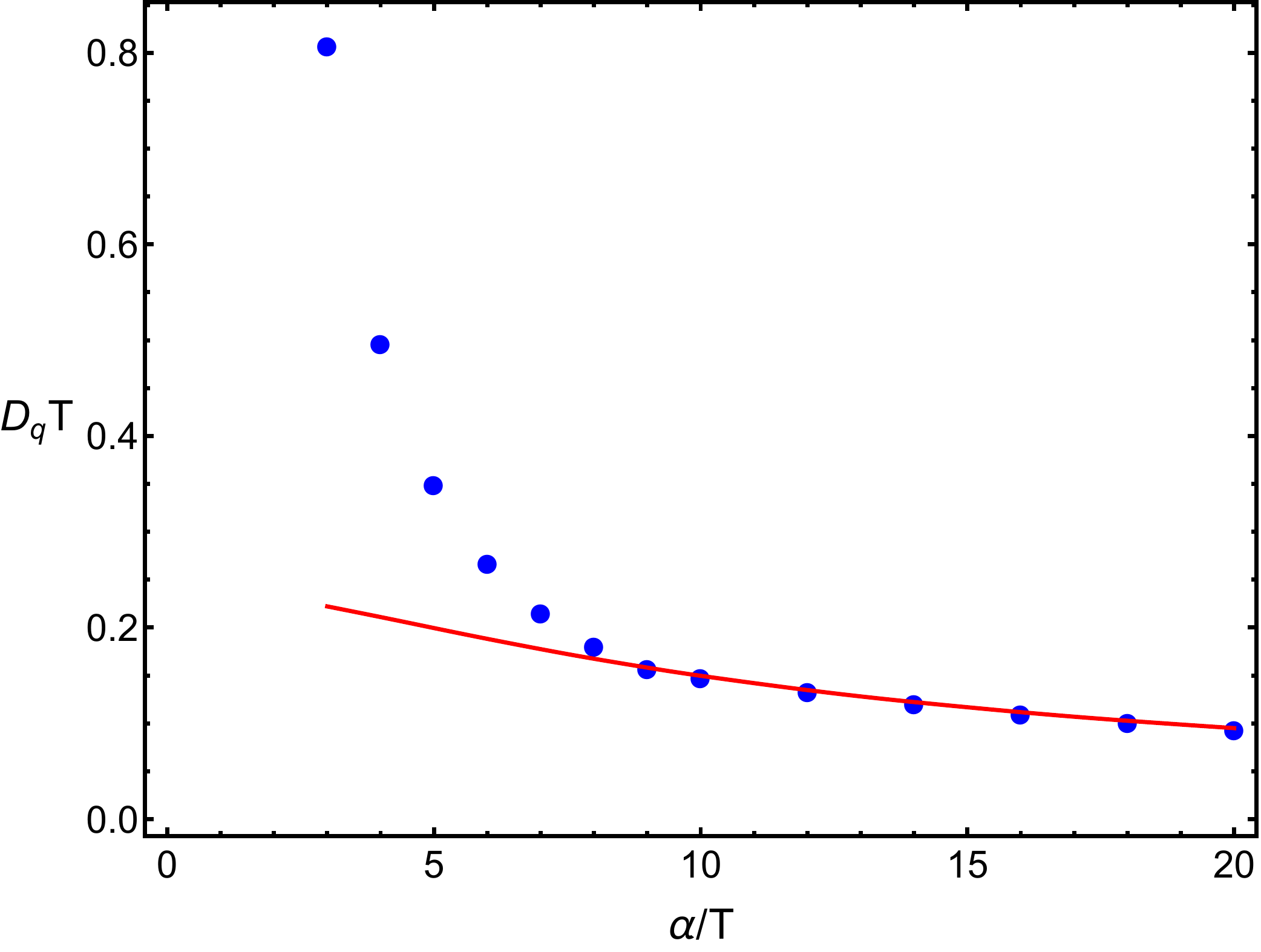}
     \caption{\textbf{Top: }The amplitudes of the various operators for the two thermoelectric diffusion modes at $T/\mu=5$. \textbf{Bottom: }The corresponding diffusion constants. The blue points are numerical data obtained from the amplitude data in the upper panel using the Fick's law~\eqref{diff:cons} as explained in appendix \ref{sec:dispaampl}. The red lines are the analytical formulas \eqref{eq:diffconsts} valid in the incoherent limit, which is reached around $\alpha/T\sim 10$. At that point, the support of the two modes decouple into energy density and charge density fluctuations. See also the figures in appendix~\ref{appen:disc}.}
     \label{new}
 \end{figure}\\
Before moving on to the broken phase, let us investigate the normal phase with broken translations further; in particular, we are interested in the two thermoelectric diffusion modes. At intermediate values of $T/\mu$ and $\alpha/T$, energy diffusion and charge diffusion are coupled and therefore their support is mixed between the various operators. This is no longer true in the so-called \text{incoherent limit} \cite{Hartnoll:2014lpa}, at which the momentum dissipation rate becomes the dominant scale in the system, \textit{i.e.} $\alpha/T,\;\alpha/\mu \gg 1$. In such a regime, we do expect the two diffusive modes to decouple again \cite{Davison:2015bea} and the corresponding diffusion constants acquire the simple values
\begin{equation}\label{eq:diffconsts}
    D_e\,=\,\frac{\kappa}{c_v}\,,\qquad D_q\,=\,\frac{\sigma}{\chi_{\rho\rho}}\,,
\end{equation}
where $\kappa$ is the thermal conductivity, $c_v$ the specific heat, $\sigma$ the electric conductivity and $\chi_{\rho\rho}$ the charge susceptibility. For the linear axion model considered in this work, the formulas for the diffusion constants above are known analytically and can be found in the literature (see for example \cite{Baggioli:2017ojd}). In fig.~\ref{new}, we repeat the analysis of fig.~\ref{fig:axnor} but this time going into the aforementioned incoherent limit. As expected, in this case, the contributions to the support of the two diffusive modes from charge and energy fluctuations crosses around $\alpha/T \sim 10$, which corresponds exactly to the location at which the diffusion constants approach the values mentioned in eq.~\eqref{eq:diffconsts}, \textit{i.e.} the edge of the incoherent limit. Going towards $\alpha/T \rightarrow \infty$, the support of the two modes becomes solely dominated by either charge or charge and energy fluctuations, as expected from their decoupling. This result is explained in more detail in appendix~\ref{appen:disc} where we show the relative coefficients in fig.~\ref{pic:incoherenapp} explicitly. Note that we extracted the diffusion constants from the amplitude data at one fixed $k/T$ using eq.~\eqref{diff:cons}. Not only that, but the amplitudes of these two diffusive modes can be derived analytically at any value of the parameters from hydrodynamics (see appendix~\eqref{eq:eigenv3}), matching perfectly the numerical results presented here.

 \begin{figure}[ht!]
     \centering
     \includegraphics[width=0.47\linewidth]{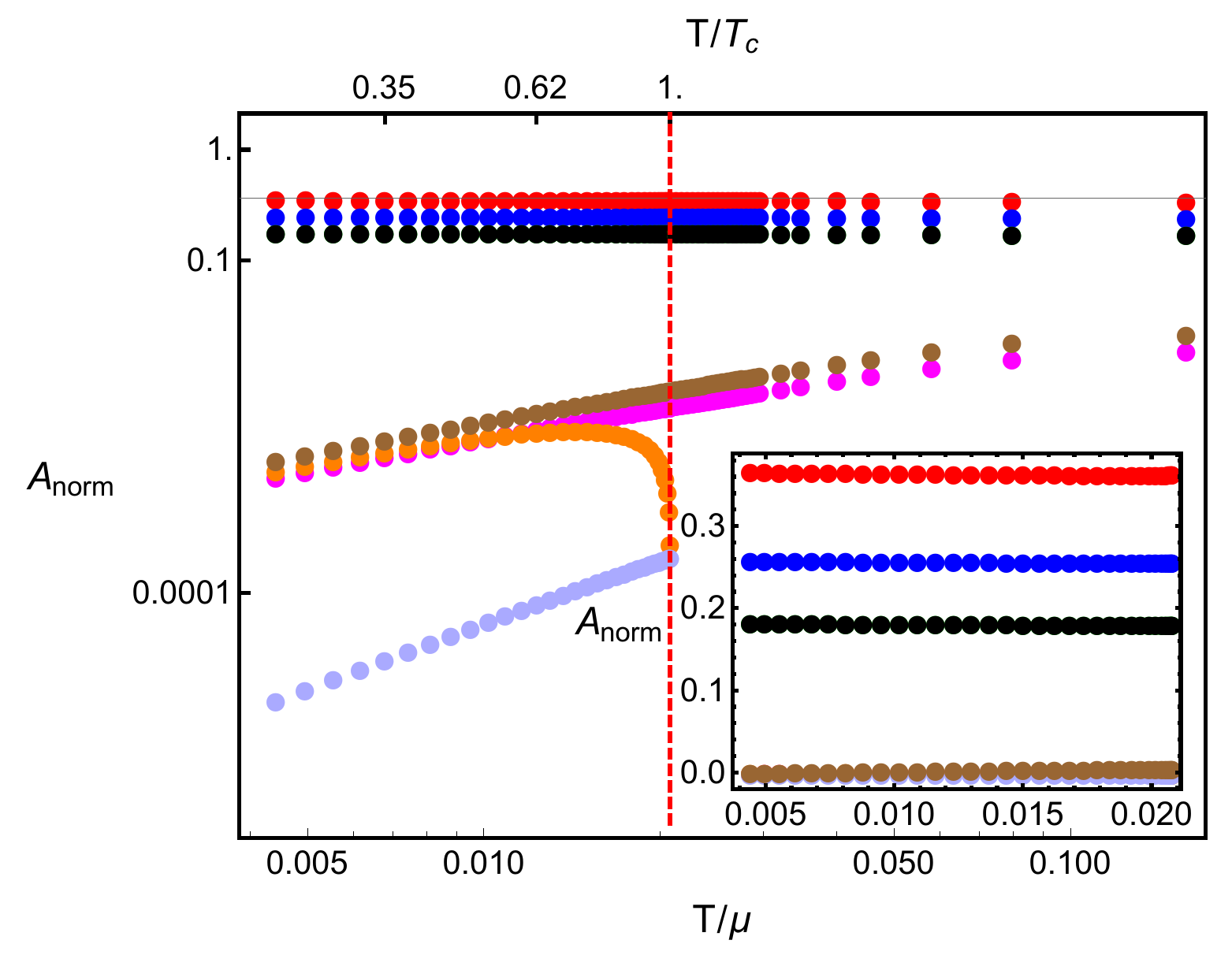}\qquad
       \includegraphics[width=0.47\linewidth]{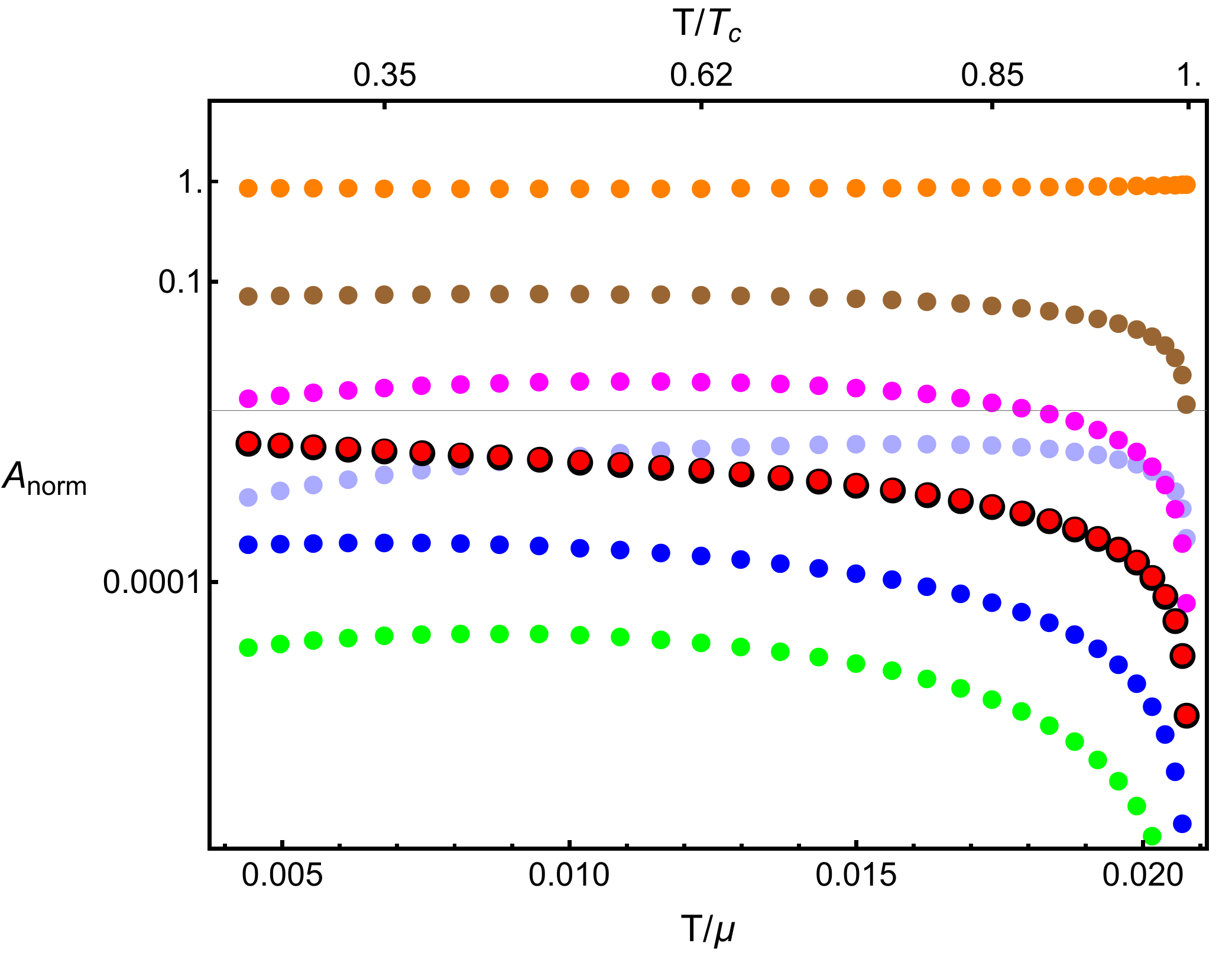}
     \caption{The amplitudes of the various operators \eqref{legends} for first (left panel) and second (right panel) sound normalized to $T$ at fixed $k/T=0.01$. The inset is a non-logarithmic presentation of the same curves to show the 
     ratio of approximately $0.5$ between pressure and energy fluctuations. The color scheme is that illustrated in \eqref{legends}.}
     \label{fig:ampsounds}
 \end{figure}
 
 \color{black}
 We now move on to the analysis of the superfluid phase of our  system. We study the amplitudes of the operators~\eqref{legends} for the two hydrodynamic modes in the longitudinal sector of this phase: first and second sound. 
 The results are plotted in fig.~\ref{fig:ampsounds}.
Notice that in this condensed phase we represent the fluctuations of the scalar operator in terms of its amplitude (the Higgs mode $\delta H$) and the time derivative of its phase which carries
information about the norm of the Goldstone boson since
in our background
$|A_\mu-i\,\partial_\mu\varphi|\sim\mu-i\,\partial_t\varphi+\mathcal{O}(\varphi^2)$.
As one can see on the left panel of fig.~\ref{fig:ampsounds},
the behavior of the amplitudes of the first sound mode is rather
featureless and dominated by energy, pressure, and
longitudinal momentum fluctuations across the
phase transition and down to the lowest temperatures.
The fluctuations of the Goldstone mode are highly suppressed, and those of charge and current decrease as T is lowered.
We show the results for second sound on the right panel of
~\ref{fig:ampsounds}. This mode is dominated by the fluctuations of the superfluid condensate (the Higgs mode).
In view of these results one can rule out a role reversal between first and second sound in which the nature of the fluctuations supporting those modes is interchanged as the temperature is lowered across the superfluid phase as in \cite{Alford:2013koa}. Finally, notice that the fluctuations of the scalar operator are not shown above the critical temperature in the left panel of fig.~\ref{fig:ampsounds}. They decouple from the other operators and therefore do not contribute to the support of the first sound mode above $T_c$.

  \begin{figure}[ht!]
     \centering
     \includegraphics[width=0.49\linewidth]{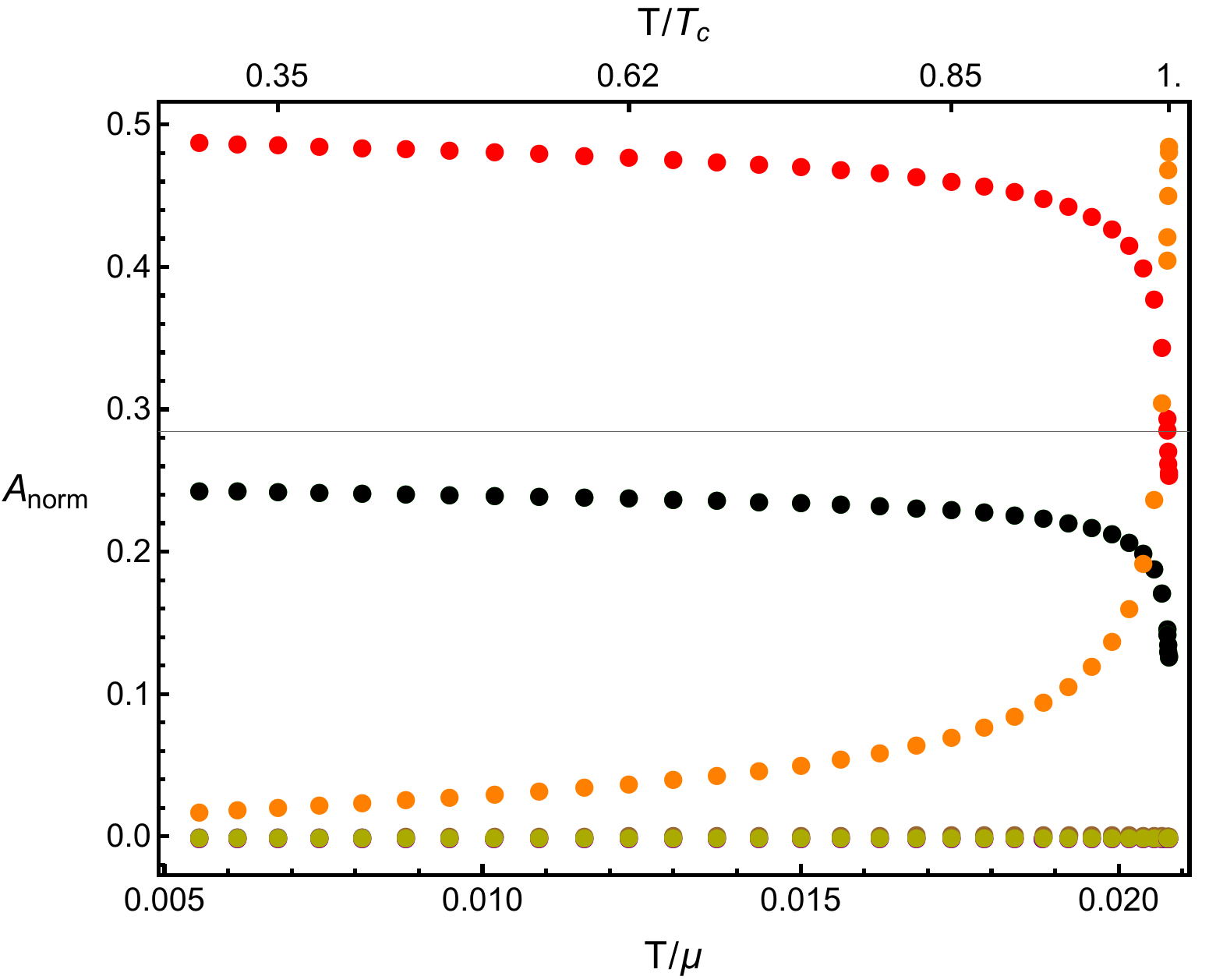}
       \includegraphics[width=0.49\linewidth]{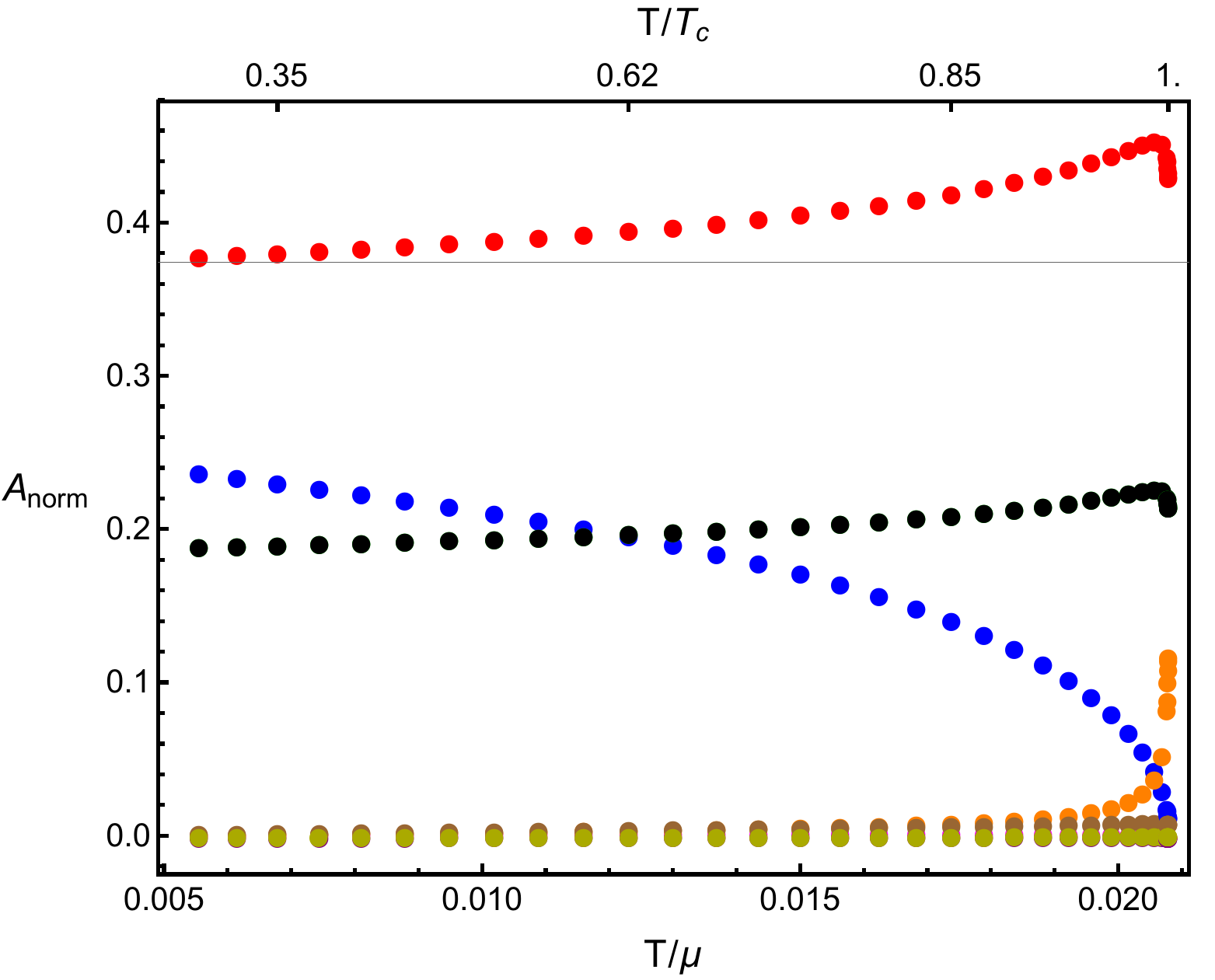}
     \caption{The amplitudes of the various operators \eqref{legends} for the diffusive mode (\textbf{left}) and $4^{th}$ sound (\textbf{right}) in the superfluid phase with momentum dissipation. We fixed $\alpha/T=6$ and $k/T=0.01$. The color scheme is that illustrated in \eqref{legends}.}
     \label{fi1}
 \end{figure}

 We can now perform the same analysis in the superfluid phase in presence of momentum dissipation. In this case, first sound is replaced by a diffusive mode and second sound is replaced by fourth.
 In fig.~\ref{fi1} we show the results for the amplitudes of the fluctuations for the diffusive mode (left) and fourth sound (right).
 Except at temperatures near the critical temperature, where the Higgs mode
 takes over, the diffusive mode is dominated by
 energy and pressure fluctuations.
 Fourth sound is always dominated by energy fluctuations.
 Interestingly,  at low temperature the ratios between energy, longitudinal momentum, and pressure fluctuations asymptote to the same ratios  observed for first sound since the speed of fourth sound tends to the speed of first sound as $T\to0$ there (while the attenuation constant tends to zero).
   Finally, we shall point out that we also observe a purely imaginary, gapped non-hydrodynamic mode which is solely supported by the scalar fluctuations. This higher quasi-normal mode might be the Higgs mode discussed in~\cite{Behrle_2018,Bhaseen:2012gg,Li:2013fhw,Pekker_2015}.
   
 \subsubsection{About the relative phases}
 As explained in detail in section \ref{sec:qnm}, the value of the amplitudes of the various fluctuations (\ref{legends}) of the different (hydrodynamic) modes is not the only information we can extract from our analysis. We can also compute the phase
 differences between the various fluctuations present in a specific (hydrodynamic) mode. We used the phase of the charge density fluctuation $\Theta_{\delta J^t}$ as reference phase and computed the phases of the other fluctuations with respect to $\Theta_{\delta J^t}$. Note that the phases are defined modulo $2\pi$. The phase differences of the various fluctuations  are listed in table~\ref{tab1}. The phases differences remain almost constant throughout the phase diagram. However, it is important to notice that the ``ideal'' phase differences as listed in~\ref{tab1} are only reached in the $k\to0$ limit. As we show in eq.~\eqref{eq:dampingdiffu} and eq.~\eqref{eq:damingattenu}, the relative phases between certain fluctuations are intimately related to damping, \textit{i.e.} for vanishing phase differences we reach ideal hydrodynamics. In other words, the relative phases give us insights about the damping in the system and we may compute the diffusion and attenuation constants, respectively, from them.
\begin{table}
\centering
\resizebox{\linewidth}{!}{\begin{tabular}{|c||c|c|c|c|c|c|c|c|}
\hline
\cellcolor{blue!10}\textbf{without momentum dissipation}& \tikz\draw[red,fill=red] (0,0) circle (.5ex); $\delta T^{tt}$ & \tikz\draw[blue,fill=blue] (0,0) circle (.5ex); $\delta T^{tx}$ & \tikz\draw[green,fill=green] (0,0) circle (.5ex); $\delta T^{xx}$ & \tikz\draw[black,fill=black] (0,0) circle (.5ex); $\delta T^{yy}$ & \tikz\draw[color5,fill=color5] (0,0) circle (.5ex); $\delta J^{x}$ & \tikz\draw[color6,fill=color6] (0,0) circle (.5ex); $\delta\mathrm{H}$ & \tikz\draw[color7,fill=color7] (0,0) circle (.5ex); $\partial_t \varphi$ & \tikz\draw[color9,fill=color9] (0,0) circle (.5ex); $\delta \phi^x$ \\
\hline
diffusive mode ($T>T_c$) & $0$ &$0$  & $\pi$ & $\pi$ & $\pi/2$ & \cellcolor{gray!10} & \cellcolor{gray!10} & \cellcolor{gray!10}\\
\hline
$1^{st}$ sound with $\mathrm{Re}(\omega)>0$ ($T>T_c$) &$\pi$  & $\pi$  & $0$ &$0$  & $\pi$ & \cellcolor{gray!10} & \cellcolor{gray!10} & \cellcolor{gray!10}\\
\hline
$1^{st}$ sound with $\mathrm{Re}(\omega)>0$ ($T<T_c$) & $\pi$ &  $\pi$&  $\pi$& $\pi$ &$0$  & $0$ & 0& \cellcolor{gray!10} \\
\hline
$2^{nd}$ sound with $\mathrm{Re}(\omega)>0$ ($T<T_c$) & $\pi/2$ &  $\pi/2$&  $\pi/2$& $\pi/2$ &$0$  & $0$ & 0 & \cellcolor{gray!10}\\
\hline\hline
\cellcolor{blue!10}\textbf{with momentum dissipation}& \cellcolor{blue!10} & \cellcolor{blue!10}  & \cellcolor{blue!10} & \cellcolor{blue!10} & \cellcolor{blue!10} & \cellcolor{blue!10} & \cellcolor{blue!10} & \cellcolor{blue!10} \\
\hline
$4^{nd}$ sound with $\mathrm{Re}(\omega)>0$ ($T<T_c$) & $\pi$ &  $\pi$&  $\pi$& $\pi$ &$0$  & $0$ & 0 & $\pi/2$\\
\hline
diffusion ($T<T_c$) & $\pi$ &$\pi/2$  & $\pi$ & $\pi$ & $-\pi/2$ & $\pi$ & $0$ & $\pi/2$\\
\hline
\end{tabular}}
\caption{The phase differences modulo 2$\pi$ with respect to $\delta J^t$ for the various fluctuations for the normal fluid ($T>T_c$) and superfluid $(T<T_c)$ phases for temperatures sufficiently far from the phase transition and in the ideal limit $k\to0$. The gray cells indicate that such a fluctuation is not present in the corresponding mode. In the case of the sound modes, the phases of the ``partner'' mode may be obtained by taking into account that the phases add up to 0 (for parity even operators) and $\pi$ (for parity odd operators), respectively. Concretely, the computation of the phases for the momentum dissipation case has been performed for $\alpha/T=6$.}
\label{tab1}
\end{table}

 \subsubsection{Momentum dependence of the amplitudes}
\begin{figure}[ht!]
     \centering
     \includegraphics[width=0.32\linewidth]{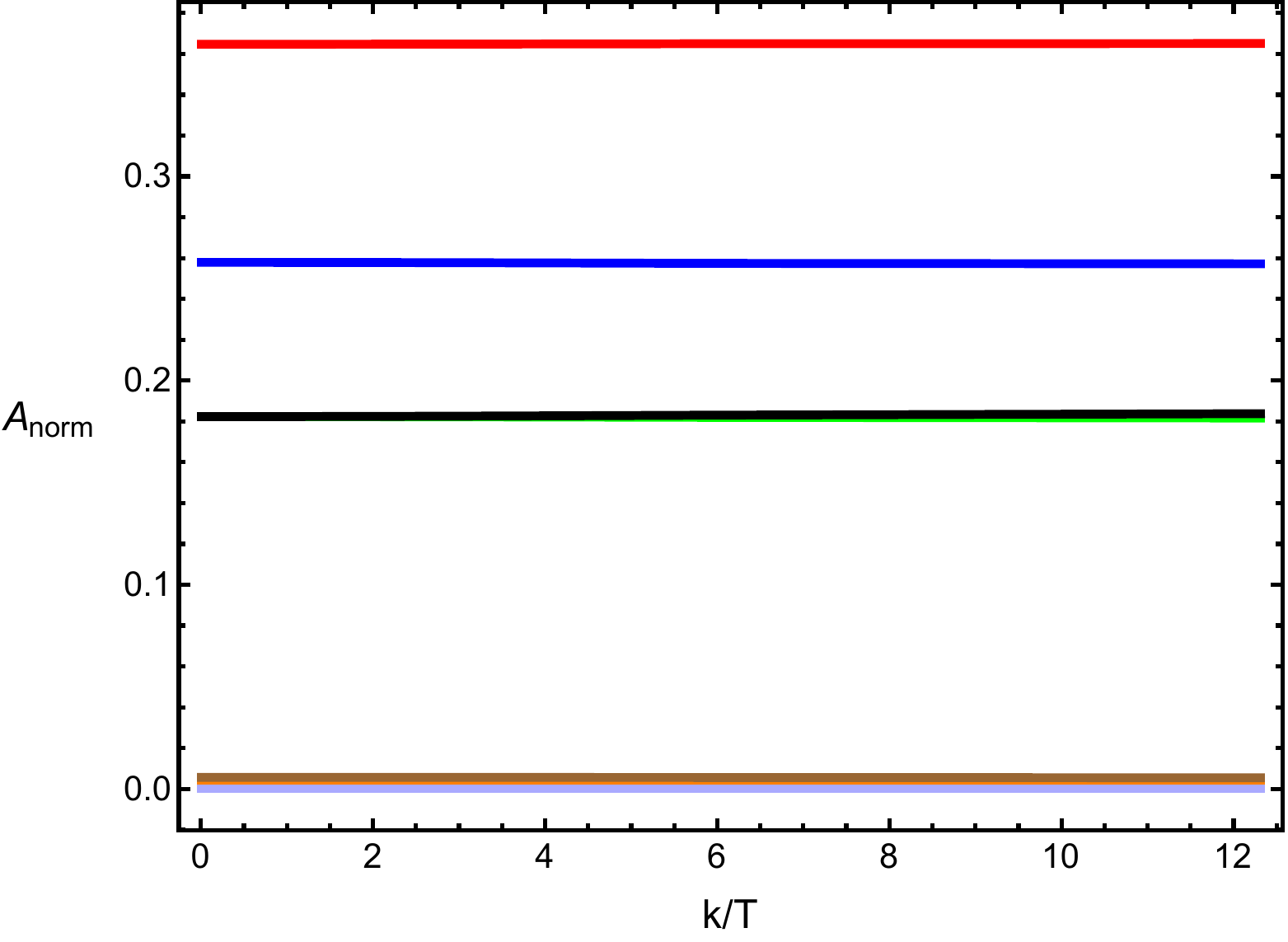}
       \includegraphics[width=0.32\linewidth]{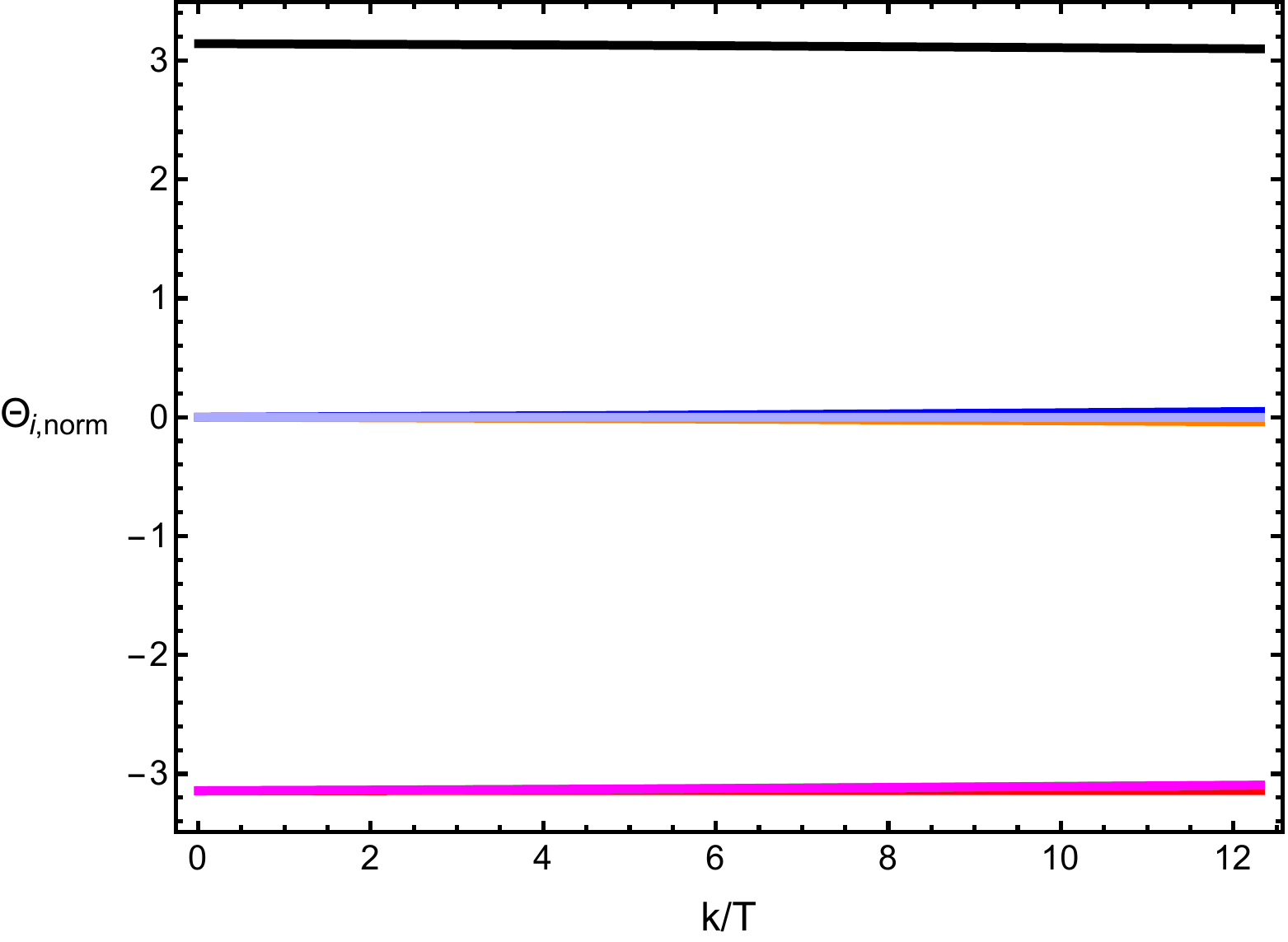}
       \includegraphics[width=0.32\linewidth]{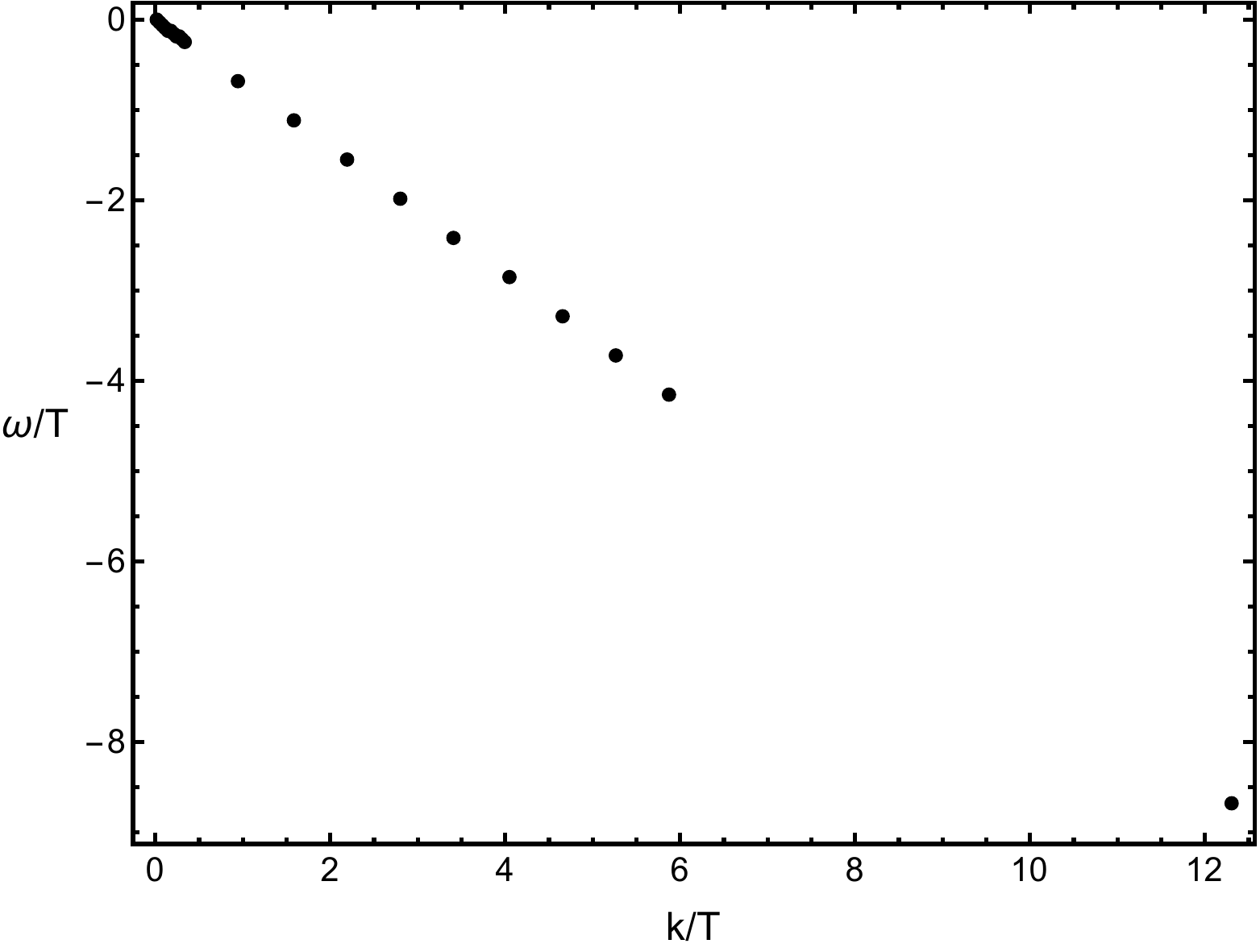}\\
       
       \vspace{0.4cm}
       
        \includegraphics[width=0.32\linewidth]{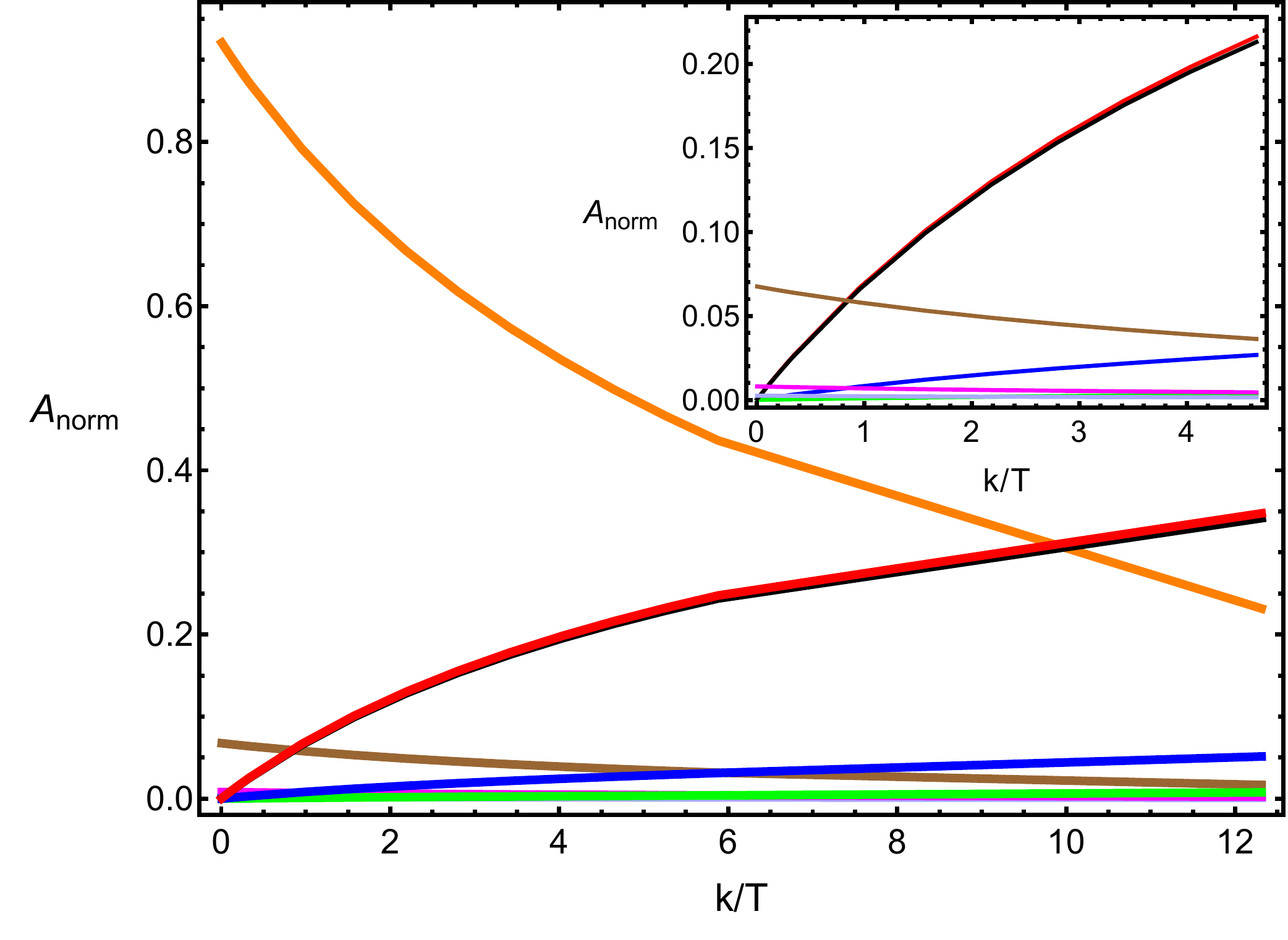}
       \includegraphics[width=0.32\linewidth]{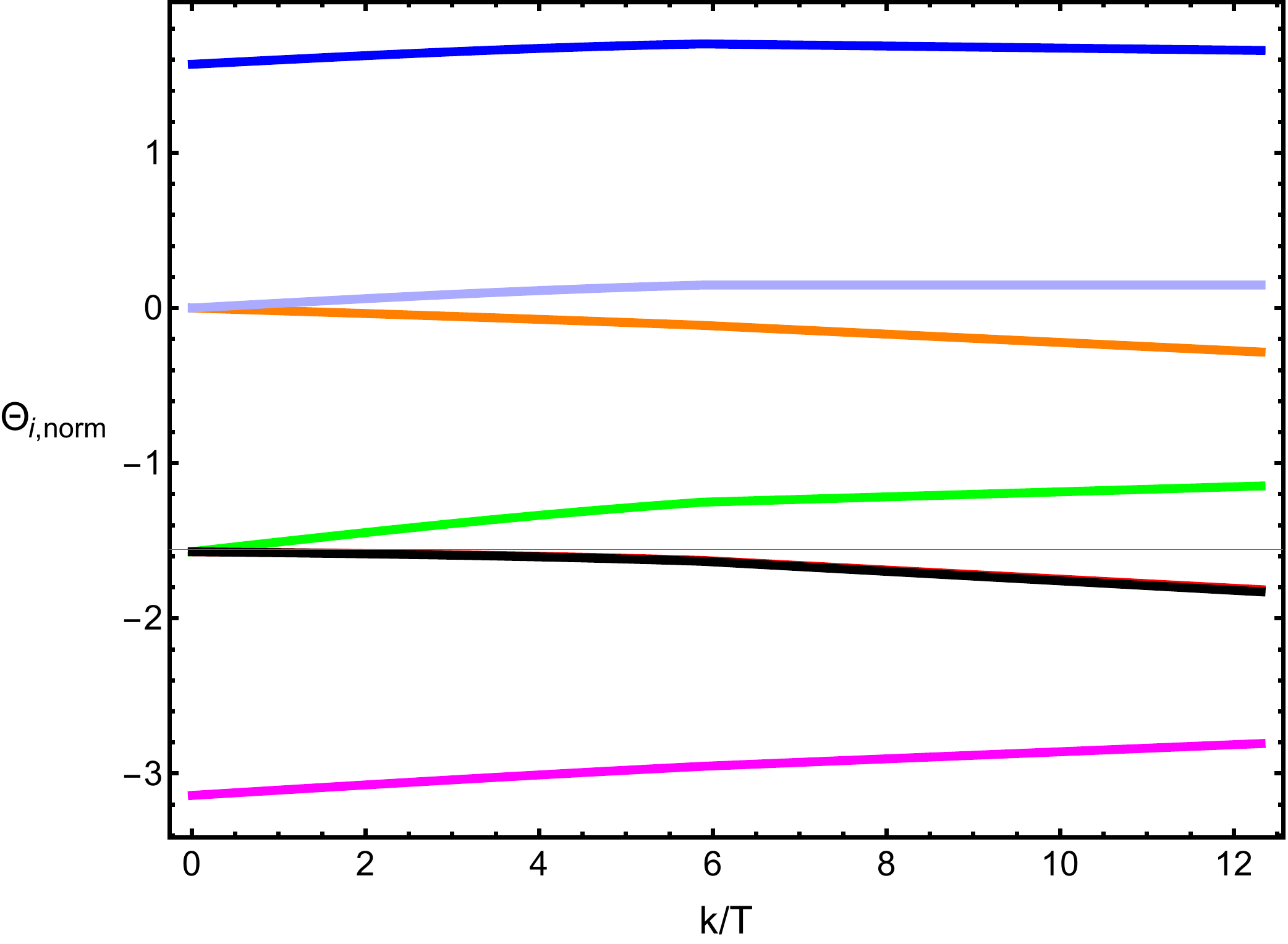}
       \includegraphics[width=0.32\linewidth]{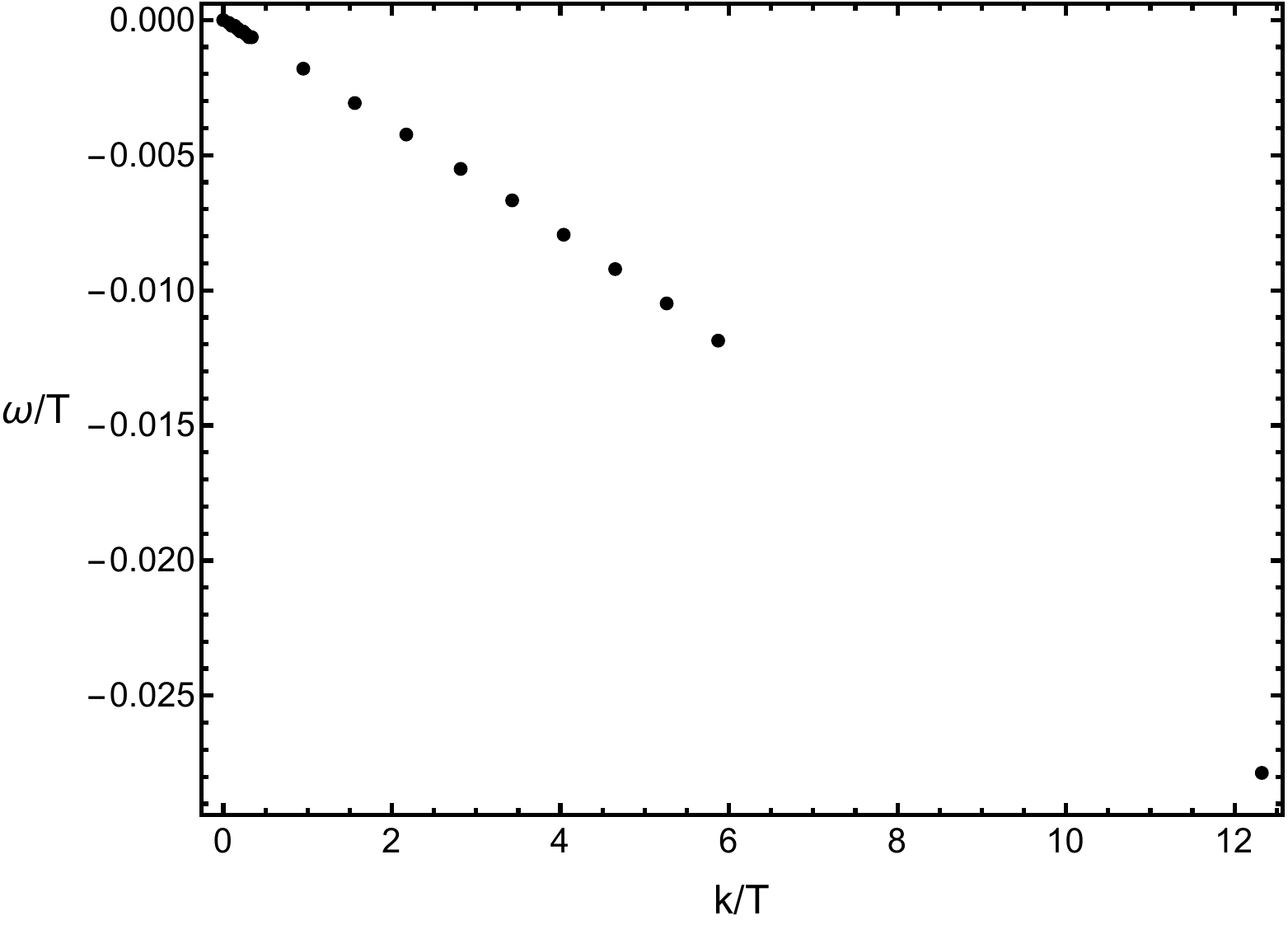}\\
       
       \vspace{0.4cm}
       
         \includegraphics[width=0.32\linewidth]{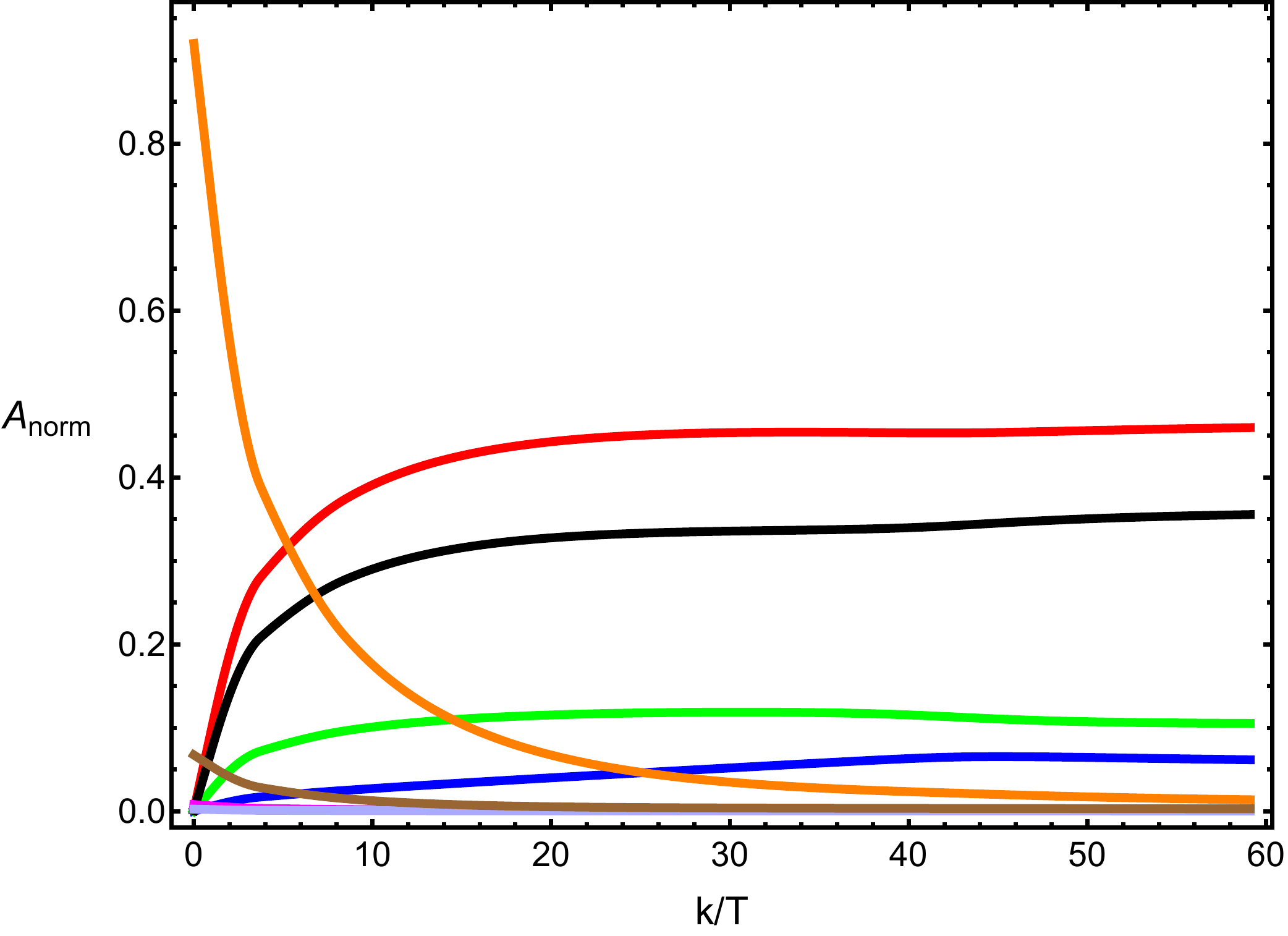}
       \includegraphics[width=0.32\linewidth]{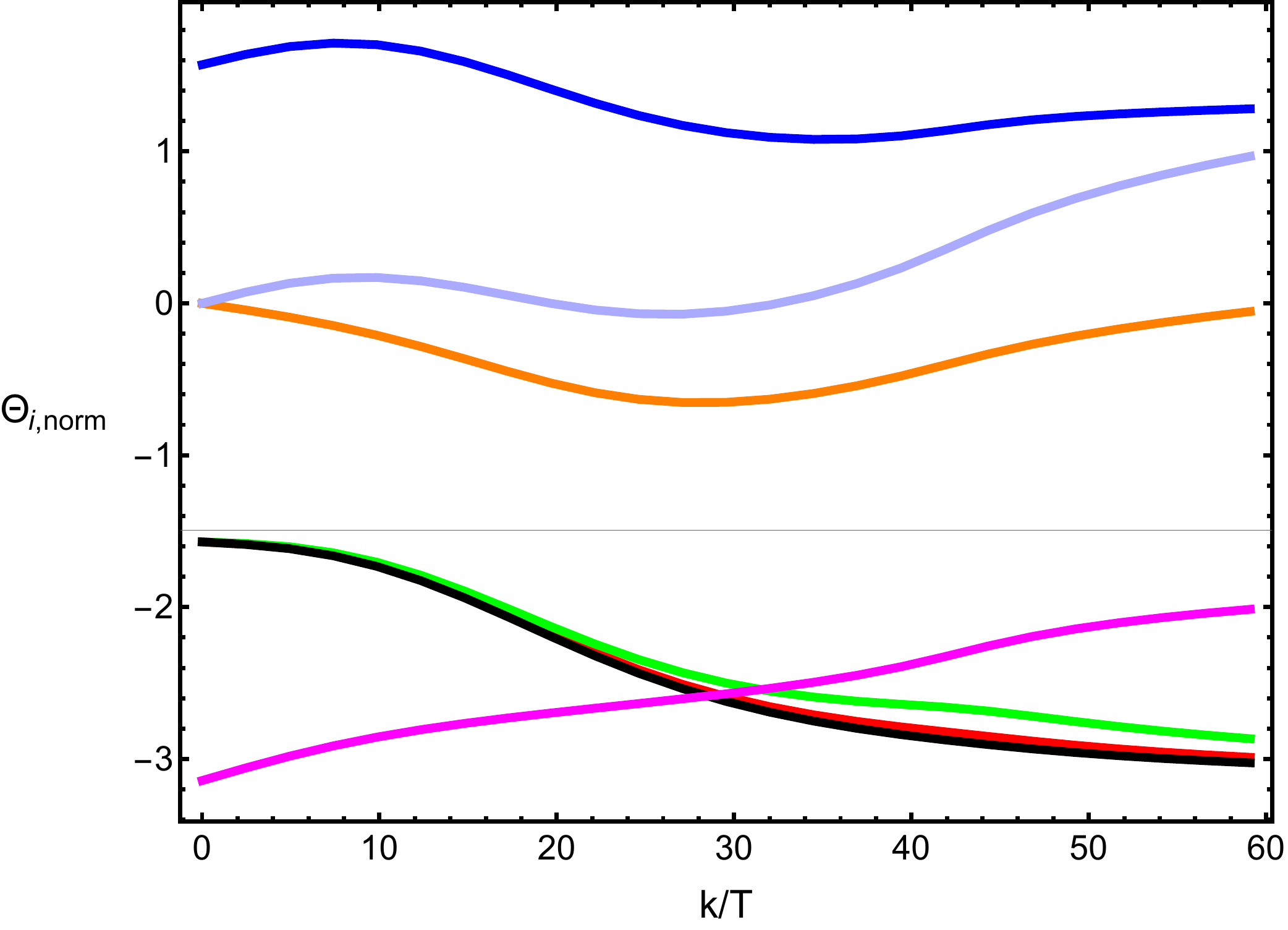}
       \includegraphics[width=0.32\linewidth]{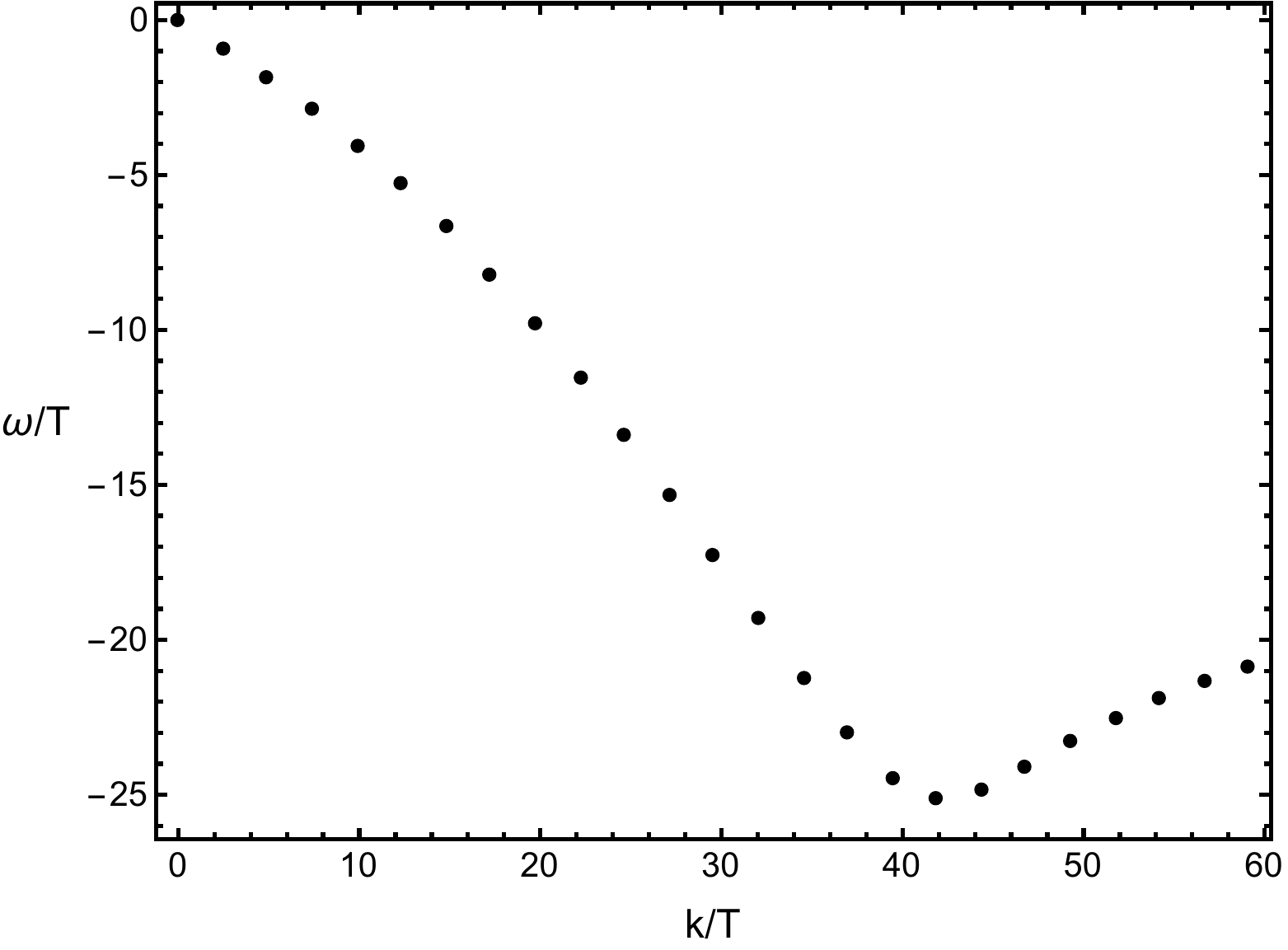}
     \caption{The momentum dependence of the normalized amplitudes \eqref{legends} at $T/\mu=0.016\, (T/T_c=0.804)$. \textbf{Top panels: }First sound. \textbf{Central panels: } Second sound; The inset shows a zoom on small momenta (and we omit plotting the amplitude of the Higgs fluctuation in orange). \textbf{Bottom panels: }Second sound beyond the linear part of the dispersion relation ($k/T \gg 1$).}
     \label{fig:boh}
 \end{figure}

So far, all the fluctuations amplitudes have been computed for a single value of momentum, $k/T=0.01$.
In this section, we briefly discuss the momentum dependence of the amplitudes presented in section \ref{ssec:amps}. The results are shown in fig.~\ref{fig:boh}.\\
For first sound, both the amplitudes and the relative phases are constant in a large range of momenta (we considered them until $k/T \approx 12$). In the case of second sound, we observe that the biggest amplitude, the Higgs fluctuations, remains dominant until $k/T \approx 10$ and the relative phases approximately constant. Away from the small momentum limit, we notice that the contribution from the fluctuations of the superfluid condensate becomes smaller and the mode is eventually dominated by the energy fluctuations which are the biggest contribution at large $k$. At the same time, at $k/T \sim 1$, we observe a crossover between the fluctuations of the charge density (which is the first subleading contribution at small $k$, even thought it accounts for less than $10\%$ of the total weight) and that of energy and longitudinal momentum which become more relevant at large wave-vector.\\
All in all, we can state that, in the hydrodynamic limit $k/T\ll1$, our results depend only minimally on the specific value of the momentum $k$.
\section{Alternative approach via complex momentum modes}
\label{sec4}
\subsection{Theory}
Quasinormal modes can be understood as the response of the system to a perturbation that happened in the past. Thus, quasinormal modes describe how the system relaxes back to equilibrium at late times. 

A different way of studying the response is by perturbing the system with a periodic source of frequency $\omega$ and looking for the corresponding wave with complex wave number $q=k+ i \kappa$. If we were to excite only one particular mode then the response would be
\begin{equation}
    \Phi^I(z,t,x) = \tilde{A}^I_n(\omega,z) e^{-i\omega t + i k_n x} e^{-\kappa_n |x|}\,.
\end{equation}
The complex momentum is now a function of the real frequency $k_n=k_n(\omega)$ and $\kappa_n=\kappa_n(\omega)$. This was investigated in \cite{Amado:2008ji} where it was shown that one can isolate the modes that decay exponentially by choosing the integration in the complex momentum plane appropriately. Only a few other works considered this choice within the holographic framework \cite{Amado:2007pv,Forcella:2014dwa,Blake:2014lva}. 

We can now proceed with the same attitude as in the previous section and investigate the amplitudes and the relative phases. Whether these are different from the ones of the quasinormal modes is an open question and it is our main motivation for this last analysis. A naive expectation would be that the phases and amplitude ratios are the same as long as the mode in question has not gone through some crossing point with another mode in the complex plane. Generally, such a crossing point has complex frequency and complex momentum so its neither visible from the quasinormal mode (real momentum) nor from the complex momentum mode (real frequency). 

 \subsection{Results}
\begin{figure}[ht!]
     \centering
     \includegraphics[width=0.45\linewidth]{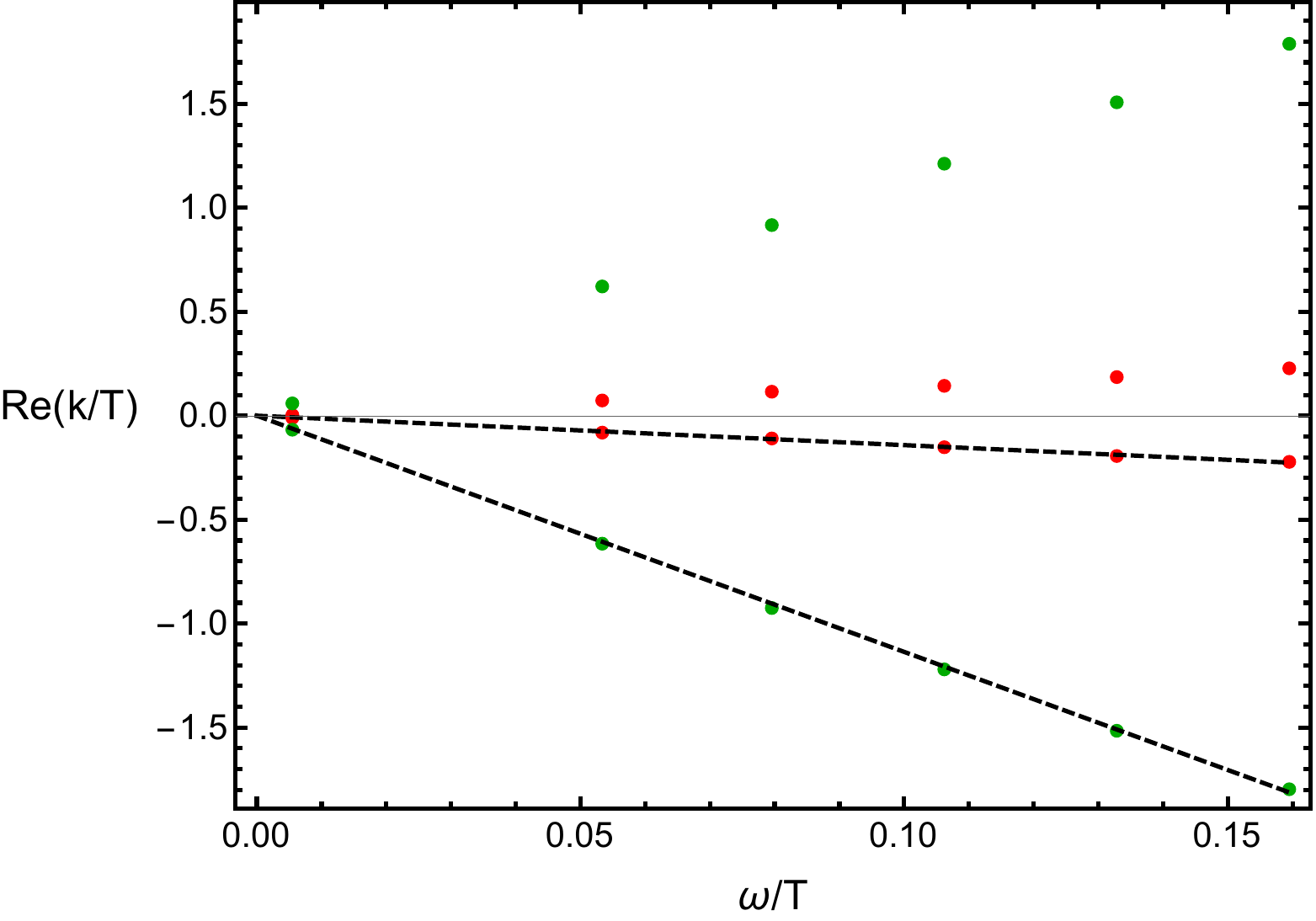}
       \includegraphics[width=0.45\linewidth]{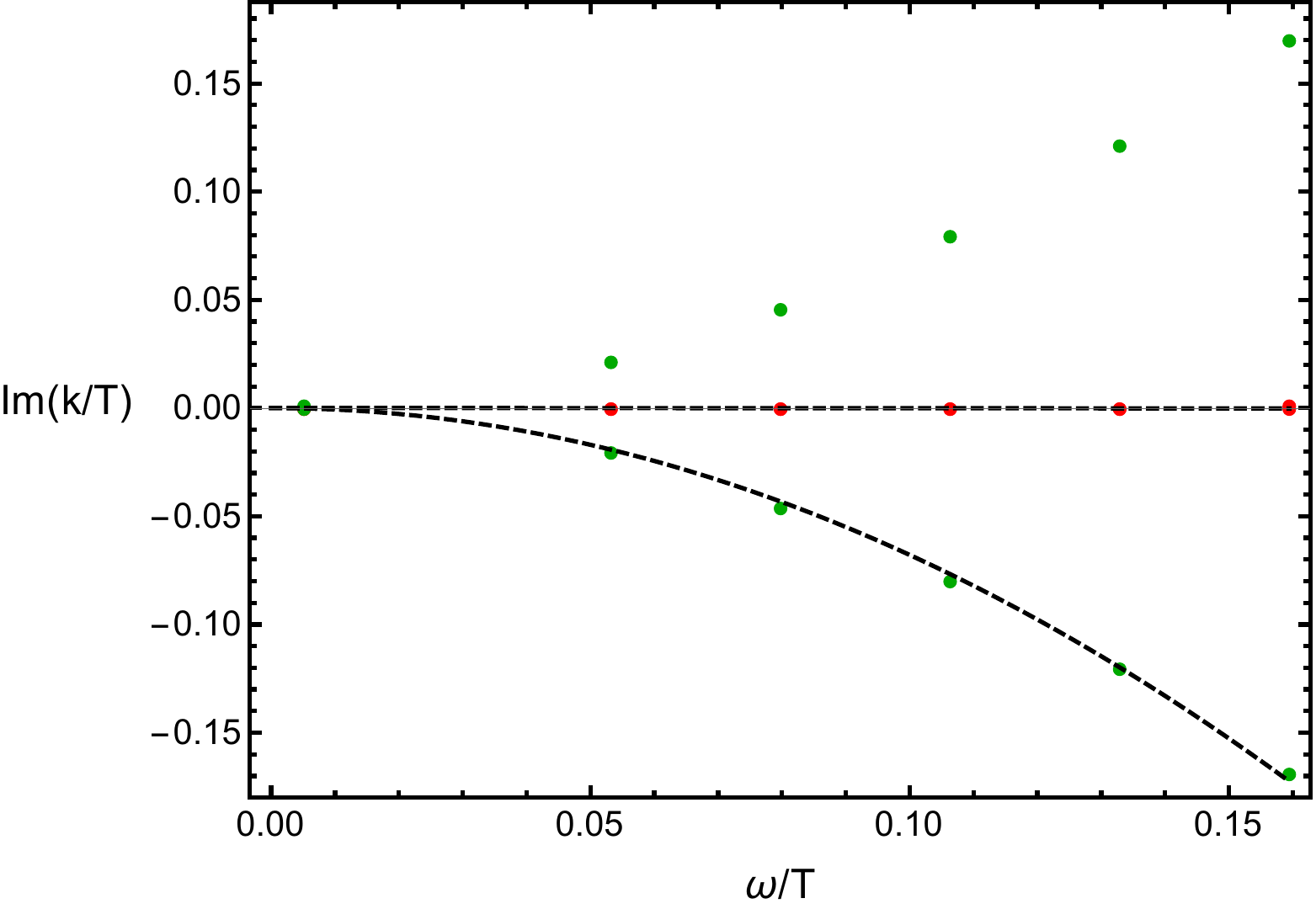}\\
       
       \vspace{0.3cm}
       
       \includegraphics[width=0.45\linewidth]{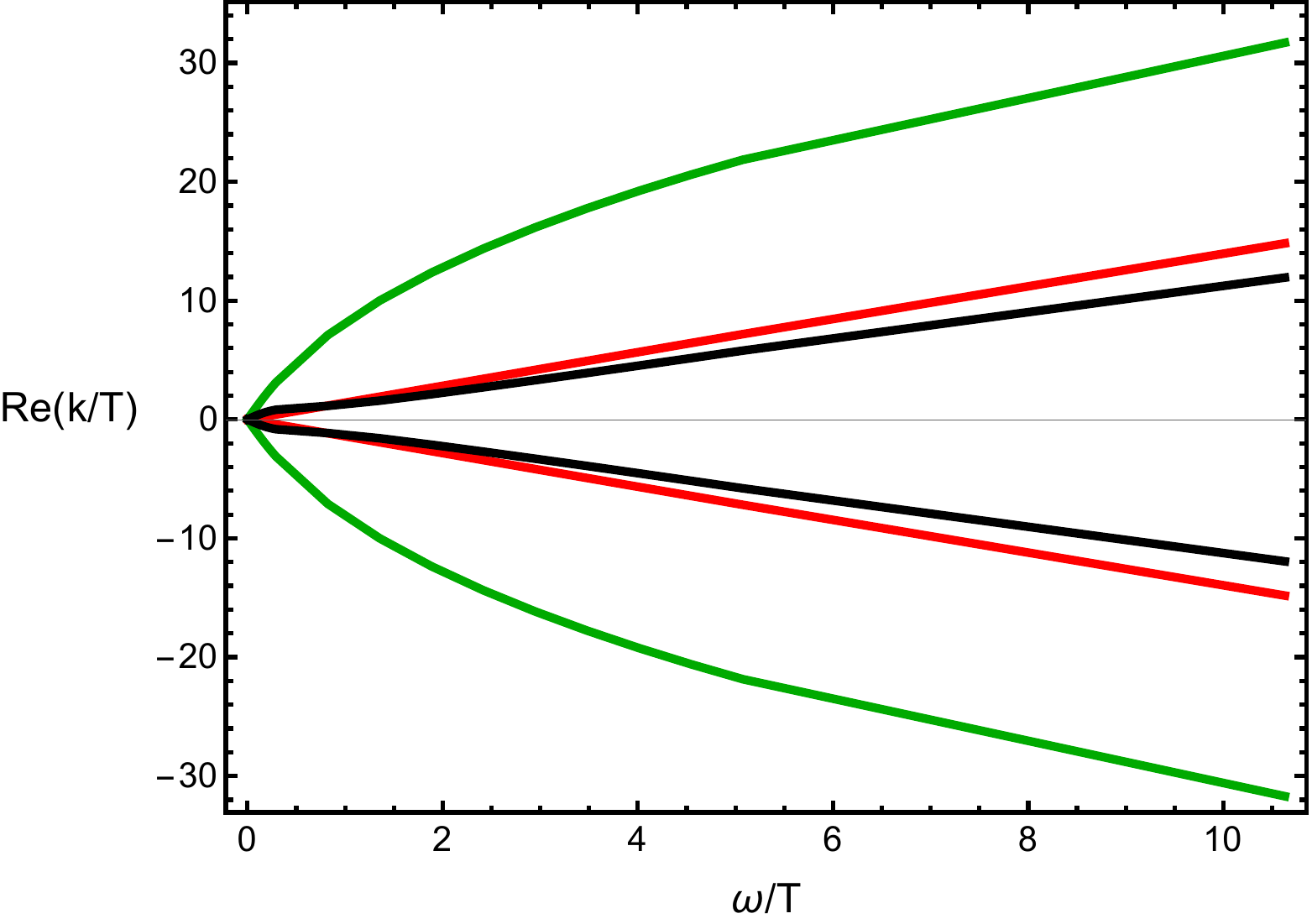}
       \includegraphics[width=0.45\linewidth]{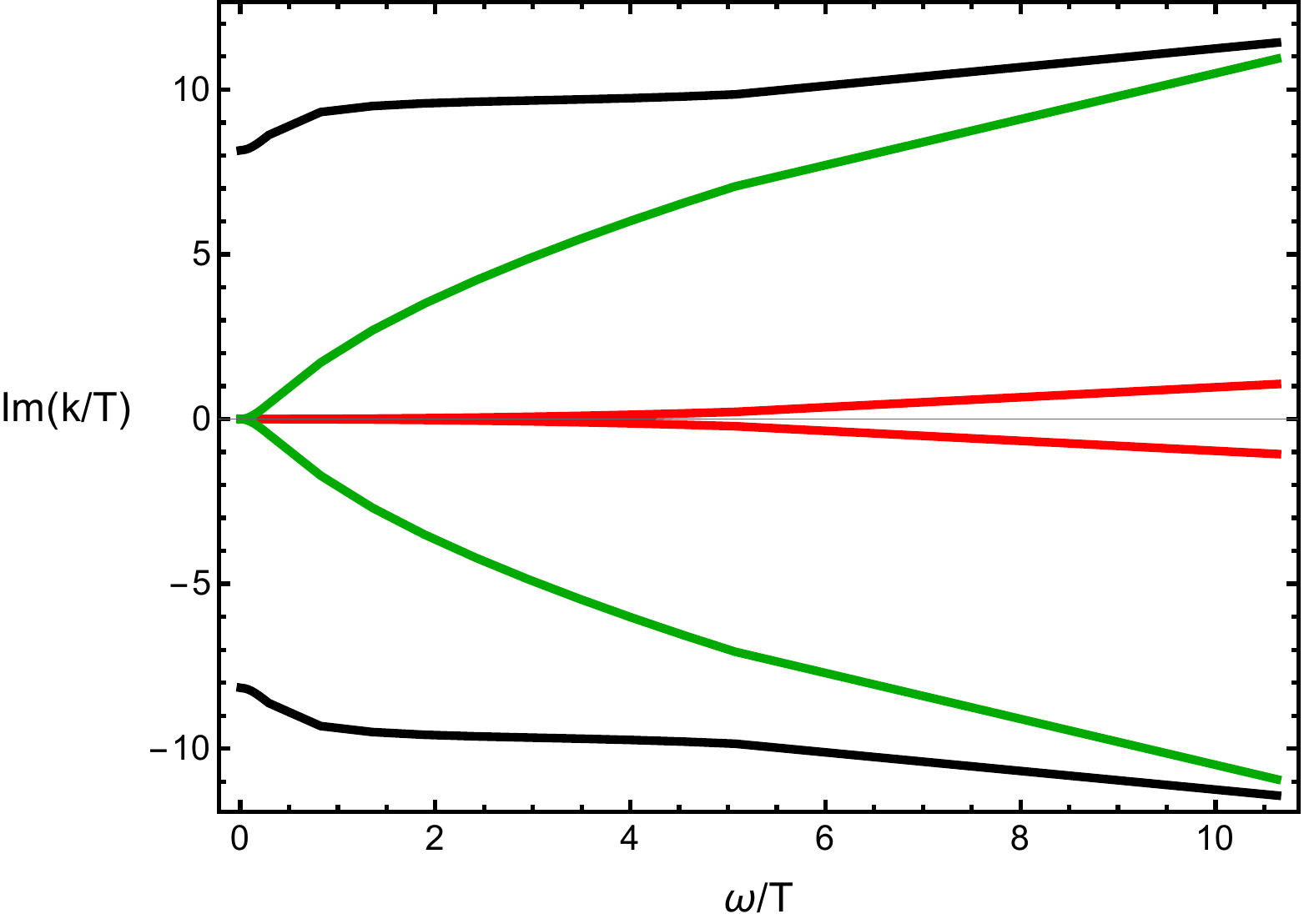}
     \caption{The three lowest modes in the superfluid phase. The temperature is fixed at $T/\mu=0.019$ ($T/T_c=0.92$). Second sound (\color{green2}green\color{black}), first sound (\color{red}red\color{black}), the former damped diffusive mode (black). The dashed lines in the top panel show the expectation from eq.~\eqref{e1k}.}
         \label{fa1}
 \end{figure}

\begin{figure}[ht!]
     \centering
      \includegraphics[width=0.45\linewidth]{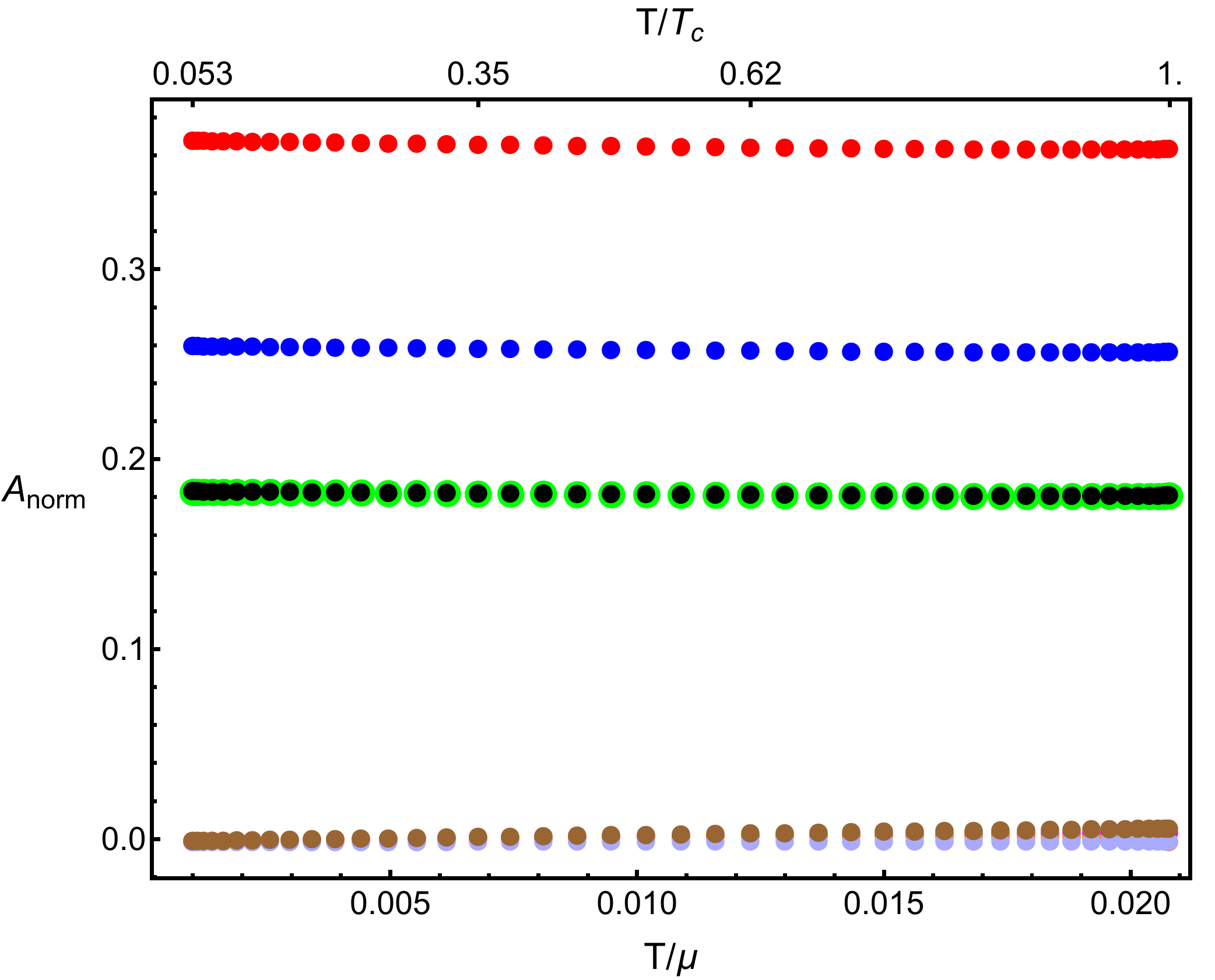}
     \includegraphics[width=0.45\linewidth]{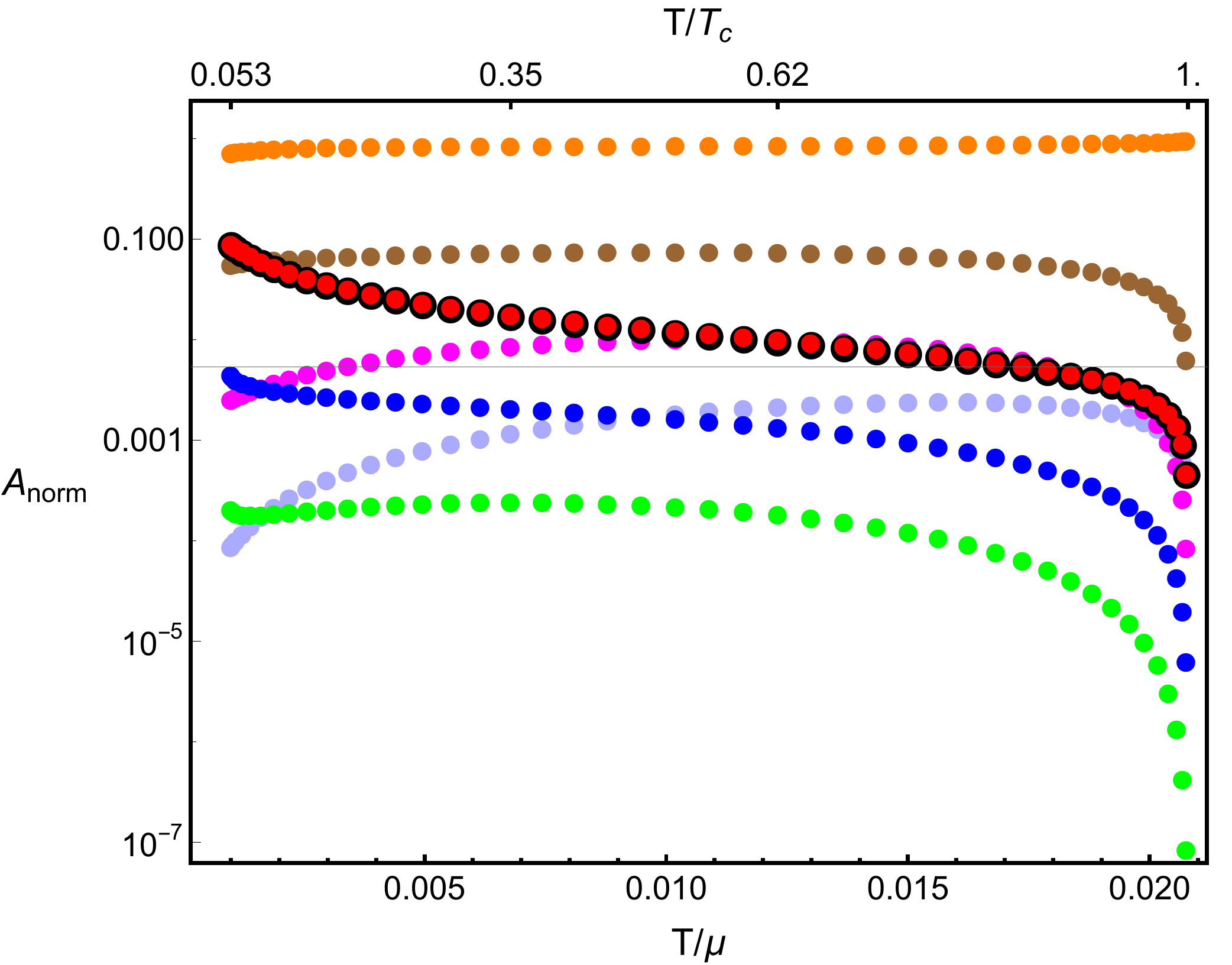}\qquad
     \caption{The amplitudes of the fluctuations \eqref{legends} for first (\textbf{left}) and second sound (\textbf{right}) at fixed $\omega/T=0.01$ in the superfluid phase ($T<T_c$) without axion fields.\label{fig:1222}}
 \end{figure}

 Before discussing the amplitudes and the phases of the various operators, let us briefly comment on the nature of the low-energy modes for complex momentum and real frequency.
 Let us start with the equation for a generic attenuated sound wave
 \begin{equation}
     \omega^2\,=\,v^2\,k^2\,-\,i\,\omega\,2\,\Gamma\,k^2\,+\,\dots\,, \label{bb}
 \end{equation}
 where as always $v,\Gamma$ are the propagation speed and the attenuation constant, respectively. Let us now consider eq.~\eqref{bb} in terms of a real-valued frequency and a complex valued momentum. At low frequency, it is easy to verify that
 \begin{align}
  & \mathrm{Re}(k)\,=\,\pm\,\frac{\omega}{v}\,+\,\mathcal{O}\left(\omega^3\right)\,,
     & \mathrm{Im}(k)\,=\,\pm\,\frac{\Gamma}{v^3}\,\omega^2\,+\,\mathcal{O}\left(\omega^4\right)\,.\label{e1k}
     \end{align}
     The same analysis can be repeated with a damped diffusive mode (such as the diffusive mode in the superfluid phase)
     \begin{equation}
         \omega\,=\,-\,i\,\gamma\,-\,i\,D\,k^2
     \end{equation}
     yielding
     \begin{align}
  & \mathrm{Re}(k)\,=\,\pm\,\frac{\omega}{2\,\sqrt{\gamma\,D}}\,+\,\mathcal{O}\left(\omega^3\right)\,,
     & \mathrm{Im}(k)\,=\,\pm\,\sqrt{\frac{\gamma}{D}}\,+\,\mathcal{O}\left(\omega^2\right)\,.\label{e2k}
     \end{align}
     The equations~\eqref{e1k} and~\eqref{e2k} represent the low-momentum behavior of the low-energy modes in this new inverted picture. As expected, our numerical results shown in fig.~\ref{fa1} are in perfect agreement with this picture.
 We can now analyze the amplitudes of the fluctuations in \eqref{legends} at constant frequency. We show them in fig.~\ref{fig:1222} for a fixed small value of (real) frequency $\omega/T=0.01$. The results for first sound are exactly identical to those obtained at complex frequency and real momentum (inset in the left panel of fig.~\ref{fig:ampsounds}). In the second sound case shown on the right panel of 
 fig.~\ref{fig:1222} (to be compared with the right panel of fig.~\ref{fig:ampsounds}), the dominant contributions
 corresponding to the condensate and charge density fluctuations are identical while the subleading contributions are slightly different between the real and imaginary frequency modes. Still the qualitative behavior in the two frameworks is very similar. 
\section{Conclusions}\label{sec5}
In this work we have revisited the holographic S-wave superfluid model \cite{Hartnoll:2008kx}
focusing on the hydrodynamic description of the low-energy modes and taking into account the backreaction on the gravity side. 
Despite the long history of this model, a complete analysis of this sort was missing.
Through this approach
we have explicitly verified the match between relativistic superfluid hydrodynamics and the holographic model, both in terms of the dispersion relations and, most importantly, 
in terms of all the first order transport coefficients. Our study has revealed that all the hydrodynamic transport coefficients are continuous at the phase transition. Moreover, differently from what \cite{Amado:2013xya} found in the probe limit, even the second sound attenuation constant is smoothly connected to the gapped scalar modes in the normal phase. We also analyzed the low energy modes for real frequency and complex momentum, finding agreement with the hydrodynamic description.
Finally, we extended the analysis to a setup where momentum
was not conserved and succesfully matched the peculiar fourth sound mode with its hydrodynamic counterpart. Interestingly, we find a non-zero normal density in the setup with broken translations, similarly to the results of~\cite{Gouteraux:2019kuy,Gouteraux:2020asq} for hyperscaling-Lifschitz type geometries.
\\

After demonstrating the validity of the hydrodynamic description of the holographic superfluid,
we carried out a study of the support of the various hydrodynamic modes in terms of the dual field theory operators. In a nutshell, we have quantified ``how much'' a single boundary field theory operator contributes to a specific hydrodynamic mode. The analysis has been performed both in the superfluid  and normal phases and also in presence of momentum dissipation. In contrast to the field theory results of~\cite{Alford:2013koa}, we do not observe a role-reversal phenomenon between first and second sound in terms of the support of those modes. From the amplitudes and phases, we can reconstruct the dispersion relations of the hydrodynamic modes. This allows us to compute the speed of sound and the attenuation/diffusion constants from our data of the amplitudes and relative phases at a fixed value of the momentum. We notice that the absolute values of the amplitudes give us insights about the speed of the sound modes, and the relative phases between the fluctuations are related to damping. Furthermore, the amplitudes and relative phases give us an (computational) easy recipe to compute the hydrodynamic dispersion relations since it is sufficient to determine them at only one point of the dispersion relation (within the radius of convergence). Beyond hydrodynamics, we observe a
purely imaginary gapped mode in the superfluid phase which is solely supported by the scalar fluctuations; this could be the Higgs mode discussed in~\cite{Behrle_2018,Li:2013fhw,Pekker_2015}.\\

In the future, it would be interesting to enlarge the analysis
in this work to scenarios like those in the following list
\begin{itemize}
    \item Consider more complex symmetry breaking patterns such as the pseudo-spontaneous breaking of the $U(1)$ symmetry (as initiated in \cite{Donos:2021pkk}).
    Even though the explicit breaking of a $U(1)$ vector symmetry is definitely not a physical option, this could serve as a toy-model to improve our understanding of the pseudo-spontaneous breaking of translations in the homogeneous models \cite{Baggioli:2014roa,Andrade:2017cnc,Alberte:2017cch,Amoretti:2016bxs,Amoretti:2017frz,Amoretti:2019kuf,Ammon:2019apj,Ammon:2020xyv} and the possible relation between phase relaxation and pseudo-spontaneous breaking \cite{Baggioli:2019abx,Ammon:2019wci,Amoretti:2018tzw,Amoretti:2019cef,Baggioli:2020haa,Baggioli:2020edn,Baggioli:2020nay}. Finally, if these two breaking mechanisms are analogous or even equivalent, one could try to tackle the phenomenology of superfluid current relaxation induced by vortices \cite{Davison:2016hno} within this effective construction.

    \item Since we investigate the match of hydrodynamics and our holographic model, it would be interesting to determine the radius of convergence of linearized superfluid hydrodynamics throughout the phase diagram following the recent studies \cite{Grozdanov:2019kge,Grozdanov2019,Withers:2018srf,Abbasi:2020ykq,Abbasi:2020xli,Jansen:2020hfd,Baggioli:2020loj,Arean:2020eus,Wu:2021mkk,Baggioli:2021ujk,Grozdanov:2021gzh,Grozdanov:2020koi,Jeong:2021zsv} in other holographic models.
   
    \item Compute the second order contributions to equilibrium transport, as initiated in~\cite{Grieninger:2021rxd,Kovtun:2018dvd}, in superfluids and systems with second order phase transitions in general.
    
    \item Work out the hydrodynamics of holographic superfluids in presence of a background superfluid velocity \cite{Arean:2010zw,Arean:2010wu,Arean:2011gz}. The excitations in the probe limit have been already presented in \cite{Amado:2013aea} and the dynamical Landau instability in \cite{Lan:2020kwn}. It would be interesting to extend them to the fully backreacted scenario and compare them to the hydrodynamic description. 
    \item Holographic superfluid models in presence of disorder have been constructed in the past years \cite{Arean:2013mta,Arean:2015sqa}. It would be interesting to study the transport of those models and compare it with our results in the simpler axion model.
\end{itemize}
We leave these questions for the near future.
\section*{Acknowledgements}
We thank Martin Ammon, Andrea Amoretti, Se\'an Gray,  Hyun-Sik Jeong, Wei-Jia Li, Yan Liu, and Ya-Wen Sun for useful discussions on related topics. We thank Blaise Gout\'eraux and Richard Davison for several useful comments on the first version of this manuscript and for useful discussions about the interpretation of our results. D.A. and S.G. are supported by the `Atracci\'on de Talento' program (2017-T1/TIC-5258, Comunidad de Madrid) and through the grants SEV-2016-0597 and PGC2018-095976-B-C21. M.B. acknowledges the support of the  Shanghai Municipal Science and Technology Major Project (Grant No.2019SHZDZX01).

\appendix

\section{The pocket dictionary}\label{app1}
In this appendix, following the standard holographic renormalization procedure, we compute the
one- and two-point functions relevant for the analysis
in the main text.
We start by computing the variation of the renormalized on-shell action. To the action \eqref{eq:action} we
add the standard counterterms
\begin{equation}
S_{\text{ct}}={1\over2\kappa_4^2} \int \dd^3x\sqrt{-\gamma}\,\left(2K-4 -|\psi|^2 - R^{(\gamma)} \right)\,.
\end{equation}
A straightforward computation leads to the following
expression for the variation of the renormalized
on-shell action (see also~\cite{Herzog:2009md})
\begin{align}
\!\delta S^{\rm o-s}_{\rm ren}&\!=\!-\!
\int\! \dd^3x\sqrt{-\gamma}\bigg\{\!n_m F^{mn}\delta A_n
+\!2n_m\left[\delta\psi g^{mn}\left(\partial_n\!+\!iq\,A_n\right)
\bar\psi\!+\!\delta\bar\psi g^{mn}\left(\partial_n\!-\!iq\,A_n\right)
\psi\right]\nonumber\\ &-\delta\gamma^{mn}\!\left(\!
K_{mn}\!-(K+2+\!{|\psi|^2\over2}+\!{R^{(\gamma)}\over2})\,\gamma_{mn}\!
-\!R^{(\gamma)}_{mn}\!
\right)+2\psi\delta\bar\psi+2\bar\psi\delta\psi
\bigg\}\bigg|_{z=\epsilon}\!\!\!\!\!\!\!,
\label{eq:dosact}
\end{align}
where $\gamma_{mn}$ denotes the induced metric at 
a $z=\epsilon$ hypersurface, while $K_{mn}$ is the
extrinsic curvature at that hypersurface, $K$
is its trace, and $R^{(\gamma)}_{mn}$ the Ricci tensor.\footnote{We write the induced metric and
extrinsic curvature in the four-indices notation,
see \textit{e.g.}~\cite{McNeesnotes}.}

In order to read the one-point functions from
\eqref{eq:dosact} we need the boundary expansions
of the bulk fields. These take the generic form
\begin{align}
    &g_{\mu\nu}=g_{\mu\nu}^{(0)}(1+o(z))+g_{\mu\nu}^{(3)}\,z^3(1+o(z))    \,,\\ &A_\mu=A_\mu^{(0)}+A_\mu^{(1)}z+o(z^2)\,,
    \qquad\psi=\psi^{(1)}z+\psi^{(2)}z^2+o(z^3)\,.\label{eq:bdryexp}
\end{align}
For the background solution (indicated with a hat symbol), we have
\begin{equation}
    \hat \psi^{(1)}=0\,,\qquad 
    \hat \psi^{(2)}=\langle\mathcal O^\psi_\text{cond}\rangle/2\,,\qquad
    \hat A_t^{(0)}=\mu\,,\qquad
    \hat A_t^{(1)}=-\rho,\qquad \hat g_{tt}^{(3)}=-u_3\,.
    \label{eq:bckbdry}
\end{equation}
Hence, we are taking the leading
contribution of the scalar towards the boundary
to be real. As usual this will correspond to picking
a real value for the symmetry-breaking condensate.

Equations  (\ref{eq:dosact}-\ref{eq:bckbdry})
result in the following expressions for the
expectation values of the dual field theory operators
up to first order in fluctuations
\begin{subequations}
\begin{align}
&\langle\Psi\rangle=
{1\over\sqrt{-\gamma_{(b)}}}{\delta S^{\rm o-s}_{\rm ren}\over\delta \bar\psi_{({\rm b})}}=
\langle\mathcal O^\psi_\text{cond}\rangle+2\psi^{(2)}(x,t)\,,\\
&\langle\bar\Psi\rangle=
{1\over\sqrt{-\gamma_{(b)}}}{\delta S^{\rm o-s}_{\rm ren}\over\delta \psi_{({\rm b})}}
=\langle\mathcal O^\psi_\text{cond}\rangle^*+2\psi^{(2)*}(x,t)\,,\\
&\langle J^t\rangle=
{1\over\sqrt{-\gamma_{(b)}}}{\delta S^{\rm o-s}_{\rm ren}\over\delta A^t_{({\rm b})}}
=
\rho - a_t^{(1)}(x,t)\,,\\
&\langle J^i\rangle=
{1\over\sqrt{-\gamma_{(b)}}}{\delta S^{\rm o-s}_{\rm ren}\over\delta A^i_{({\rm b})}}
=a_i^{(1)}(x,t)\,,\quad (i=x,y)\,,\\
&\langle T_{tt}\rangle= 
{2\over\sqrt{-\gamma_{(b)}}}{\delta S^{\rm o-s}_{\rm ren}\over\delta \gamma^{tt}_{({\rm b})}}=
2u_3
-2h_{tt}^{(3)}(x,t)\,,\\
&\langle T_{xx}\rangle= 
{2\over\sqrt{-\gamma_{(b)}}}{\delta S^{\rm o-s}_{\rm ren}\over\delta \gamma^{xx}_{({\rm b})}}=
u_3-h_{tt}^{(3)}(x,t)+3h_{yy}^{(3)}(x,t)\,,\\
&\langle T_{yy}\rangle= 
{2\over\sqrt{-\gamma_{(b)}}}{\delta S^{\rm o-s}_{\rm ren}\over\delta \gamma^{yy}_{({\rm b})}}=
u_3-h_{tt}^{(3)}(x,t)+3h_{xx}^{(3)}(x,t)\,,\\
&\langle T_{tx}\rangle=-3h_{tx}^{(3)}(x,t)\,,\qquad
\langle T_{ty}\rangle=-3h_{ty}^{(3)}(x,t)\,,\qquad
\langle T_{xy}\rangle=-3h_{xy}^{(3)}(x,t)\,,
\end{align}
\label{eq:onepoints}
\end{subequations}
where the energy density is given by $\varepsilon=2\,u_3$.
The subindex (b) stands for the finite value of the different fields towards the boundary, hence once stripped of the powers of $z$
(\textit{e.g.} $\delta\gamma_{mn(b)}=\delta \gamma_{mn}/z^2$).

It is worth pointing out that the boundary expansions
\eqref{eq:bdryexp} fulfill the necessary relations
for the relevant Ward Identities to be satisfied.
First, they obey the constraint
\begin{equation}
    h_{xx}^{(3)}(x,t)+h_{yy}^{(3)}(x,t)=0\,,
\end{equation}
guaranteeing that $\langle T^\mu_\mu\rangle =0$, as required by conformal invariance.
As for the Ward Identity $\langle \nabla^\mu T_{\mu\nu}\rangle =0$, that results from translational
invariance, it boils down to the following
constraints:
\begin{subequations}
\begin{align}
&2 \omega h_{tt}^{(3)}(\omega,k)+3k h_{tx}^{(3)}(\omega,k)=0\,,\\
&-3\omega h_{tx}^{(3)}(\omega,k)+k(-h_{tt}^{(3)}(\omega,k)+3h_{yy}^{(3)}(\omega,k))=0\,,
\end{align}
\label{eq:transward}
\end{subequations}
where we have assumed an harmonic spacetime dependence
of the fluctuations as in~\eqref{eq:harmonic}.
Finally, the Ward Identity $\langle\nabla_\mu J^\mu\rangle=0$, stemming from the global $U(1)$ symmetry, is satisfied
once
\begin{equation}
k\,a_x^{(1)}+\omega\, a_t^{(1)} =0\,,
\label{eq:u1ward}
\end{equation}
holds.
By solving the equations \eqref{eq:eoms} towards the boundary we have checked that indeed the constraints
(\ref{eq:transward}, \ref{eq:u1ward}) are 
satisfied.

From the Ward identities and the tracelessness of the energy-momentum tensor, we find the following relations between the expectation values of the fluctuations
\begin{align}
   & \langle\delta T^{tt}\rangle=\frac{1}{1-\omega^2/k^2}\,\langle \delta T^{yy}\rangle=-\frac{k}{\omega}\,\langle \delta T^{tx}\rangle=\frac{k^2}{\omega^2}\,\langle \delta T^{xx}\rangle,\label{eq:wardrel1}\\& \langle \delta J^t\rangle=\frac k\omega\,\langle \delta J^x\rangle,\label{eq:wardrel2}
\end{align}
where the $\delta$ in the expectation values indicates that we are referring to the fluctuations.
\\
Regarding the axion fields introducing
momentum dissipation, the boundary expansion
for the axion fluctuations reads we have (see \cite{Andrade:2013gsa} for the derivation)
\begin{equation}
    \phi^I\,=\,\phi_{(0)}^I\,+\,\ldots\,+\,\phi_{(3)}^I\,z^3
\end{equation}
Hence, in addition to the expectation values \eqref{eq:onepoints},
we have the expectation value of the axion fluctuation~\cite{Andrade:2013gsa}
\begin{equation}
    \langle\, \phi^I \rangle\,=\,6\,\phi_{(3)}^I\,.
\end{equation}
for the operator dual to $\phi^I$. Since the axions break translational invariance, the Ward identities discussed above do no longer hold at the level of the fluctuations. For a discussion see~\cite{Kim:2016hzi}.

Finally, by computing
the one-point functions at sources on from \eqref{eq:dosact}, we can
determine the form of the correlators relevant
for the Kubo-formulas employed in the main
text:
\begin{subequations}
\begin{align}
&\langle\Psi\Psi\rangle={\psi^{(2)}\over \psi^{(1)}}   +i(q\mu-\omega)\,,\qquad \langle T_{xy} T_{xy}\rangle=u_3-3{h_{xy}^{(3)}
\over h_{xy}^{(0)}}
+i\omega\left(
{k^2\over2}-\omega^2\right),\\
&
\langle J^x J^x\rangle = {a_x^{(1)}\over a_x^{(0)}}+i\omega
\,,\qquad
\langle J^x J^t\rangle =-{a_x^{(1)}\over a_t^{(0)}}+i k\,.\label{eq:twopointscurr}
\end{align}
\label{eq:twopoints}
\end{subequations}
\section{Dispersion Relations from Amplitudes}\label{sec:dispaampl}
From the amplitudes computed at a certain value of the momentum, we might extract the diffusion constants and the hydrodynamic dispersion relations. The dispersion relation of the purely diffusive mode in the normal phase is for example given by eq.~\eqref{eq:normaldiff}
\begin{equation}
    \omega=-i\,D_q\,k^2.
\end{equation}
Using the Ward identity eq.~\eqref{eq:wardrel2}, we find
\begin{equation}
\langle\delta  J^x\rangle=-i\,D_q\,k\,\langle \delta J^t\rangle    
\end{equation}
or in other words, the diffusion constant is given by
\begin{equation}
D_q   =\frac{i}{k}\frac{\langle\delta  J^x\rangle}{\langle\delta  J^t\rangle}\label{diff:cons}
\end{equation}
which is nothing else than the very well-known Fick's law.
The imaginary unit in eq.~\eqref{diff:cons} simply implies that the amplitudes have a phase difference of $\pi/2$
\begin{equation}
D_q   =\frac{1}{k}\frac{|\langle\delta  J^x\rangle|}{|\langle\delta  J^t\rangle|}\,e^{i(\Theta_{\delta J^x}-\Theta_{\delta J^t}+\pi/2)}\overset{D_q\in \mathbb R}{=}\frac{1}{|k|}\frac{|\langle\delta  J^x\rangle|}{|\langle\delta  J^t\rangle|}\sin(\Theta_{\delta J^t}-\Theta_{\delta J^x}),\label{eq:dampingdiffu}
\end{equation}
where we required the diffusion constant to be real in the last step. Requiring the diffusion constant to be real and positive  and thus the cosine to vanish fixes the relative amplitudes to $\pi/2$ mod $2\pi$. From this equation it is obvious, that damping is intimately related to a phase difference between the fluctuations. We will see the same for sound waves. Note that the momentum dependence of the expectation values is more general, \textit{i.e.} they also contain all higher orders in $k$. 

For a sound mode with speed $v_s$, attenuation constant $\Gamma_s$ and real momentum $k$, we notice with eq. \eqref{eq:wardrel1} that
\begin{equation}
   - v_s+i\,\Gamma_s/2\, k= \frac{\langle\delta  T^{tx}\rangle}{\langle\delta  T^{tt}\rangle}.\label{eq:disp}
\end{equation}
In the case, where we know the speed of sound (first sound), we can simply take the absolute value of this equation and solve for the attenuation constant
\begin{equation}
\Gamma_s=\frac{2}{k}\,\sqrt{\left|\frac{\langle\delta  T^{tx}\rangle}{\langle\delta  T^{tt}\rangle}\right|^2-v_s^2}.
\end{equation}
The easiest way to compute the speed is to use the information about the relative phases and decompose the ratio of the expectation values as
\begin{equation}
   \frac{\langle\delta  T^{tx}\rangle}{\langle\delta  T^{tt}\rangle} =\frac{|\langle\delta  T^{tx}\rangle|}{|\langle\delta  T^{tt}\rangle|}\, e^{i\, (\Theta_{\delta T^{tx}}-\Theta_{\delta T^{tt}})}
\end{equation}
and we thus find for the speed of sound according to eq.~\eqref{eq:disp} and Euler's formula
\begin{equation}
    v_s= \frac{|\langle\delta  T^{tx}\rangle|}{|\langle\delta  T^{tt}\rangle|}\,\cos\left(\Theta_{\delta T^{tt}}-\Theta_{\delta T^{tx}}+\pi\right).\label{eq:speedfromamplitude}
\end{equation}
Note that in the case of first sound we observe a ratio of approximately 0.7 between $|\langle\delta T^{tx}\rangle|$ and $|\langle\delta T^{tt}\rangle|$ (see for example figure \ref{fig:ampnormal} or \ref{fig:ampsounds}) which yields a conformal speed of $1/\sqrt{2}$.
The attenuation constant is then simply given by
\begin{equation}
\Gamma_s=\frac{2}{k}\,\frac{|\langle\delta  T^{tx}\rangle|}{|\langle\delta  T^{tt}\rangle|}\,\sin\left(\Theta_{\delta T^{tt}}-\Theta_{\delta T^{tx}}+\pi\right).\label{eq:damingattenu}
\end{equation}
Instead of the Ward identity between $\langle\delta  T^{tt}\rangle$ and $\langle\delta  T^{tx}\rangle$, we can of course derive similar relations with $\langle\delta  T^{xx}\rangle$ and $\langle\delta  T^{yy}\rangle$ using the tracelessness of the energy-momentum tensor. This is in particular important in the case of 4th sound, where the Ward identity associated with energy momentum conservation no longer holds. Independently, we may also use eq.~\eqref{eq:wardrel2} to express the speed and attenuation constant in terms of the expectation values. 
\section{Eigenvectors in Hydrodynamics}
\label{appendixtodo}
In this appendix, we discuss two simple examples
within the hydrodynamic framework. They serve the
purpose of putting our results for the amplitudes
in section~\ref{sec:tomog} into context and
connecting them to the existing literature.

\subsection{The purely diffusive mode in normal fluids}
Let us consider the longitudinal sector of a conformal charge fluid with conserved momentum.
Following the same notations as \cite{Davison:2015taa}, we may write the Fourier transformed hydrodynamic equations of motion in the form
\begin{equation}\label{mnm}
    \begin{pmatrix}
    -i\omega&i k&0\\
    ik\,\beta_1&\frac{k^2\pi s}{4\,(\mu\, \rho+sT)}-i\omega&0\\
    k^2\,\alpha_1\,\alpha_Q&\frac{ik\,\rho}{\mu\rho+sT}&-i\omega-\frac{k^2\,\alpha_1\alpha_Q\,(\mu\rho+sT)}{\rho}
    \end{pmatrix}\cdot\begin{pmatrix}
    \delta T^{tt}\\ \delta T^{tx}\\ \delta J^t
    \end{pmatrix}=\begin{pmatrix}
    0\\0\\0
    \end{pmatrix},
\end{equation}
where $\alpha_1=
 \left(\frac{\partial\mu}{\partial\epsilon}\right)_\rho-\frac{\mu}{T}\left(\frac{\partial T}{\partial\varepsilon}\right)_\rho, \,\beta_1=\left(\frac{\partial P}{\partial \epsilon}\right)_\rho, \,\alpha_Q=\sigma$. Note that $\delta T^{tt}$ and $\delta T^{tx}$ are not independent variables but related by the Ward identities~\eqref{eq:wardrel1}. Instead of an independent hydrodynamic variable, $\delta T^{tx}$ should be viewed as an auxiliary variable to reduce the generalized eigenvalue problem to first order in $\omega$ as explained below eq.~\eqref{mae}.

Solving the hydrodynamic equations as a generalized eigenvalue problem with respect to the frequency $\omega$, we find a pair of sound modes and a purely diffusive mode in agreement with the results in the main text, \textit{i.e.}
\begin{equation}
    \omega_{1,2}=\pm \sqrt{\beta_1}k-i\,\frac{\pi\,s}{8\,(sT+\mu\rho)}k^2+o(k^3)\,,\quad \omega_3=i \frac{\alpha_1\,(sT+\mu\rho)\,\alpha_Q}{\rho}\,k^2+o(k^3)\,,
\end{equation}
together with their corresponding eigenvectors
\begin{equation}
    v_{1,2}=\begin{pmatrix}
    \frac{\mu\rho+sT}{\rho}\\\mp\sqrt{\beta_1}(\frac{\mu\rho+sT}{\rho})+i\frac{\pi s}{8\rho}k\\1
    \end{pmatrix}+o(k^2)\,,\quad v_3=\begin{pmatrix}
    0\\0\\1
    \end{pmatrix}+o(k^2).\label{eq:eigenv3}
\end{equation}
Importantly, this means that the eigenvector corresponding to the purely diffusive mode, $v_3$, consists only of $\langle\delta J^t\rangle$ to first order in $k$, exactly as we observe in the left plot of fig.~\ref{fig:ampnormal}. Furthermore, we see that the ratio of the contributions to first sound from $\langle\delta T^{tt}\rangle$ and $\langle\delta J^t\rangle$ is given by $\frac{\mu\rho+sT}{\rho}$, exactly as we observe in the middle plot of fig.~\ref{fig:ampnormal}.

\begin{figure}[ht!]
     \centering
      \includegraphics[width=0.6\linewidth]{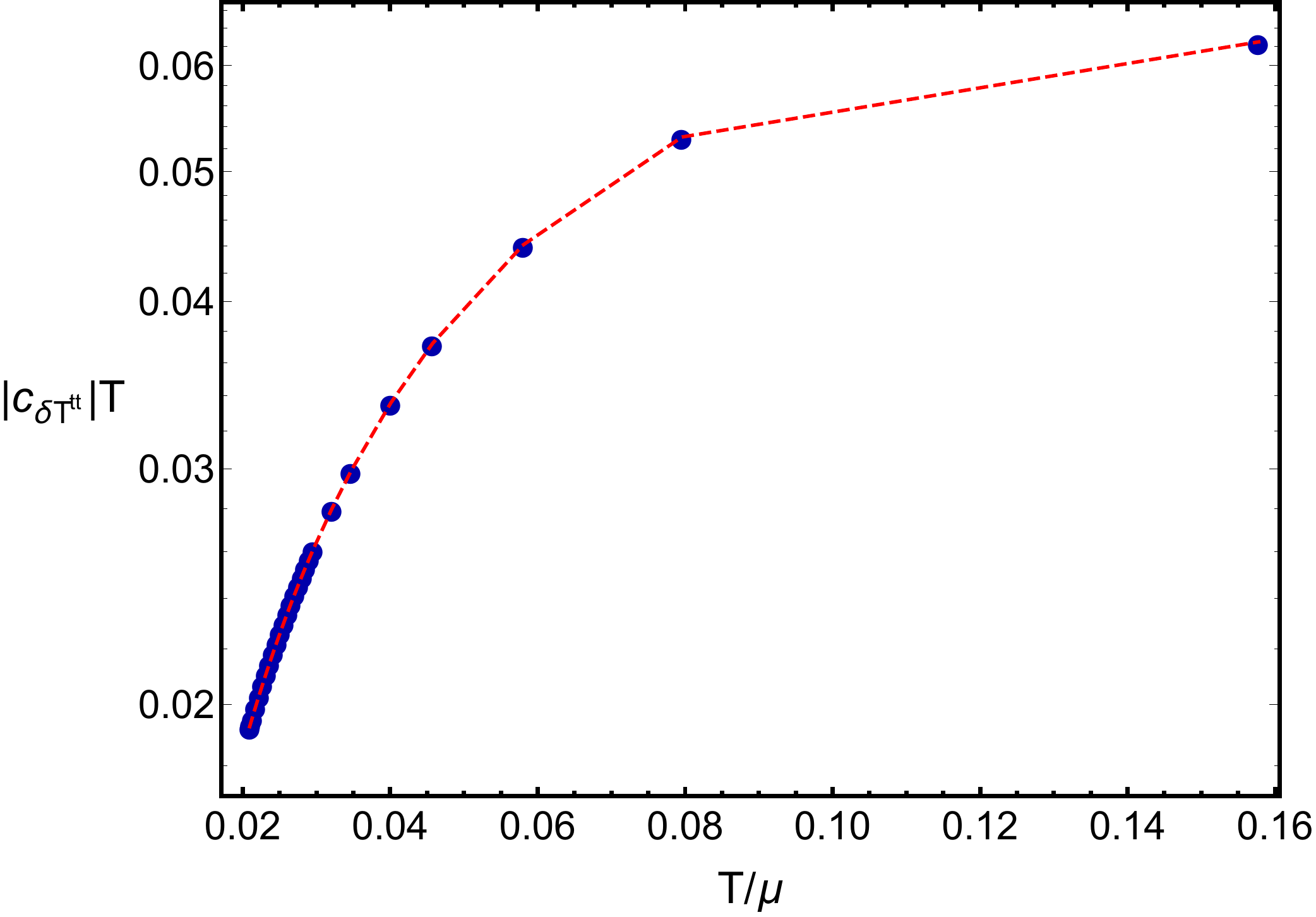}
     \caption{The contribution $c_{\delta T^{tt}}$ (eq.~\eqref{coeffdef}) of the $\delta T^{tt}$ fluctuations to the incoherent current $\tilde v_3$. The blue points are numerically extracted by inverting the matrix of the amplitudes of the different modes presented in the main text as in eq.~\eqref{eq:relation}. The red dashed line is the analytic result $\frac{\rho\,T}{sT+\mu\rho}$.}
     \label{proof}
 \end{figure}

Let us now explain why our results are not in contradiction with those in~\cite{Davison:2015taa} and how they relate to them. In~\cite{Davison:2015taa} it was stated that the purely diffusive mode is carried by the incoherent current $\delta Q^\text{diff}\equiv\delta J^t-\frac{\rho}{sT+\mu\rho}\delta T^{tt}$. To get to this result, we have to perform a basis transformation and consider
\begin{equation}
   \begin{pmatrix}
    \tilde v_1&\tilde v_2&\tilde v_3
    \end{pmatrix}=  \begin{pmatrix}
    v_1& v_2& v_3
    \end{pmatrix}^{-1}\begin{pmatrix}\delta T^{tt}\\\delta T^{tx}\\\delta J^t\end{pmatrix}\,. \label{eq:relation}
\end{equation}
Notice that, in this basis,
the diffusive mode corresponds to $\tilde v_3=\delta J^t-\frac{\rho}{sT+\mu\rho}\delta T^{tt}$. In the language of \cite{Davison:2015taa}, $\tilde v_3$ is the incoherent current. Note that the relative coefficient in $\tilde v_3$ is exactly the inverse of the relative coefficient between $\delta T^{tt}$ and $\delta J^t$ in the sound mode. For simplicity, we define the coefficients $c_I$ appearing in the decomposition of the various eigenvectors in terms of the basis of operator fluctuations using:
\begin{equation}
   \tilde v_n\,=\,\sum_I\,c_I\,\delta \mathcal{O}^I \,,\label{coeffdef}
\end{equation}
where $n$ labels the eigenvectors and $I$ the basis of operators. The coefficients $c_I$ are exactly the quantities discussed in \cite{Davison:2015taa} and displayed in figure \ref{proof}, \ref{pic:incoherenapp} and \ref{pic:relcoef}.
 
 The remaining vectors in \eqref{eq:relation} are
 \begin{align}
     &\tilde v_{1,2}=\frac{\rho  \left(\mp\frac{i \pi  k s}{\sqrt{\beta_1}}+8 \mu  \rho +8 s T\right)}{16 (\mu  \rho +s T)^2}\delta T^{tt}\mp\frac{\rho  \left(\pi ^2 k^2 s^2+128 \beta_1 (\mu  \rho +s T)^2\right)}{256 \beta_1^{3/2} (\mu  \rho +s T)^3}\delta T^{tx}+\ldots
 \end{align}
 and they correspond to the first sound mode. Using the Ward Identity relating $\delta T^{tt}$ and $\delta T^{tx}$ and assuming that the modes corresponding to $\tilde v_{1,2}$ (sound modes) are not excited, \textit{i.e.} $\tilde v_1,\tilde v_2=0$, one immediately obtains the condition $\delta T^{tt}=0$. This implies that only charge can fluctuate and leads to $\tilde v_3=\delta J^t$, which matches exactly our result presented in the main text shown in fig.~\ref{fig:ampnormal}.
In a nutshell, the main physical difference between the approach of \cite{Davison:2015taa} and ours lies in considering the response of the system
 to different perturbations.
 While in section~\ref{sec:tomog} we studied the support of the different hydrodynamic modes when only one of those modes is switched on,
 ref.~\cite{Davison:2015taa} considered the question
 for a generic perturbation in which all modes
 are excited. Obviously, as shown in eq.~\eqref{eq:relation}, the information of the two approaches can be extracted from the same original hydrodynamic matrix (\textit{e.g.} eq.~\eqref{mnm}). As a proof of that, we can invert the matrix of the amplitudes of the eigenvectors, project it on the basis of the fluctuations and obtain exactly the $\frac{\rho}{sT+\mu\rho}$ factor derived in \cite{Davison:2015taa}. The results of this procedure are shown in fig.~\ref{proof}. \\
 
 \subsection{The diffusive modes in a normal fluid with momentum dissipation}\label{appen:disc}
 We can perform the same analysis for the hydrodynamic modes in the longitudinal sector when momentum is dissipated at a rate $\sim \alpha/T$. This section may be viewed as a supplemental discussion of the data presented in fig.~\ref{new}. With momentum dissipation, the only conserved quantities are energy and electric charge. This leads to a $2
 \times 2$ hydrodynamic system in terms of energy and charge fluctuations $\delta T^{tt}, \delta J^t$ which, in matrix formalism, takes (in the conventions of~\cite{Hartnoll:2014lpa}) the following form
 \begin{equation}\label{easy}
    \begin{pmatrix}
 \frac{k^2 (\chi (\bar\alpha  \mu +\bar\kappa)-\zeta (\mu  \sigma +\bar\alpha  T))}{\chi c_\mu-T \zeta^2}-i \omega  & \frac{k^2 \left(\bar\alpha  \left(c_\mu T-\chi \mu ^2\right)-\bar\kappa (\chi \mu +T \zeta)+\mu  \sigma  (c_\mu+\mu  \zeta)\right)}{\chi c_\mu-T \zeta^2} \\
 \frac{k^2 (\bar\alpha  \chi-\sigma  \zeta)}{\chi c_\mu-T \zeta^2} & \frac{k^2 (\sigma  (c_\mu+\mu  \zeta)-\bar\alpha  (\chi \mu +T \zeta))}{\chi c_\mu-T \zeta^2}-i \omega 
    \end{pmatrix}\cdot\begin{pmatrix}
    \delta T^{tt}\\  \delta J^t
    \end{pmatrix}=\begin{pmatrix}
    0\\0
    \end{pmatrix}.
\end{equation} \begin{figure}[ht!]
     \centering
     \includegraphics[width=0.43\linewidth]{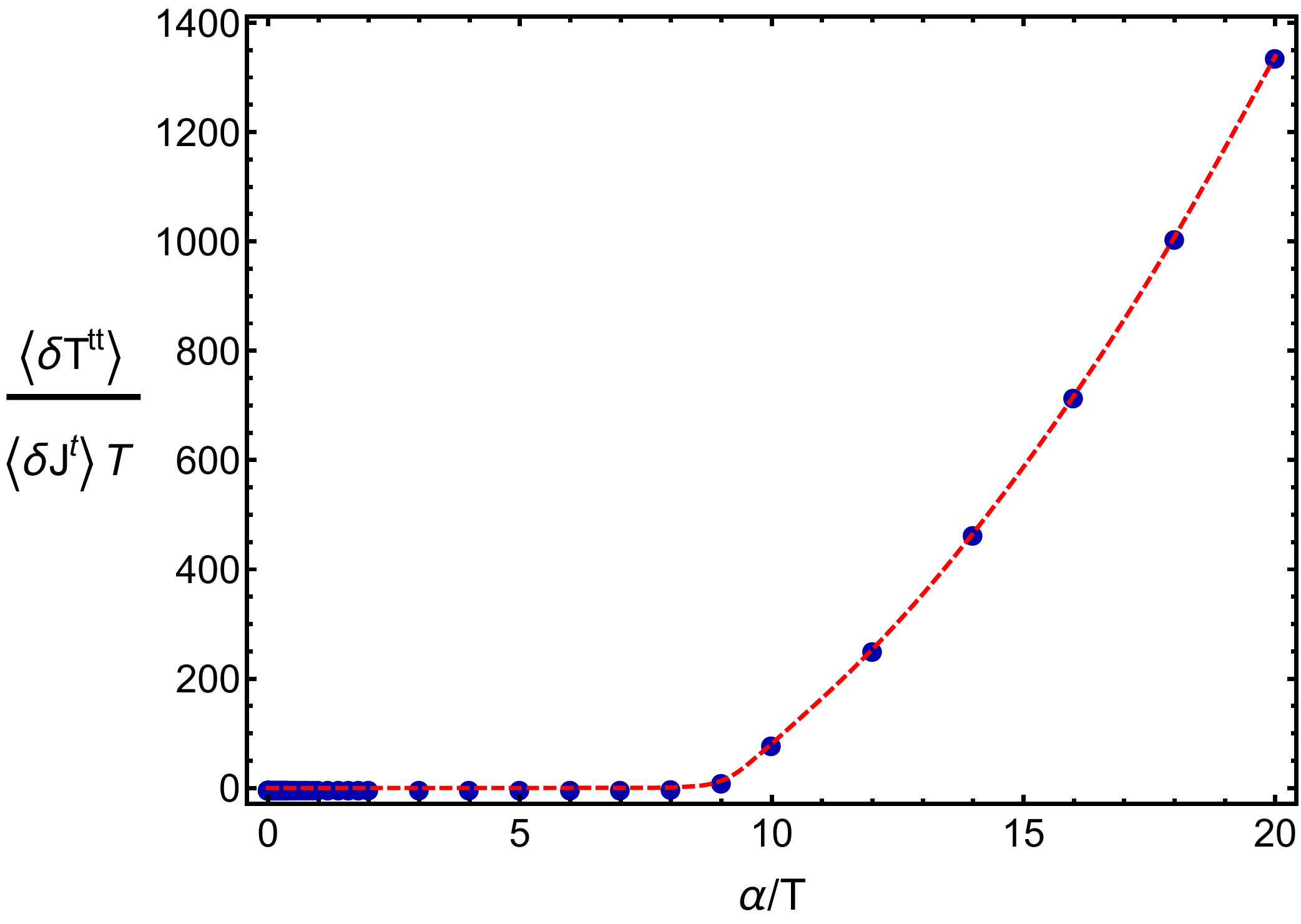}\qquad
      \includegraphics[width=0.43\linewidth]{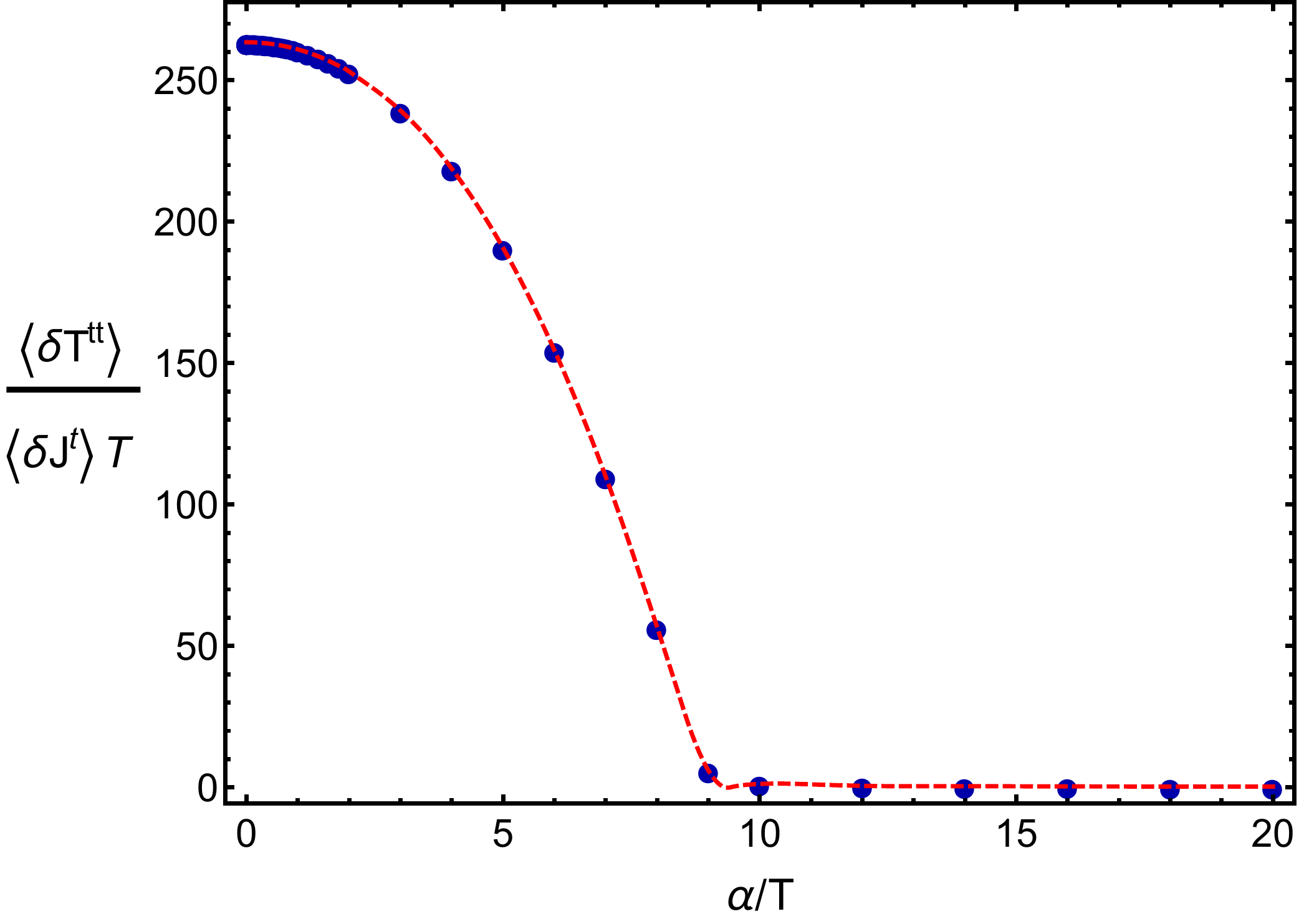}
     
     \vspace{0.2cm}
     
        \includegraphics[width=0.43\linewidth]{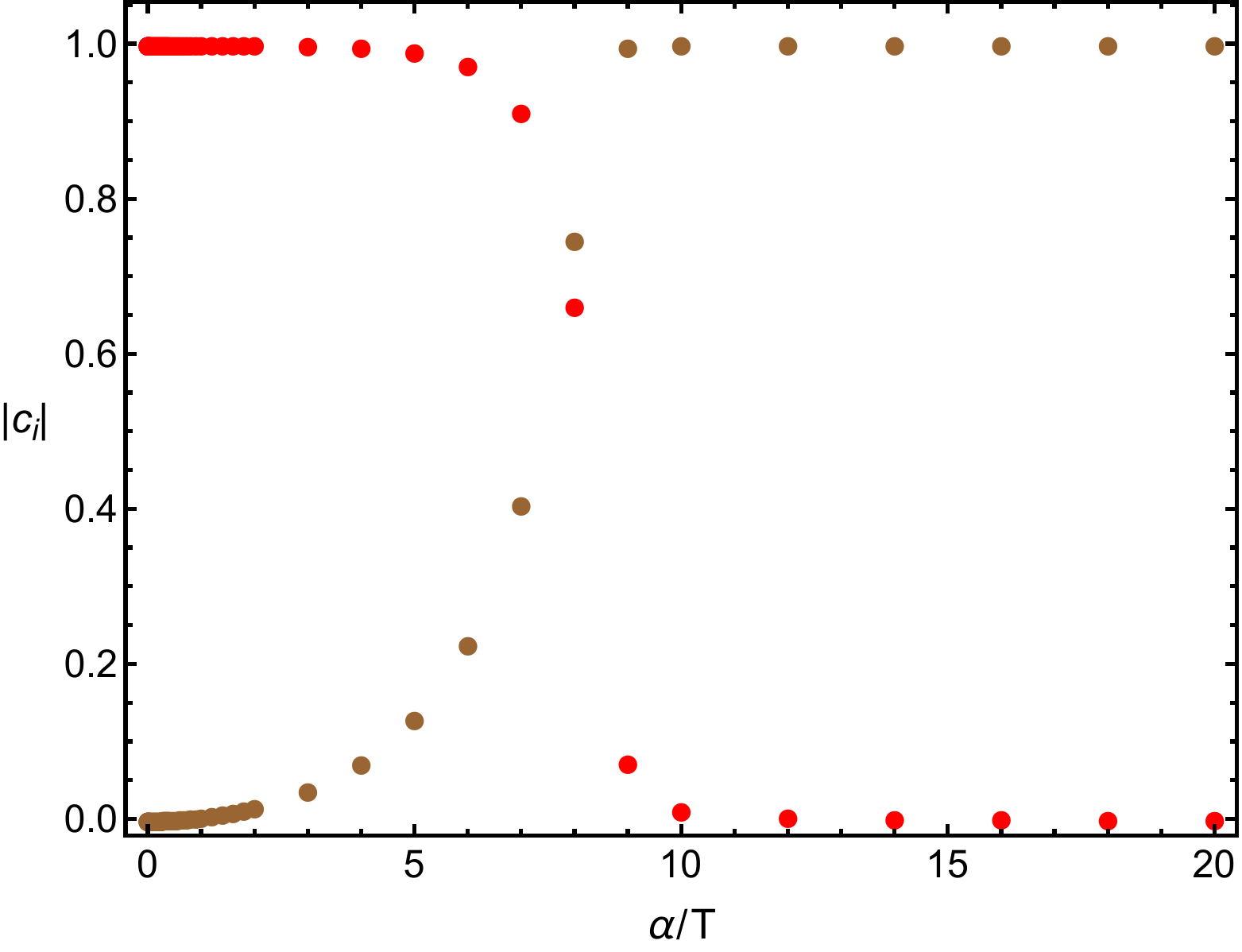}\qquad
      \includegraphics[width=0.43\linewidth]{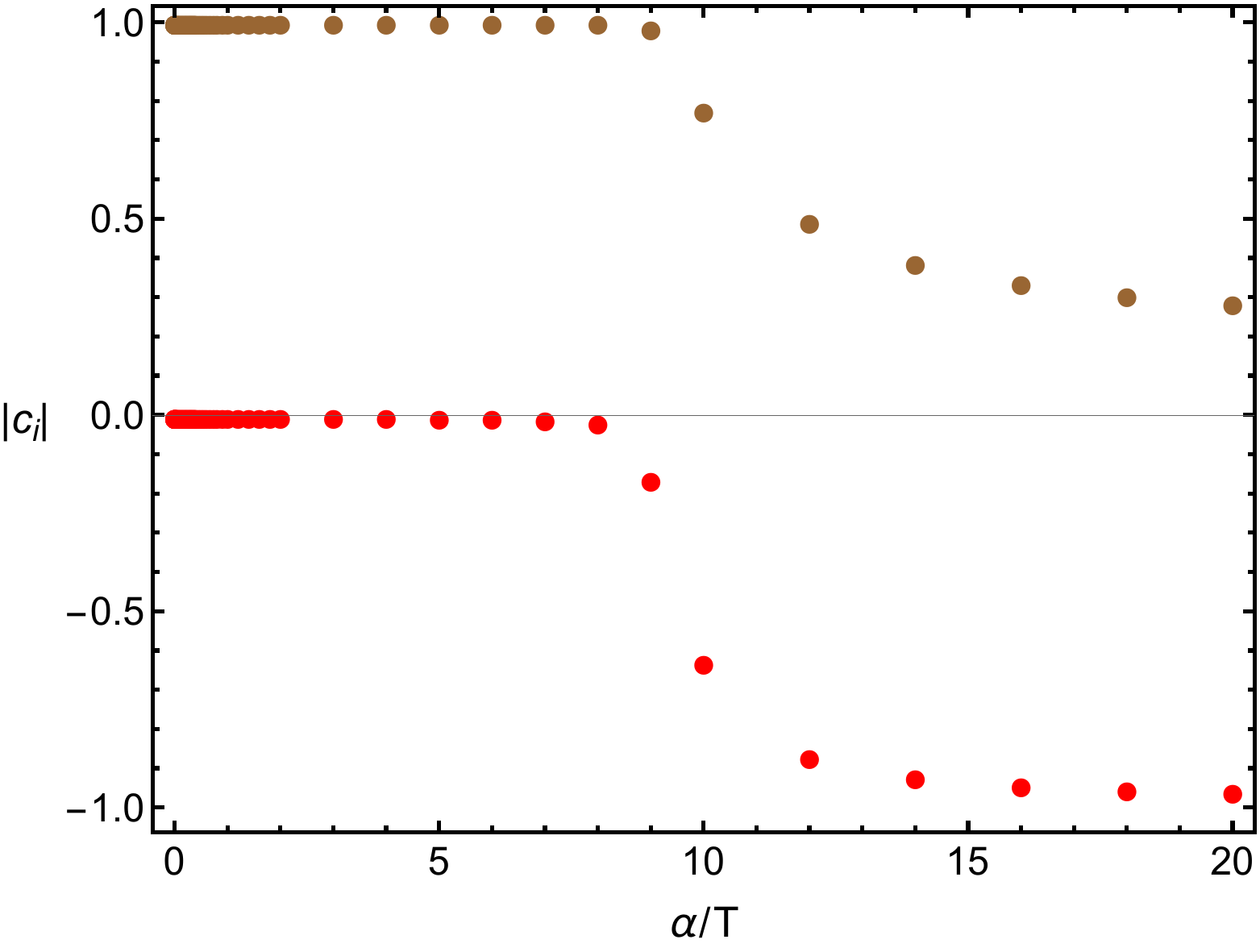}
     \caption{\textbf{Top: }The relative amplitudes of $\langle \delta T^{tt}\rangle$ and $\langle \delta J^t\rangle$ for the two thermoelectric diffusion modes at $T/\mu=5$ (in blue, corresponding to the data in fig.~\ref{new}) and the hydrodynamic prediction for the ratio (dashed red line). \textbf{Bottom:} The normalized, dimensionless coefficients $c_{\delta J^t}\,T^2$ and $c_{\delta T^{tt}}\, T^3$ in the basis of the eigenvectors from eq.~\eqref{coeffdef}. In the incoherent limit $\alpha\gg\mu$, $\delta J_+\to \delta  J^t$ and $\delta J_-\sim \delta T^{tt}+\gamma_-\,\alpha\delta J^t$. This implies that $\delta J^-$ is purely $\delta J^t$ and the other mode is still a mixture of $\delta T^{tt}$ and $\delta J^t$. The relative coefficient of the bottom right plot is presented in fig.~\ref{pic:relcoef}.}
     \label{pic:incoherenapp}
 \end{figure} 
We have defined: $\sigma$ the electric conductivity, $\chi\equiv \partial \rho/\partial \mu$ the charge susceptibility, $\bar{\kappa}$ the thermal conductivity, $\bar{\alpha}$ the thermoelectric conductivity, $\zeta=\partial s /\partial \mu$ and the specific heat $c_\mu=T \partial s/\partial T$. All these coefficients are known analytically for the linear axion model, see for example \cite{Baggioli:2017ojd}.\\
At this point, we can perform the same analysis as before and compute the eigenvectors $J_\pm$ for the matrix eq.~\eqref{easy}. The matrix is diagonal in the basis of the eigenvectors $\delta J_\pm=a_\pm\,(\delta T^{tt}+\gamma_\pm\,\alpha \,\delta J^t),$ where $a_\pm$ is just an overal normalization and $\gamma_\pm=-\frac{3\epsilon}{4\alpha\,\rho}\left(1\pm\sqrt{1+\frac{16\alpha^2\,\rho^2}{9\epsilon^2}}\right)$~\cite{Davison:2015bea}, with $\epsilon$ being the energy density and $\alpha$ the dimensionful strength of momentum dissipation. We present the amplitudes and coefficients $c_I$ in fig.~\ref{pic:incoherenapp}. In fig.~\ref{pic:relcoef}, we explicitly verify that the relative coefficient matches the analytic expression of~\cite{Davison:2015bea}.\begin{figure}[ht!]
     \centering
      \includegraphics[width=0.6\linewidth]{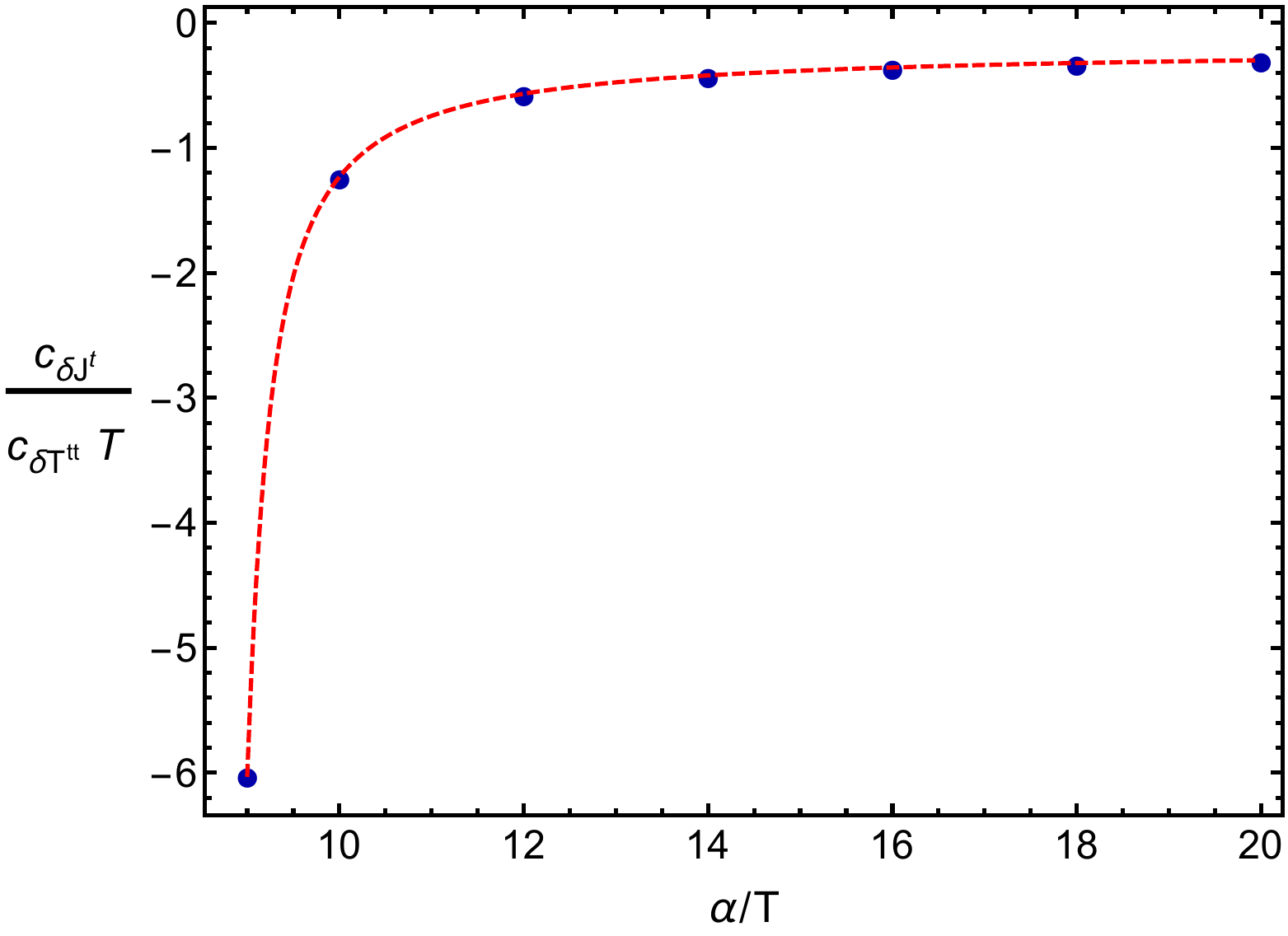}
     \caption{The relative coefficient of $\delta J^t$ to $\delta T^{tt}$ in the basis of the eigenvectors $J_\pm$ in the incoherent limit (cf. eq.~\eqref{coeffdef}). The blue points are numerically extracted by computing the coefficients in the basis of the eigenvectors of the data presented in the top right panel of fig.~\ref{pic:incoherenapp}. The red dashed line is the analytic result for $\delta J_-\sim \delta T^{tt}+\gamma_-\,\alpha\,\delta J^t$.}
     \label{pic:relcoef}
 \end{figure}
\color{black}

\section{Numerical methods}
\label{app:numerics}
 In this section, we briefly describe the numerical methods we employed to solve the differential equations in this work following~\cite{Grieninger:2020wsb,Ammon:2016fru,Grieninger:2017jxz} (for a detailed introduction see~\cite{Boyd1989ChebyshevAF}). Both, the background equilibrium solutions as well as the solutions to the linearized perturbations about the equilibrium state are obtained by means of pseudo-spectral methods. The idea of spectral methods is to approximate the numerical solution in terms of basis functions on a discretized grid. Throughout this work, we choose Chebychev polynomials as basis functions and a Chebychev-Lobatto grid to discretize the radial direction. Spectral methods solve the equations of motions globally which is a big advantage compared to shooting methods or finite-differences where we have to vary the initial conditions on one side of the domain until we find the desired boundary values at the other end of the interval.  More importantly, spectral method are highly accurate and have a fast convergence rate and are thus perfectly suited for problems in numerical holography.
 
 Computing the static background and two-point functions in terms of pseudo-spectral methods is fairly standard and we thus only comment on how to obtain the quasi-normal modes. After obtaining the static background solution, we want to investigate time- and space dependent (linearized) fluctuations about that background to compute (a) the quasi-normal modes and (b) the retarded Green's functions for the transport coefficients. The coupled ordinary differential equations for the linearized fluctuations are generally of the form
  \begin{equation}
(\bm{C}(k)\,\omega^2+\bm{A}(k)\,\omega-\bm{B}(k))\,\bm{x}(\omega,k)=0,\ \ \ (\bm{F}(\omega)\,k^2+\bm{D}(\omega)\,k-\bm{E}(\omega))\,\bm{x}(\omega,k)=0\label{mae}
 \end{equation}
 where $\bm{A}$ to $\bm{F}$ are differential operators and the vector $\bm{x}$ consists of all the fluctuations $\bm{x}=\{h_{tt},\,h_{tx},\,h_{xx},\,h_{yy},\,a_t,\,a_x,\,\delta \psi_1,\, \delta \psi_2,\,\delta\phi_x\}$.
 We may recast both problems in terms of a generalized eigenvalue problem with respect to the quasi-normal frequency $\omega$ or the momentum $k$ by introducing auxiliary functions of the form $\tilde{\mathfrak{f}}= \omega\, \mathfrak{f}$ and $\tilde{\mathfrak{f}}= k\, \mathfrak{f}$ so that we find 
   \begin{equation}
(\tilde{\bm{A}}\,\omega-\tilde{\bm{B}})\,\tilde{\bm{x}}_1=0,\ \ \ (\tilde{\bm{D}}\,k-\tilde{\bm{E}})\,\tilde{\bm{x}}_2=0
 \end{equation}
 where 
 \begin{equation}
     \tilde{\bm{x}}_1=\{h_{tt},\,h_{tx},\,h_{xx}+h_{yy},\,h_{xx}-h_{yy},\,a_t,a_x,\,\delta \psi_1,\, \delta \psi_2,\,\delta\phi_x,\, \tilde{h}_{xx}+ \tilde{h}_{yy}\}
 \end{equation}
 and
 \begin{equation}
     \tilde{\bm{x}}_2=\{h_{tt},\,h_{tx},\,h_{xx},\,h_{yy},\,a_t,a_x,\,\delta \psi_1,\, \delta \psi_2,\,\delta\phi_x,\,\tilde{h}_{tt},\,\tilde{h}_{yy},\,\tilde{a}_{t},\,\delta \tilde{\psi}_1,\, \delta \tilde{\psi}_2\}\,.
 \end{equation}
 The tilde in $\tilde{\bm{x}}_1$ and $\tilde{\bm{x}}_2$ denote the auxiliary functions. 
 
 By solving the generalized eigenvalue problem, we find for each eigenvalue $\omega_n$ (or $k_n$) an eigenvector $\tilde{\bm{x}}$. Finally, to resolve the strong gradients at the horizon at low temperatures accurately we use up to 350 gridpoints in order to satisfy the equations of motion and constraints with accuracy better than $10^{-12}$.\\ The Chebychev coefficients of the slowest convergent solution drop below $10^{-13}$ in this case. 
\bibliographystyle{JHEP}
\bibliography{sound}

\end{document}